\documentclass[12pt,preprint]{aastex}
\usepackage{color}
\usepackage{graphicx}
\usepackage{amsmath}

\shorttitle{BH X-ray Polarization in Thermal State}
\shortauthors{Schnittman \& Krolik}

\begin{document}

\title{X-ray Polarization from Accreting Black Holes: II. The Thermal State}

\author{Jeremy D.\ Schnittman}
\affil{Department of Physics and Astronomy,
Johns Hopkins University\\
Baltimore, MD 21218}
\email{schnittm@pha.jhu.edu}

\and
\author{Julian H.\ Krolik}
\affil{Department of Physics and Astronomy,
Johns Hopkins University\\
Baltimore, MD 21218}
\email{jhk@pha.jhu.edu}

\begin{abstract}

We present new calculations of X-ray polarization from black hole (BH)
accretion disks in the thermally-dominated state, using a Monte-Carlo
ray-tracing code in full general
relativity. In contrast to many previously published studies, our
approach allows us to include returning radiation that is deflected by
the strong-field gravity of the BH and scatters off of the disk before
reaching a distant
observer.  Although carrying a relatively small fraction of the total
observed flux, the scattered radiation tends to be highly polarized
and in a direction perpendicular to the direct radiation. For
moderately large spin parameters $(a/M \gtrsim 0.9)$, this scattered
returning radiation dominates the polarization signal at energies above the
thermal peak, giving a net rotation in the polarization angle of
$90^\circ$. We show how these new features of the polarization 
spectra from BHs in the thermal state may be developed into a powerful tool for
measuring BH spin and probing the gas flow in the innermost
disk. In addition to determining the emission profile, polarization
observations can be used to constrain other properties of the
system such as BH mass, inclination, and distance.
New instruments currently under development should be able to exploit this
tool in the near future.

\end{abstract}

\keywords{black hole physics -- accretion disks -- X-rays:binaries}

\section{INTRODUCTION}\label{intro}

A recent flurry of new mission proposals has renewed interest in
X-ray polarization from a variety of astrophysical sources. The
Gravity and Extreme Magnetism SMEX ({\it GEMS}) mission\footnote{\tt
  heasarc.gsfc.nasa.gov/docs/gems}, which has
recently been approved for phase A study in
the latest round of NASA Small Explorer proposals, should be able to
detect a degree of polarization $\delta \lesssim 1\%$ for a flux of a few
mCrab \citep{black:03, bellazzini:06}. A similar detector for
the International X-ray Observatory ({\it IXO}) could achieve
sensitivity roughly $10\times$ greater ($\lesssim 0.1\%$ degree of
polarization; Jahoda et al.\ 2007, Costa et al.\ 2008). A large number
of galactic and extra-galactic sources are expected to be polarized
at the $\delta \gtrsim 1\%$ level, including black holes, magnetars,
pulsar wind nebulae, and active galactic nuclei. In this paper, we focus on
accreting stellar-mass black holes (BHs) in the thermal state, which are
characterized by a broad-band spectrum peaking around 1 keV. The
typical level of polarization from these sources should be a few
percent in the $1-10$ keV range, depending on BH spin and the
inclination angle of the accretion disk. 

Symmetry arguments demand that in the flat-space (Newtonian) limit, the
observed polarization from the disk must 
be either parallel or perpendicular to the BH/disk rotation axis.
However, the effects of relativistic beaming,
gravitational lensing, and gravito-magnetic frame-dragging can combine to give
a non-trivial net rotation to the integrated polarization
vector.  Early work exploring these effects
\citep{stark:77,connors:77,connors:80}
showed that they create changes in the angle and degree of polarization
that are strongest for higher photon energy.  The reason for this trend
is that the temperature in a standard accretion disk increases as the BH is
approached, and that is, of course, where the relativistic effects are
strongest.  Since those first efforts, there have been studies of the predicted
polarization from more general accretion geometries, including UV and X-ray
emission from AGN disks \citep{laor:90,matt:93} and ``lamp post''
models for irradiating the accretion disk with a non-thermal source above the
plane \citep{dovciak:04}.  Quite recently, \citet{dovciak:08}
investigated the effect of atmospheric optical depth on the
polarization signal, and \citet{li:08} applied the original
calculations of thermal X-ray polarization to the problem of measuring
the inclination of the inner accretion disk. 

Despite this relatively large body of literature dedicated to X-ray
polarization from BH accretion disks, to the best of our knowledge,
all previous work has modeled the relativistic effects by calculating
the transfer function along geodesics between the observer and
emitter. By its very nature, this method precludes the possibility of
including the effects of returning radiation---photons emitted from one
part of the disk and bent by gravitational lensing so that they are
absorbed or reflected from another part of the disk \citep{cunningham:76}.
Using new methodology, in this paper we show the contribution from this
returning radiation is very important to the observed
polarization. The details of our method are described in
\citet{schnittman:08} (hereafter ``Paper I''). The most important
feature of our approach is that the photons are traced {\it from} the
emitting region in all directions, either returning to the disk,
scattering through a corona, getting captured by the BH, or eventually
reaching a distant observer. It should be noted that a sort of
hybrid technique was used in \citet{agol:00} to estimate the effects
of returning radiation on polarization by situating an observer in the
disk plane and shooting rays backwards to see how much flux was
incident from other regions of the disk, but this approach cannot
account for {\it multiple} scattering events, where the returning
radiation passes over the BH more than once.

The inclusion of returning radiation leads to the most important of the new
results presented in this paper, namely a transition between
horizontal- and vertical-oriented polarization as the photon energy
increases. At low energies we reproduce the ``Newtonian'' result of
a semi-infinite scattering atmosphere emitting radiation weakly polarized
in a direction parallel to the emission surface, an orientation we
call {\it horizontal} polarization \citep{chandra:60}. At higher
energy, corresponding to 
the higher temperature of the inner disk, a greater fraction of the
emitted photons returns to the disk and is then scattered to the
observer. These scattered photons have a high degree of polarization
and are aligned parallel to the disk rotation axis ({\it vertical}),
as projected onto the image plane. At the transition point between
horizontal and vertical polarization, the relative contributions of
direct and reflected photons are nearly equal, and no net polarization is
observed. 

Since the effects of returning radiation are greatest for
photons coming from the innermost regions of the disk, the predicted
polarization signature is strongly dependent on the behavior of gas
near and inside the inner-most stable circular orbit (ISCO). Thus,
polarization observations could be used to measure the spin of the
black hole and also constrain the dissipation profile of the inner
disk. For systems where the BH mass, distance, and disk
inclination are known, a single polarization observation can
constrain the emission profile at least well as fitting the thermal
continuum spectrum \citep{gierlinski:01,davis:06,shafee:06}. When
these system parameters are not known {\it a priori}, polarization
can be used in many cases to measure them simultaneously with the
emission profile (albeit with lower confidence than for the
sources with known priors).

The structure of this paper is as follows: after briefly describing
our methods in Section \ref{methods}, we review in Section \ref{direct}
some of the classical results of earlier polarization
calculations from thermal disks, including only the radiation that
reaches the observer directly from the disk. In Section \ref{returning},
we include the effects of returning radiation, focusing on new features
in the polarization spectrum. Sections \ref{results} and
\ref{parameters} contain a
detailed discussion of our results and how they depend on model
parameters, and in Section \ref{discussion} we conclude. 

\section{METHODOLOGY}\label{methods}

The inner disk of a stellar-mass BH binary can be well described
locally as a plane-parallel, electron scattering-dominated atmosphere
\citep{shakura:73,novikov:73}.
In the local fluid frame, the classical result of \citet{chandra:60}
applies: the polarization vector of the emitted
radiation is oriented parallel to the disk plane and normal to
the direction of propagation. The degree of polarization varies from
zero for photons emitted normal to the disk surface up to $\sim 12$\%
for an inclination angle of $90^\circ$. In addition to the
polarization effects, the scattering of the outgoing flux causes
limb darkening, effectively focusing the emitted radiation in the
direction normal to the disk surface. Both the degree of polarization
and the limb
darkening/brightening factors are tabulated as a function of emission
angle in Table XXIV in \citet{chandra:60}.

For the local emission spectrum, we assume a diluted black-body
spectrum with hardening factor $f=1.8$ \citep{shimura:95}.  As
demonstrated by \citet{davis:05}, this is a reasonable approximation
for disks around stellar-mass black holes because the temperature
is so high that there is little opacity due to anything but Compton
scattering and free-free absorption.  For a
locally emitted flux $\mathcal{F}$, the effective temperature is defined as
\begin{equation}\label{T_eff}
T_{\rm eff} \equiv \left(\frac{\mathcal{F}}{\sigma}\right)^{1/4},
\end{equation}
where $\sigma$ is the Stefan-Boltzmann constant. The spectral
intensity of the diluted emission is then given by
\begin{equation}\label{Inu_dilute}
I_\nu = \frac{1}{f^4}B_\nu(f\, T_{\rm eff}),
\end{equation}
with $B_\nu$ the black-body function. For the emitted flux function
$\mathcal{F}(R)$, we consider two different models: that given by
\citet{novikov:73} for a zero-stress inner boundary condition at the
ISCO, and a quasi-Newtonian
expression $\mathcal{F}(R) \sim R^{-3}$ that extends the classical scaling at
large radius all the way down to the horizon. Recent global
magneto-hydrodynamic (MHD)
simulations suggest that reality lies somewhere in between these two
models \citep{noble:08}. Outside of the ISCO, the gas follows
circular, planar geodesic orbits as in \citet{novikov:73}. Inside the
ISCO, the fluid continues along plunging trajectories with constant
energy and angular momentum, resulting in a rapid decline of surface
density with decreasing radius.

As described in Paper I, at each point in the disk, multi-wavelength
photon packets are emitted in all directions, with appropriate intensity
weights and polarizations as a function of emission angle. We use a
Monte Carlo algorithm with typically $10^5-10^6$ rays traced per
radius, with 500 radial bins spaced logarithmically between the
horizon and $R=100M$.
Since the photon trajectories originate at the disk, the
observer's inclination angle is not specified; rather, all photons
are followed until they get captured by the BH, return to the disk
plane, or reach ``infinity,'' in practice a spherical hypersurface at
large radius. At that point, the photons are binned by observer
inclination, effectively simulating all inclinations
simultaneously. By calculating the angle that each photon makes with
the detector sphere, we are also able to create an image of the
source, much like a classical pin-hole camera. 

Each photon
packet contains a complete broad-band spectrum initially described by
equation (\ref{Inu_dilute}). Since Thomson scattering
in the disk atmosphere is frequency-independent over the range of
relevant X-ray energies ($\sim 0.1-10$ keV for thermal emission from a
solar-mass BH), the photon packet can be described by a
single polarization amplitude and direction. 

Following \citet{connors:80}, we characterize the polarization of each
photon packet by the normalized Stokes parameters:
\begin{subequations}\label{X_Y}
\begin{eqnarray}
X &=& Q/I, \\
Y &=& U/I,
\end{eqnarray}
\end{subequations}
and
\begin{subequations}\label{delta_psi}
\begin{eqnarray}
\delta &=& (X^2+Y^2)^{1/2}, \\
\psi &=& \frac{1}{2}\tan^{-1}(Y/X),
\end{eqnarray}
\end{subequations}
where $\psi$ is the angle of polarization, ranging from
$-90^\circ$ to $+90^\circ$, and $\delta \le 1$ is the degree of
polarization, invariant along a geodesic.

The polarization vector $\mathbf{f}$ is a normalized space-like
4-vector perpendicular
to the photon propagation direction: $\mathbf{k}\cdot\mathbf{f}=0$ and
$\mathbf{f}\cdot\mathbf{f}=1$. It
is parallel-transported along the photon's null geodesic, conserving
the complex-valued Walker-Penrose integral of motion $\kappa$
\citep{walker:70}.  Analogous to the way that Carter's constant \citep{carter:68}
can be used to constrain four-velocity components,
$\kappa$ can be used to reconstruct the polarization vector at
any point along the geodesic.  The two orthonormality conditions
stated above, along with the real and imaginary parts of $\kappa$,
give a total of four equations for the four components of the
polarization vector $\mathbf{f}$. When a ray reaches an
observer at infinity, the polarization vector is projected onto
the detector plane defined by the basis vectors $\mathbf{e}_1$ and
$\mathbf{e}_2$ to give the polarization angle:
$\mathbf{f}=[0,\cos\psi\, \mathbf{e}_1+\sin\psi\, \mathbf{e}_2]$ (in
our convention, $\mathbf{e}_2$ is parallel to the projected rotation
axis of the BH and disk). From $\delta$ and $\psi$, as measured by the
detector, we can reconstruct the normalized Stokes parameters $X$ and $Y$ via
equation (\ref{delta_psi}). The observed spectrum $I_\nu$
is propagated along the geodesic path as well, containing within it the
emitted spectrum, modified by all relativistic effects as well as any
energy change that may occur when returning radiation scatters off disk
electrons.  $I_\nu$ provides the absolute normalization that, when
applied to $X$ and $Y$ allows $Q_\nu$ and $U_\nu$ to be recovered
for each photon bundle reaching the detector. Integrating over all
photon bundles gives the polarization degree $\delta_\nu$ and angle
$\psi_\nu$ as a function of energy (see Paper I for details). These are
ultimately the quantities that we wish to be able to measure and use
to probe the accretion geometry of the BH. 

\section{DIRECT RADIATION}\label{direct}

\begin{figure}
\caption{\label{direct_image} Ray-traced image of direct radiation
  from a thermal disk. The observer is located at an inclination of
  $75^\circ$ relative to the BH and disk rotation axis, with the gas
  on the left side of the disk moving towards the observer, which
  causes the characteristic increase in intensity due to relativistic beaming.
  The black hole has
  spin $a/M=0.9$, mass $M=10 M\odot$, and is accreting at $10\%$ of
  the Eddington limit with a Novikov-Thorne zero-stress emissivity
  profile, giving peak temperatures around 1 keV. The observed intensity is color-coded on a
  logarithmic scale and the energy-integrated polarization vectors are
  projected onto the image plane with lengths proportional to the
  degree of polarization.}
\begin{center}
\includegraphics*[52,450][410,720]{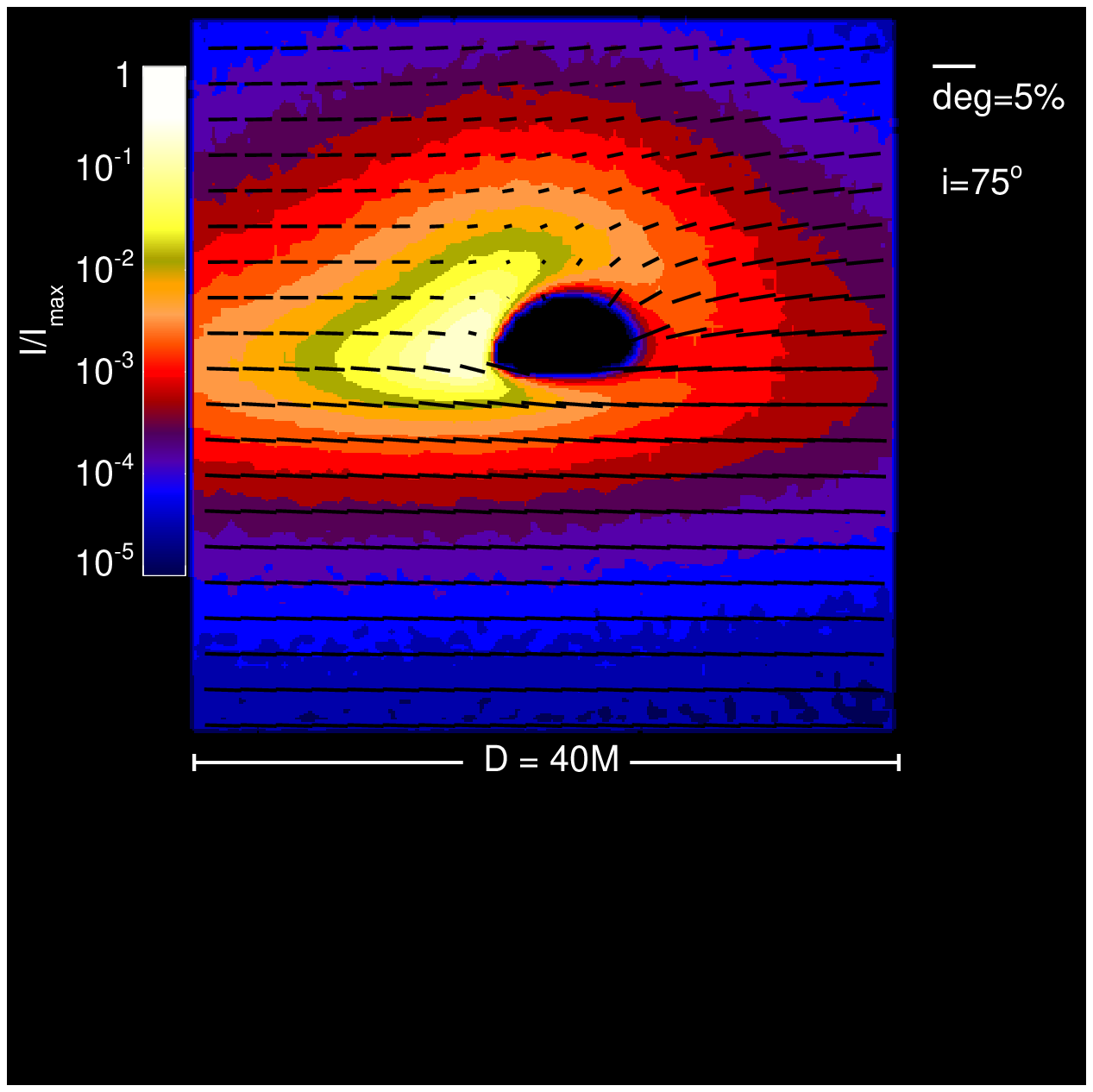}
\end{center}
\end{figure}

In Figure \ref{direct_image}, we show a simulated image of a
Novikov-Thorne accretion disk around a black hole
with spin parameter $a/M=0.9$ and luminosity $0.1 L_{E}$,
corresponding to a disk whose X-ray spectrum peaks around 1 keV for a
BH mass of $10 M_\odot$. With the observer at an
inclination of $i=75^\circ$, significant relativistic
effects are clearly apparent. The increased intensity on the left side of
the disk is due to special relativistic beaming of the gas moving
towards the observer, and the general relativistic light bending makes
the far side of the disk appear warped and bent up towards the
observer. Superposed on top of the intensity map is the polarization
signature, represented by small black vectors whose lengths are
proportional to the degree of polarization observed from that local
patch of the disk.  Far from the
black hole, the polarization is essentially given by the classical
result of \citet{chandra:60} for a scattering-dominated atmosphere:
horizontal orientation with $\delta \approx 4\%$ when $i=75^\circ$;
nearer to the black hole, a variety of relativistic effects alter
the polarization.

While the fundamental calculation of polarization from a thermal BH
accretion disk is not at all new \citep{connors:80}, to the best of
our knowledge this is the first published
polarization map as seen projected in the image plane. The usefulness
of such a map is primarily as a tool to understand how relativistic
effects determine the integrated polarization at different energies.

The two most prominent relativistic effects are gravitational lensing
and special relativistic beaming, both lowering the net level of
polarization seen by the observer. Gravitational lensing causes the
far side of the disk to appear warped up towards the observer, and
thus have a smaller effective inclination.  Relativistic beaming causes
photons emitted normal to the disk plane in the fluid frame to travel
forward in the direction of the local orbital motion when seen by
a distant observer; the result is a smaller effective
inclination and thus degree of polarization.  On the other hand,
photons emitted at high inclination angle in the local fluid frame
but against the direction of orbital motion are ultimately seen
by observers located at lower inclination, and therefore raise the polarization
in that direction.  Naturally, these relativistic effects are most
important close to the black hole, where the gas is also hottest and
the photons have the highest energies. All these effects are clearly visible in
Figure \ref{direct_image}, which shows a smaller degree of
polarization where the beaming is greatest (yellow region of high
intensity in the left of the image) and the lensing is strongest (just
above the center of the image). At the same time, the gas moving away
from the observer on the right of the image has an enhanced level of
polarization because the observer sees photons emitted at a larger
inclination in the fluid frame.

\begin{figure}
\caption{\label{direct_pol} Polarization degree and angle from a thermal
  disk as a function of observed photon energy, including only direct
  radiation. The BH parameters are as in Fig.\ \ref{direct_image}
  (dashed curves) and also include identical calculations corresponding to a
  non-spinning BH (solid curves). The angle of polarization is
  measured with respect to the horizontal axis in the image plane with
  point symmetry through the origin: $\psi = \psi-180^\circ$. }
\begin{center}
\scalebox{0.8}{\includegraphics{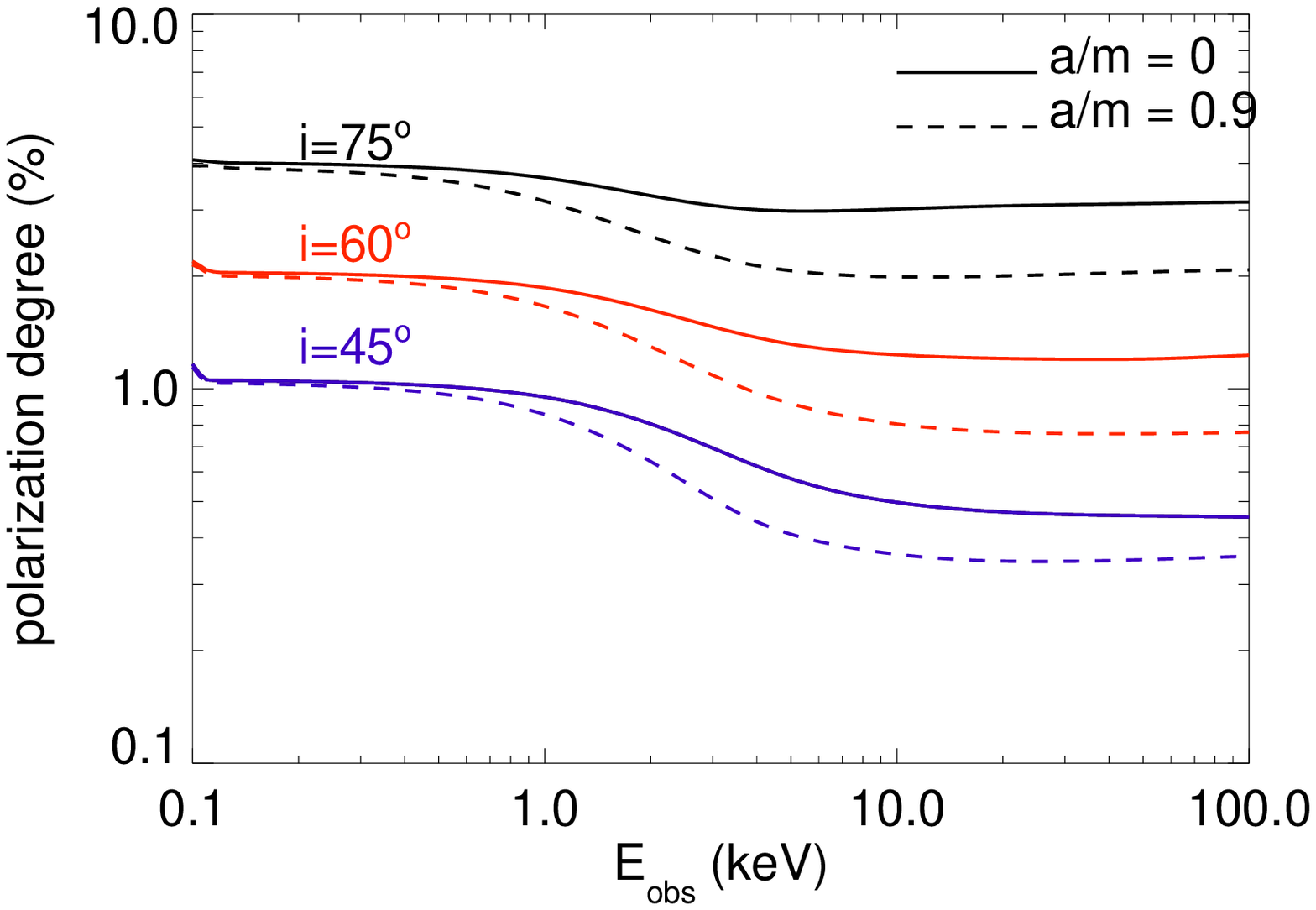}}
\scalebox{0.8}{\includegraphics{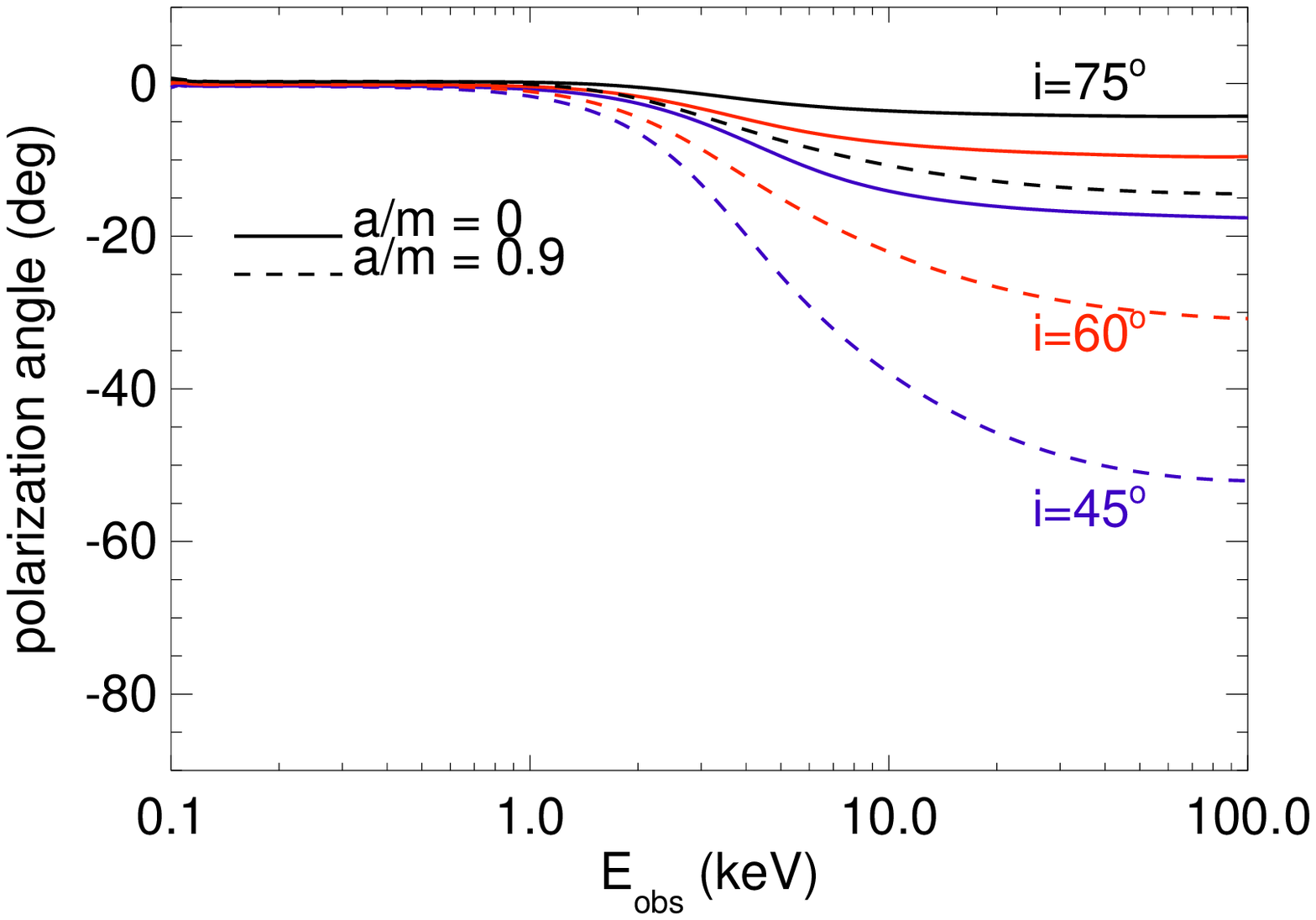}}
\end{center}
\end{figure}

Integrating over the entire disk, we can calculate the angle and
degree of polarization as a function of energy. Plotted in Figure
\ref{direct_pol}, these results essentially reproduce those of
\citet{connors:80}, who performed a similar calculation using semi-analytic
transfer functions from the observer to the disk. Again we see that at
low energies (dominated by emission from large disk radii), the
polarization is given by the
Chandrasekhar result, and is a function only of the disk inclination
angle. At higher energies, we begin to probe the inner regions of the
accretion disk and see the relativistic effects described above that lower the
total degree of polarization. In addition to reducing the polarization
amplitude, the strong gravity near the black hole also rotates the
angle of polarization due to the parallel transport: since the 
polarization vector $\mathbf{f}$ must remain perpendicular to the
photon momentum $\mathbf{k}$, as the geodesic path bends around the
BH, the polarization angle must also rotate. For example, the
high-energy photons beamed towards the observer 
in Figure \ref{direct_image} are also bent by the BH's
gravity, initially moving to the left before curving back to the
observer, thus producing the rotated polarization vectors seen just to
the left of the image center (the same lensing effect rotates the red-shifted
photons from the right side of the disk, but in an opposite direction). 

The angle of polarization as a function of energy is plotted in the 
lower panel of Figure \ref{direct_pol}.
For an inclined accretion disk with the black hole axis
projected onto the vertical, or y-axis, $\psi=0$ corresponds to
horizontal polarization parallel to the disk surface. From
equation (\ref{delta_psi}), we see there is a point symmetry in the
definition of $\psi$, giving $\psi=\psi \pm 180^\circ$. 
As can be seen in the simulated image of Figure \ref{direct_image},
the individual polarization vectors are rotated in the inner disk, due
largely to gravitational lensing and, to a lesser degree,
frame-dragging around a spinning black hole. While individual photons
can experience significant rotation, the net result is a modest
rotation of the total observed polarization in the clockwise
direction, giving $\psi < 0$ at higher energies. From Figure
\ref{direct_pol}, it is clear this effect is more pronounced for small
inclinations, where the low amplitude of polarization makes it easier
to ``overcome'' the classical result with additional relativistic
effects. Similarly, the polarization rotation is greater for more
rapidly spinning black holes, where the accretion disk extends closer
in towards the horizon and probes a stronger gravitational field. In
this way, it has previously been proposed that the polarization degree
and angle as a function of energy could be used to infer the spin of
the black hole in the thermal state \citep{connors:80, laor:90,
  dovciak:08}.

\section{RETURNING RADIATION: QUALITATIVE DISCUSSION}\label{returning}

When returning radiation is included, although little
changes in terms of the total observed spectrum, the polarization picture
(Fig.~\ref{total_image}) changes significantly---in much of
the inner disk, the integrated polarization rotates by $90^\circ$,
even though none of the model's physical parameters has
been changed at all!

This effect can be understood qualitatively in very simple
fashion (see also \citet{agol:00}).  For most reasonable stellar-mass
BH accretion disk models, the opacity in the inner disk is dominated
by electron scattering \citep{shakura:73, novikov:73}, so
returning radiation in the $\sim 1-10$ keV energy range should scatter
off the disk with negligible absorption. Detailed atmosphere
calculations (S. Davis, private communication) show that
the photospheric region is strongly scattering dominated for $R
\lesssim 50 M$ whenever $L/L_{\rm Edd} \gtrsim 0.01$. If these
calculations were altered to take into account magnetic
contributions to vertical support, which should be substantial at these
altitudes in this regime \citep{hirose:09},
absoroption opacity would likely
be even weaker (however, when the central mass
is much larger, as in an AGN, the inner disk temperature is
much lower, giving significantly greater absorptive opacity
in the X-ray band).  Moreover, the typical photon scatters only
once or twice before permanently departing the disk.  Its outgoing
polarization is therefore very sensitive to the electron
scattering cross section's polarization-dependence:
\begin{equation}\label{dsigma1}
\left(\frac{d\sigma}{d\Omega}\right)_{\rm pol} = 
   r_0^2 \left| \mathbf{f}_i \cdot \mathbf{f}_f\right|^2 ,
\end{equation}
where $r_0$ is the classical electron radius.  In other words,
scattering is strong only when the polarization direction changes
little.  This is possible for both senses of linear polarization
when the change in photon direction is small, but when the scattering
is nearly perpendicular, the new photon wave-vector becomes
nearly parallel to one of the initial polarization directions.
Thus, for nearly-perpendicular scattering, the outgoing light
can be close to $100\%$ polarized.

For observers at high inclination angles, such as in
Figure~\ref{total_image}, returning radiation photons initially emitted
from the far side of the disk (top of the image) are reflected off the
near (bottom) side with a relatively small scattering angle,
maintaining a moderate horizontal polarization as in Figure~\ref{direct_image}.
On the other hand, photons emitted from the left side of the disk can be
bent back to the right side (or {\it vice versa}), and then scatter at roughly $90^\circ$ to
reach the observer, thereby aquiring a large vertical polarization
component. 
Although relatively small in total flux, this latter contribution
can dominate the polarization because it is so strongly polarized.
A flux component of only $10\%$, when $100\%$ polarized,
contributes more to the polarized flux than does a $90\%$ contribution
that is only $5\%$ polarized.

Because high-energy photons
from the hotter inner parts of the disk experience stronger
gravitational deflection, they are more likely to return to the
disk than the low-energy photons emitted at larger radii. We find
that, for each value of the spin parameter, there is a
characteristic ``transition radius,'' within which the returning
radiation dominates and produces net vertical polarization. Outside 
this point, the direct radiation dominates and
produces horizontal polarization. This transition
radius is in fact only weakly dependent on $a/M$, ranging from
$R_{\rm trans} \approx 7M$ for $a/M=0$ to $R_{\rm trans} \approx 5M$ for
$a/M=0.998$. The location and shape of the observed polarization swing
can be used to infer the radial temperature profile near the
transition radius (see Sec.\ \ref{parameters} below for more
details). While the majority of the returning obviously originates from
the inner-most disk, we find that most of it also scatters inside
$R\lesssim 10M$ as well (where the atmosphere opacity is completely
dominated by electron scattering), due to the additional focusing
effects of gravitational lensing.

\section{RETURNING RADIATION: QUANTITATIVE RESULTS}\label{results}

\begin{figure}
\caption{\label{total_image} Ray-traced image of radiation from a
  thermal disk, as in Fig.\ \ref{direct_image}, but here including
  returning radiation. Photons emitted from the inner disk get bent by
  the BH and scatter off the opposite side of the disk towards the distant
  observer.}
\begin{center}
\includegraphics*[52,450][410,720]{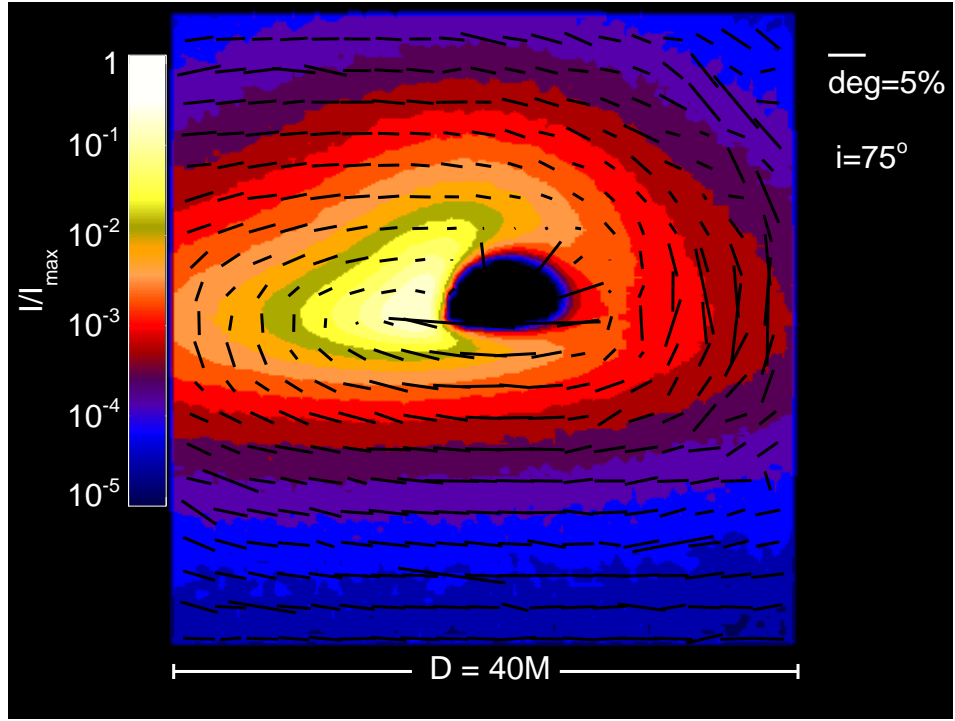}
\end{center}
\end{figure}

Polarization maps like that shown in Figure \ref{total_image} provide
a useful picture for understanding the physical origin of the local
polarization features, but are less useful in providing quantitative
predictions for future observations. Since vertically- and
horizontally-polarized photons add up to give zero net polarization,
it is not clear from Figure \ref{total_image} alone what the
integrated polarization signal should be. Furthermore, Figure
\ref{total_image} shows only the energy-integrated polarization at
each point in the disk, masking the important energy dependence of
polarization seen in Figure \ref{direct_pol}. 

\begin{figure}
\caption{\label{total_pol_a0} Intensity spectrum and polarization degree
  and angle from a thermal disk, including direct and returning
  radiation, for a BH with $a/M=0$, $M=10M_\odot$, $L/L_{\rm
  Edd}=0.1$, and a Novikov-Thorne emission profile. In the left column
  we plot the
  observed flux, in the middle the degree of polarization, and on the
  right the angle of polarization. In each plot, the flux is divided
  into contributions from the direct (dotted curves), reflected
  (dashed curves), and total (solid curves). From top to bottom, the
  observer inclination is $45^\circ$, $60^\circ$, and $75^\circ$.}
\begin{center}
\scalebox{0.3}{\includegraphics{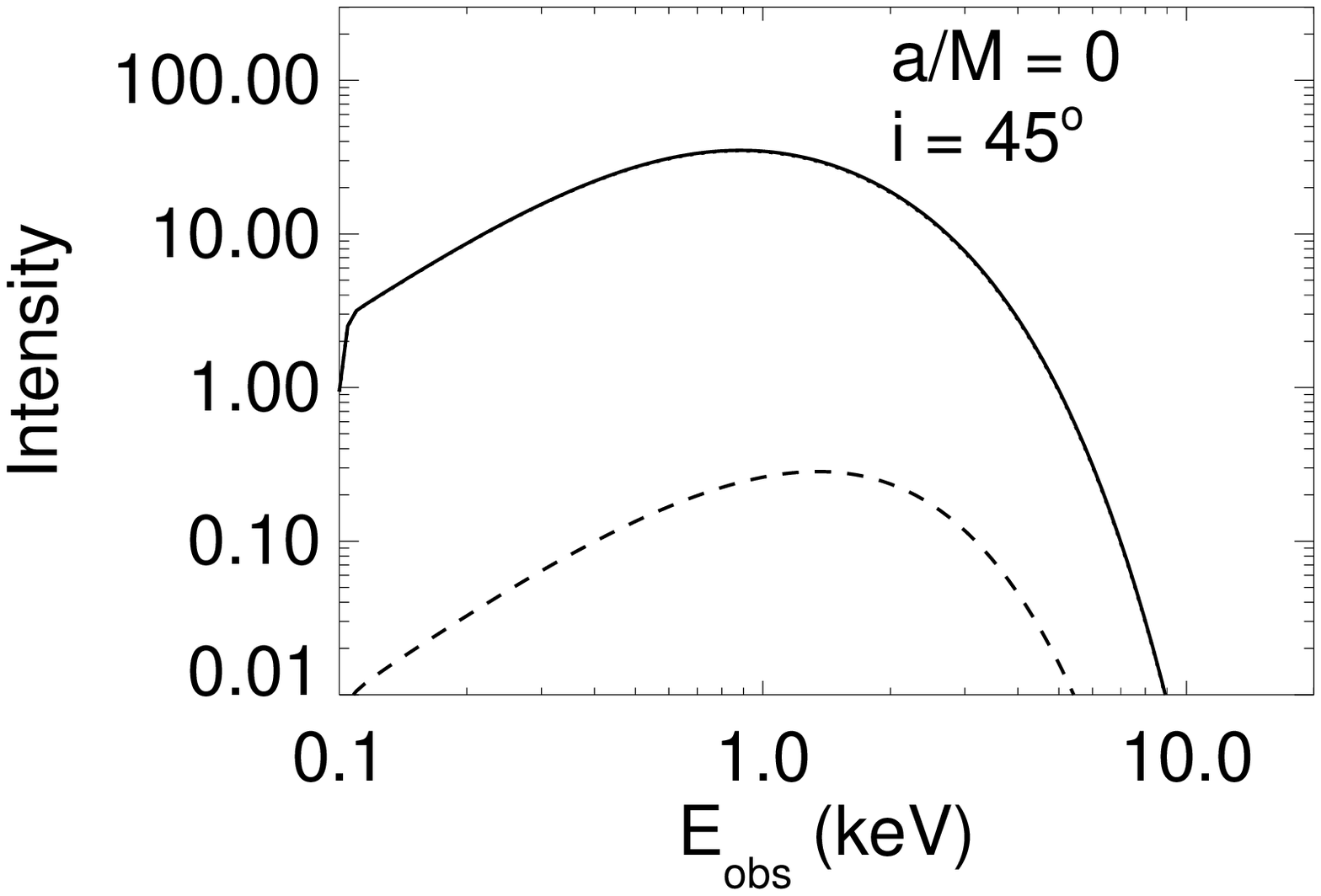}}
\scalebox{0.3}{\includegraphics{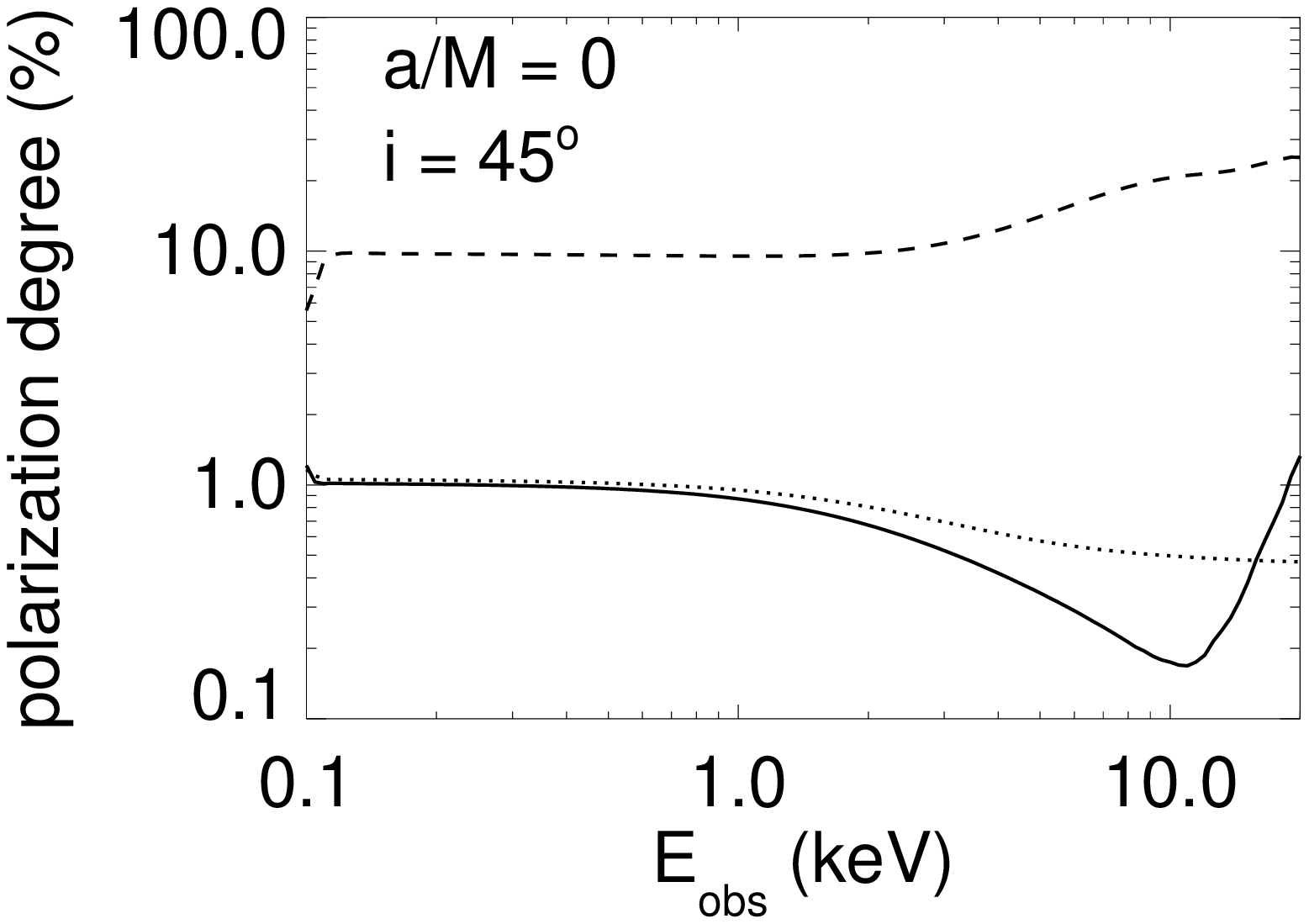}}
\scalebox{0.3}{\includegraphics{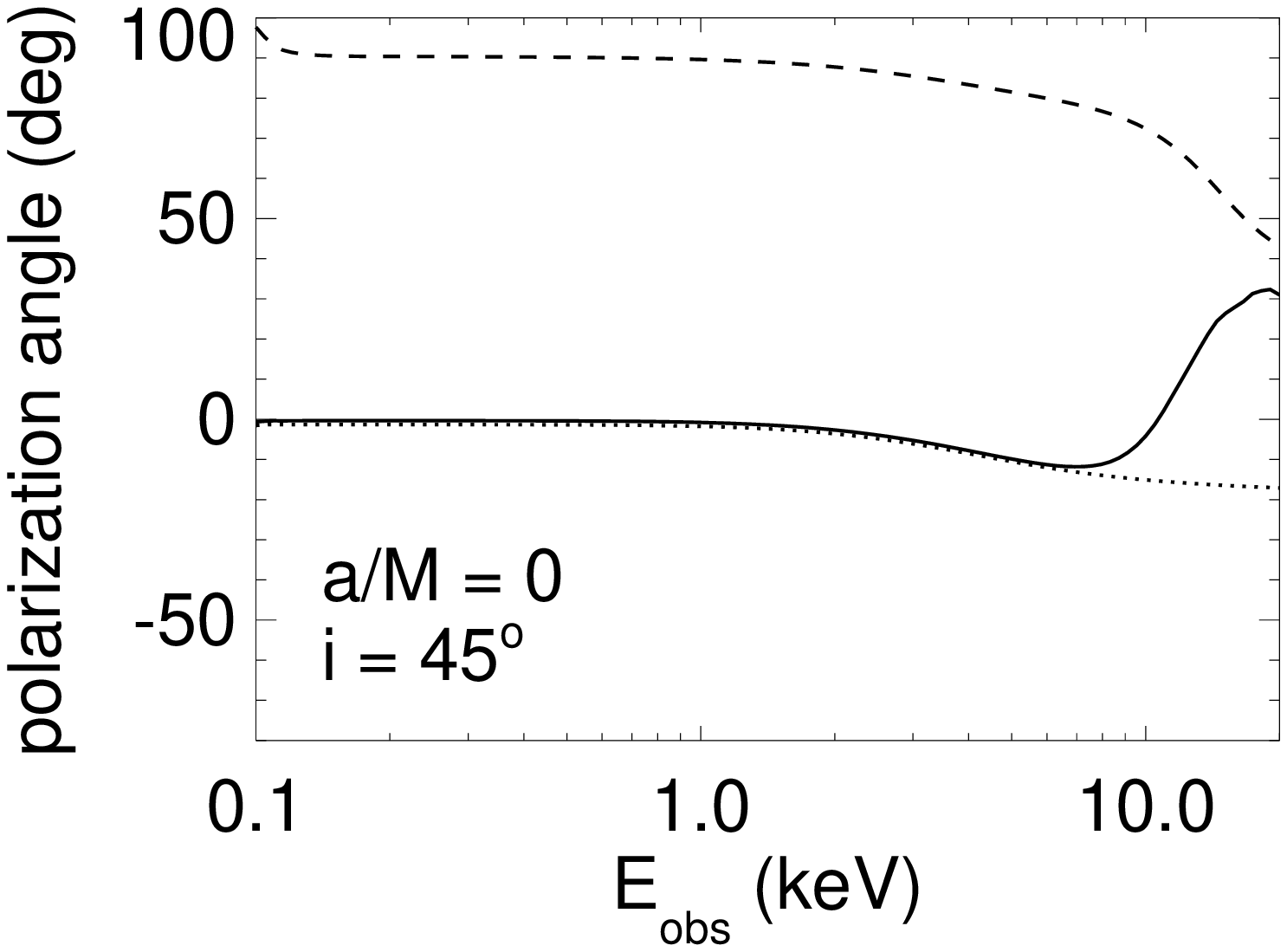}}\\
\scalebox{0.3}{\includegraphics{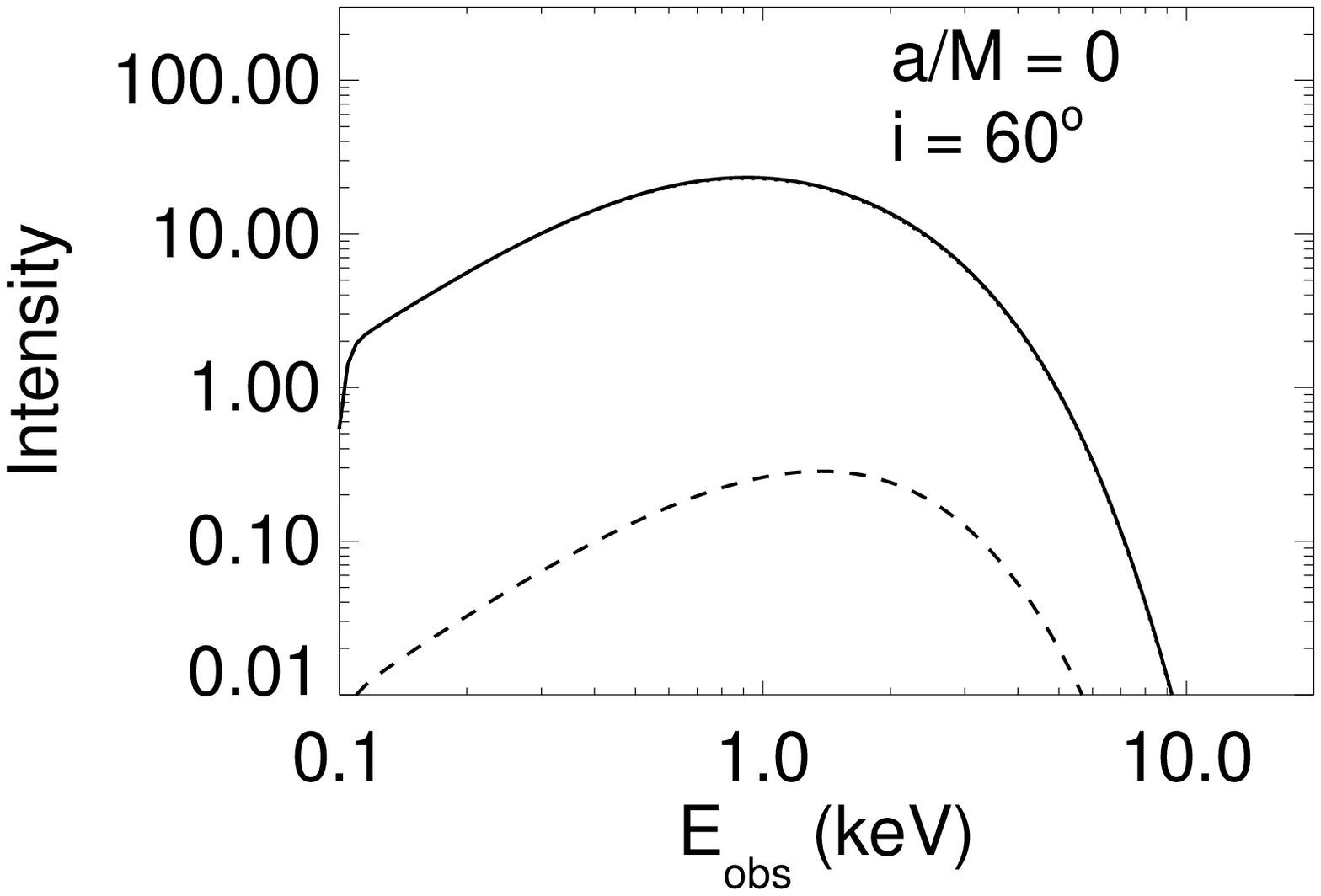}}
\scalebox{0.3}{\includegraphics{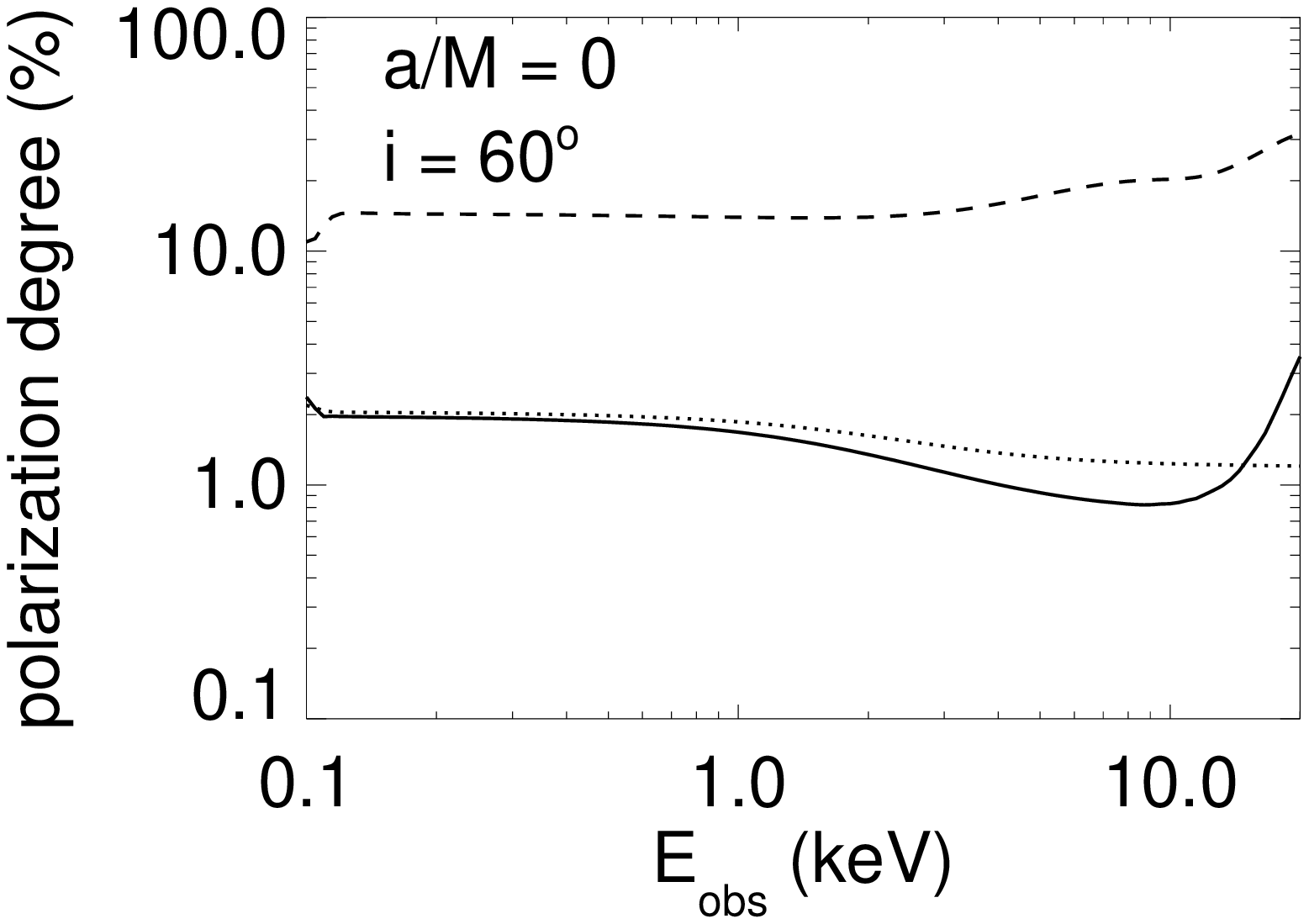}}
\scalebox{0.3}{\includegraphics{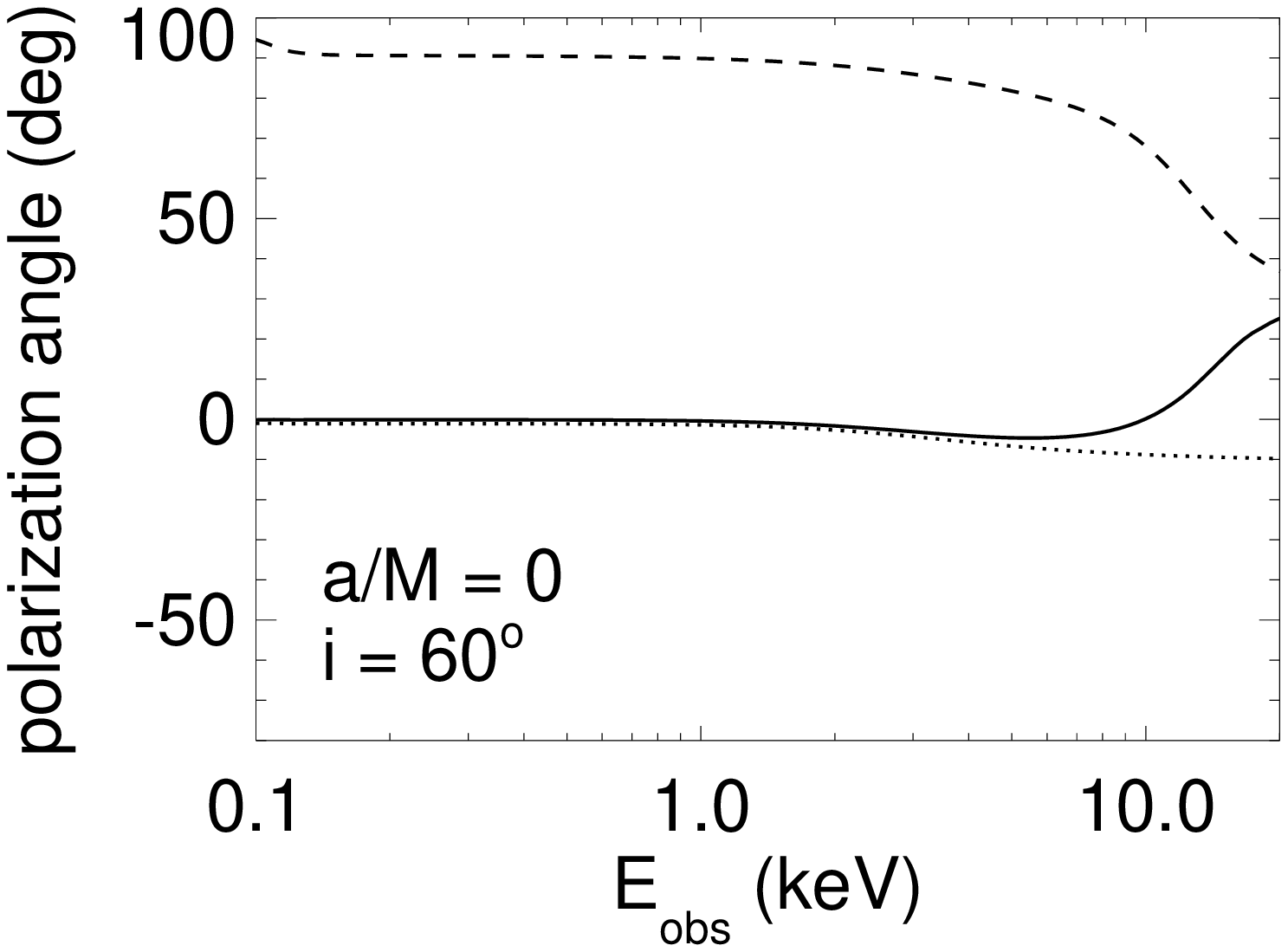}}\\
\scalebox{0.3}{\includegraphics{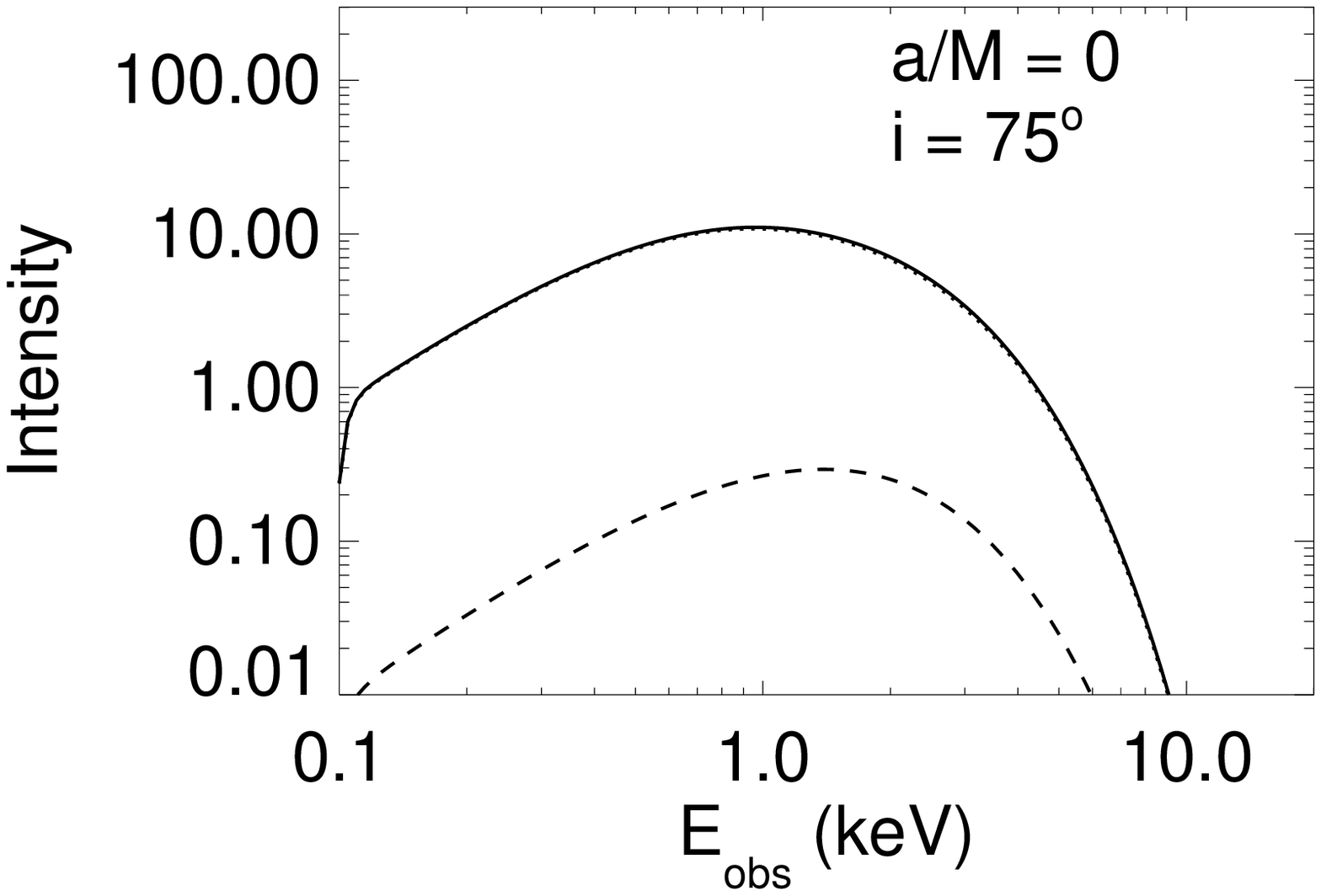}}
\scalebox{0.3}{\includegraphics{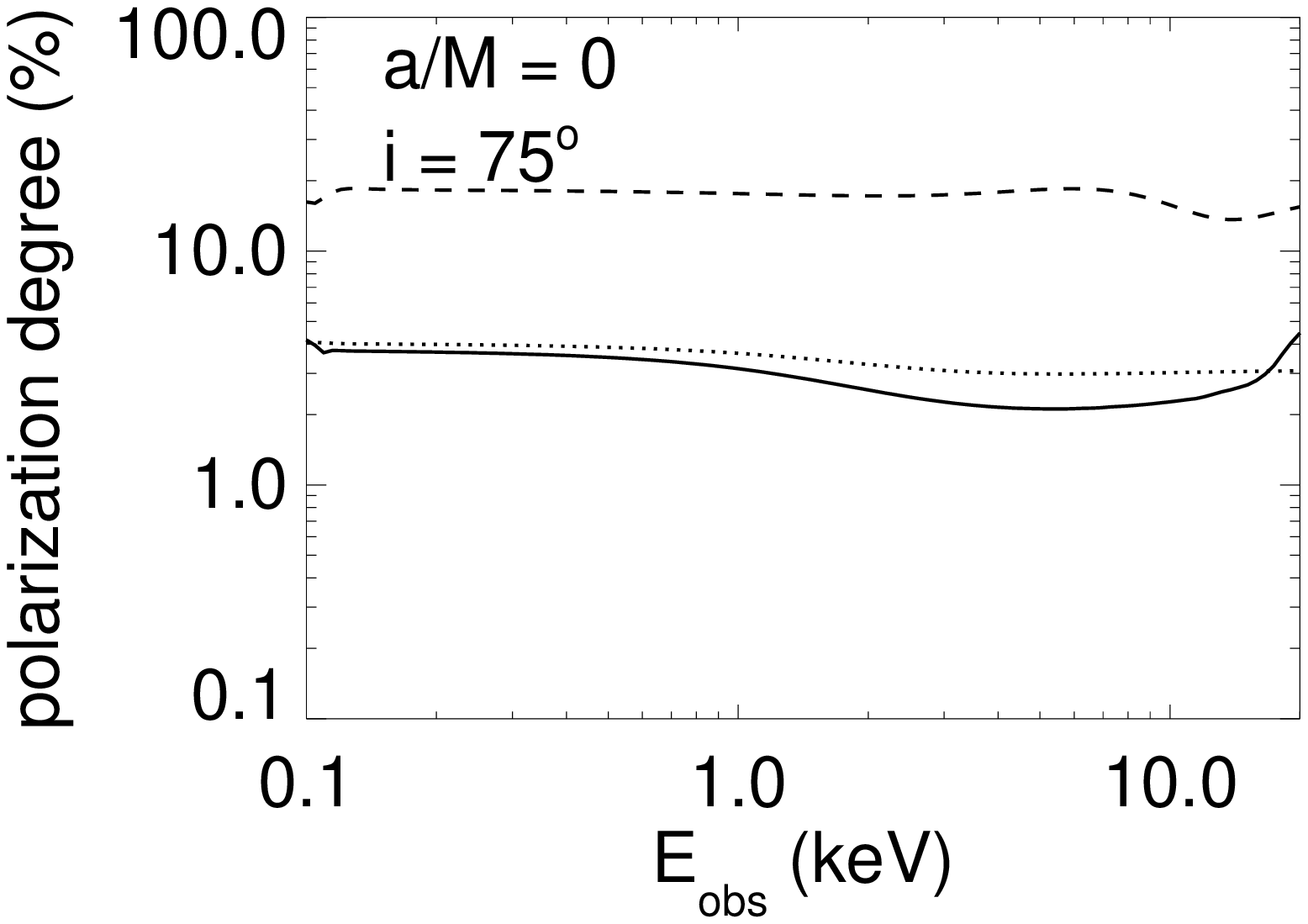}}
\scalebox{0.3}{\includegraphics{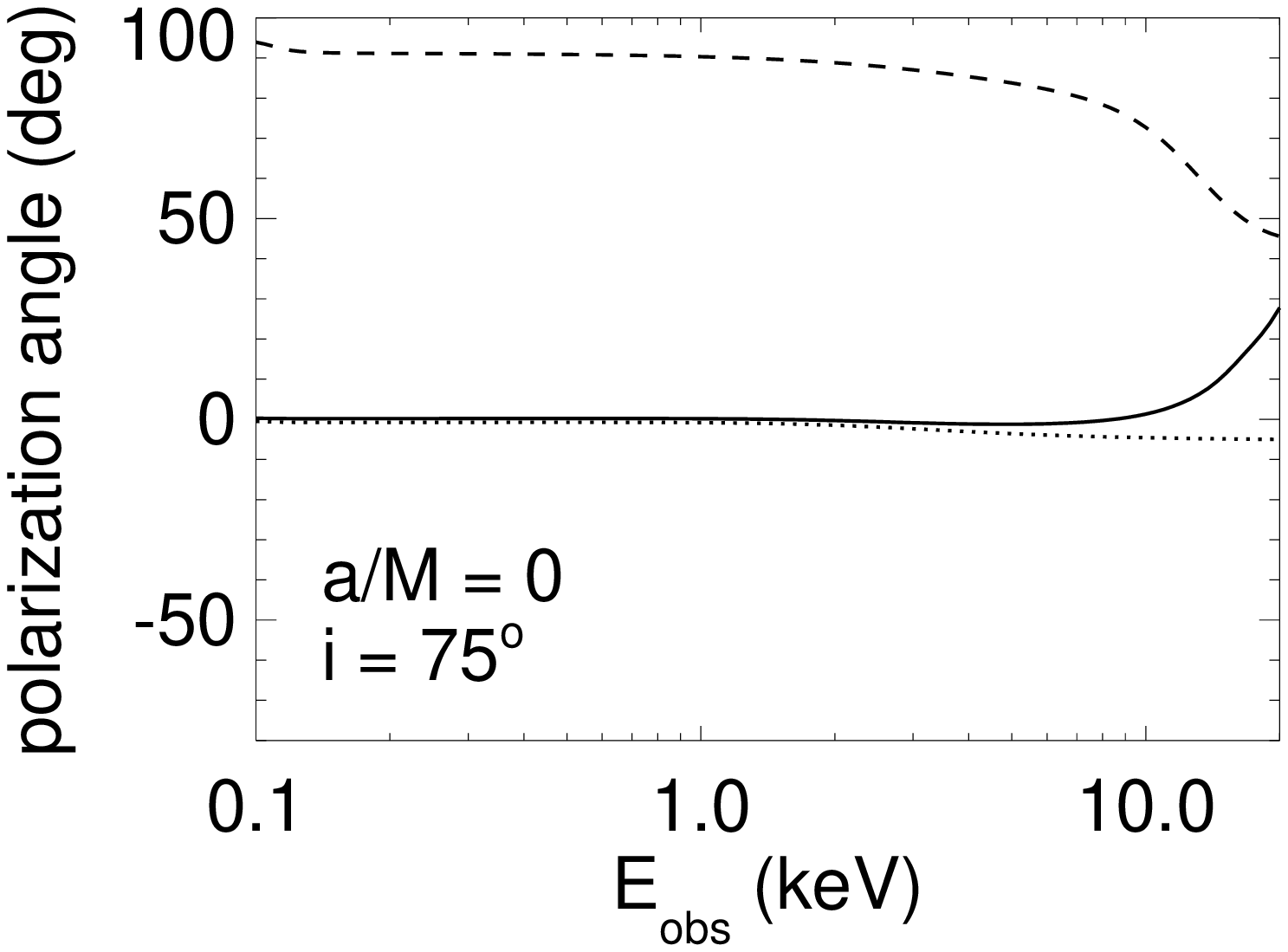}}
\end{center}
\end{figure}

\begin{figure}
\caption{\label{total_pol_a9} Intensity spectrum and polarization degree
  and angle from a thermal disk, as in Figure \ref{total_pol_a0}, but for
  a Kerr BH with $a/M=0.9$.}
\begin{center}
\scalebox{0.3}{\includegraphics{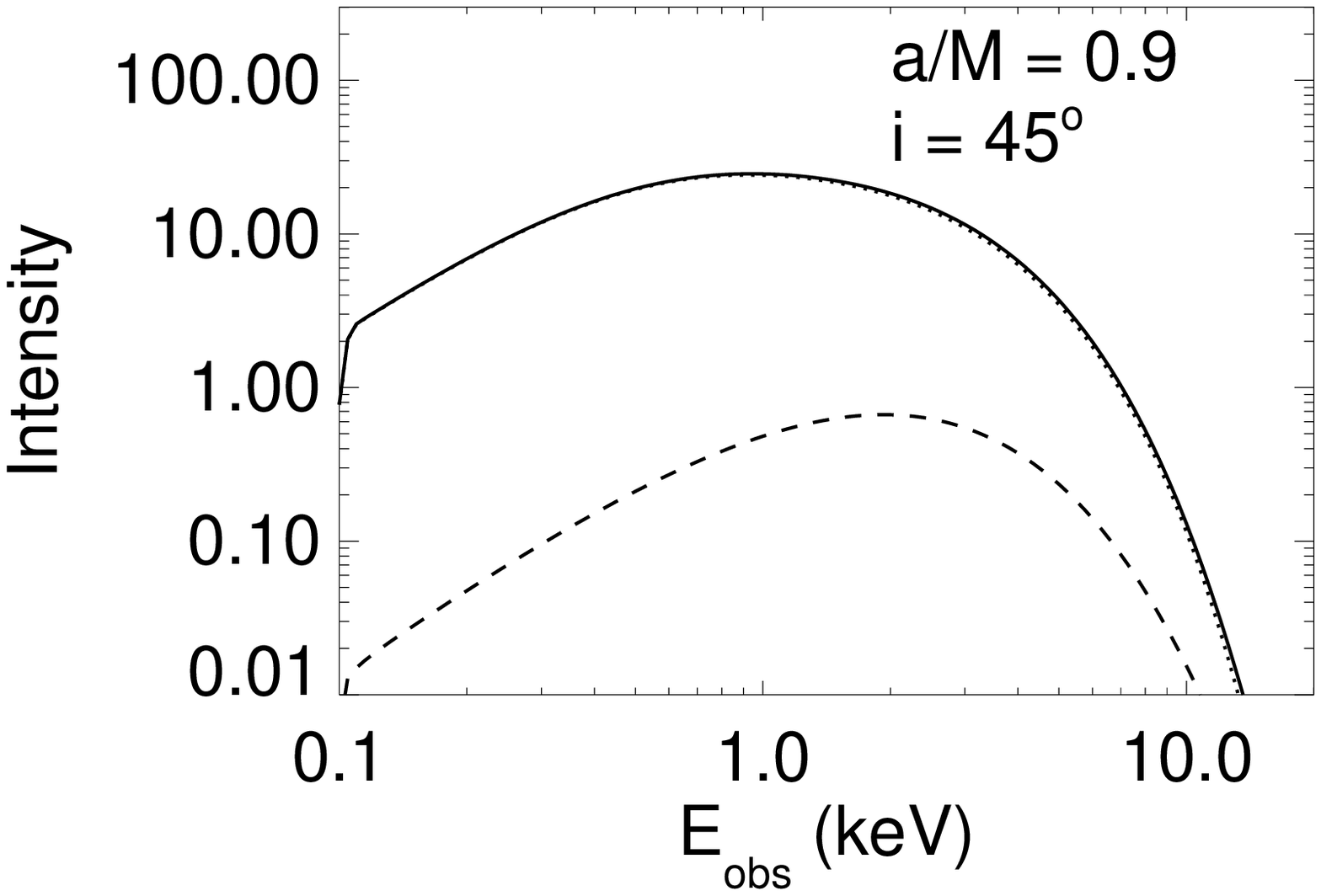}}
\scalebox{0.3}{\includegraphics{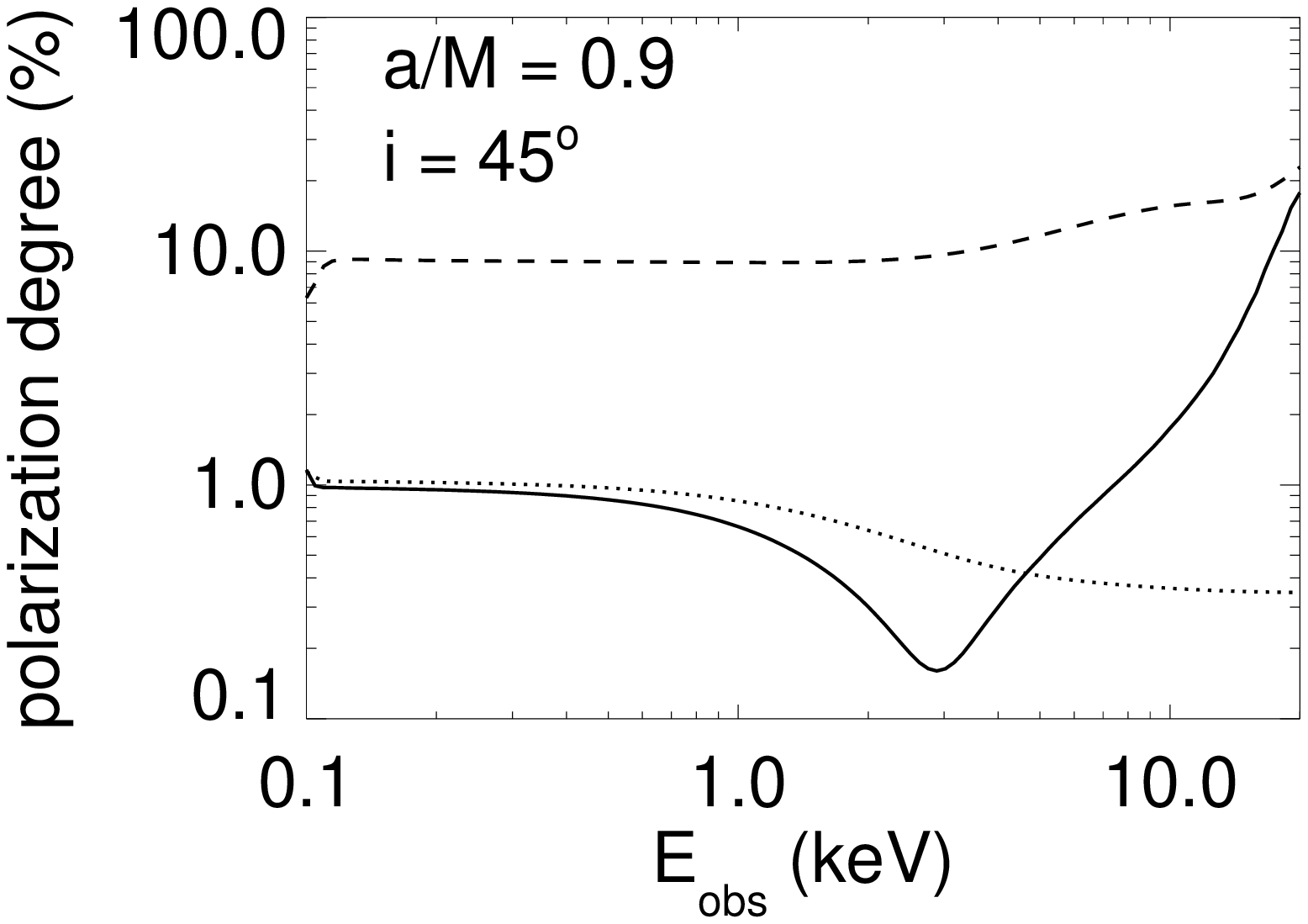}}
\scalebox{0.3}{\includegraphics{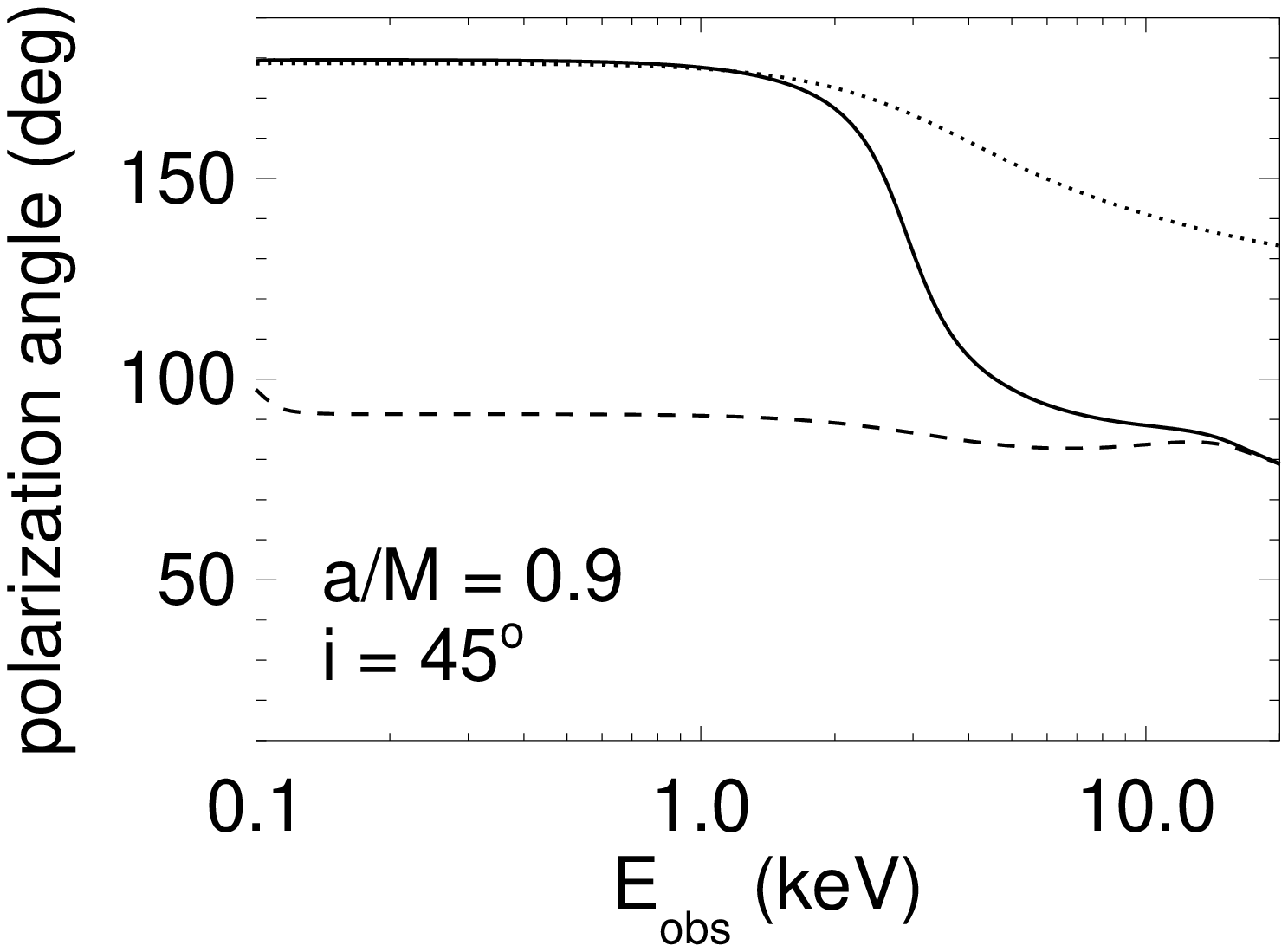}}\\
\scalebox{0.3}{\includegraphics{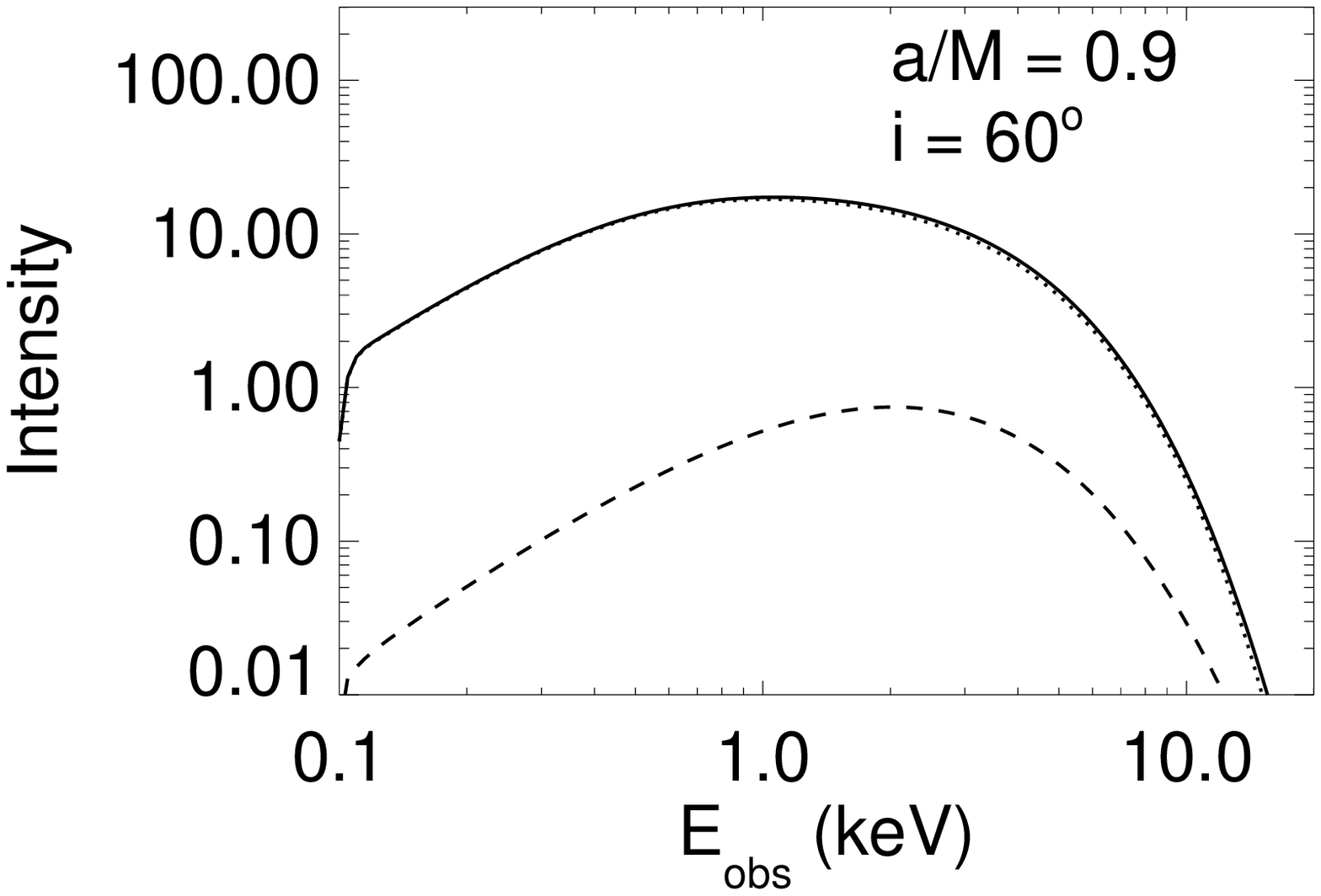}}
\scalebox{0.3}{\includegraphics{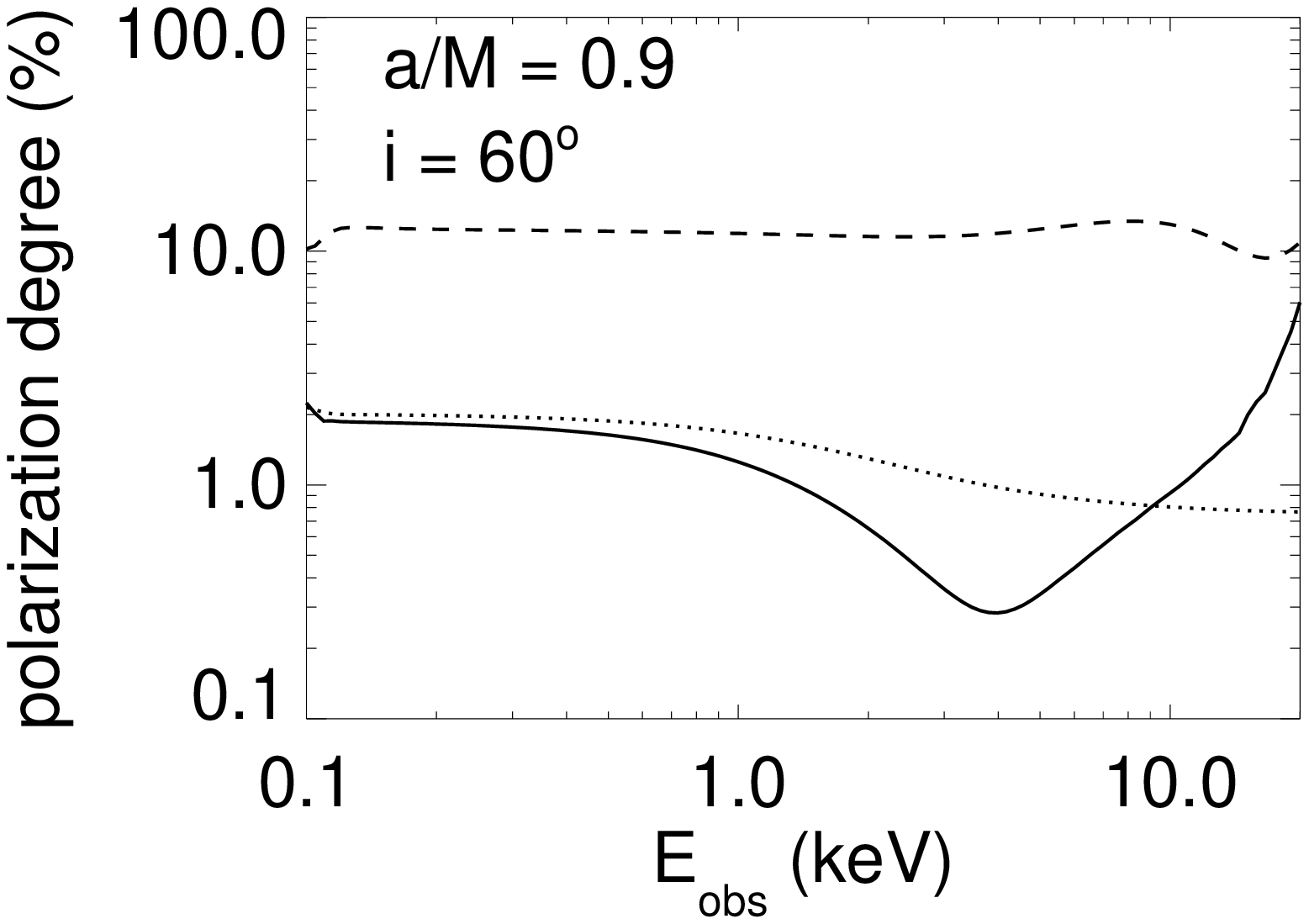}}
\scalebox{0.3}{\includegraphics{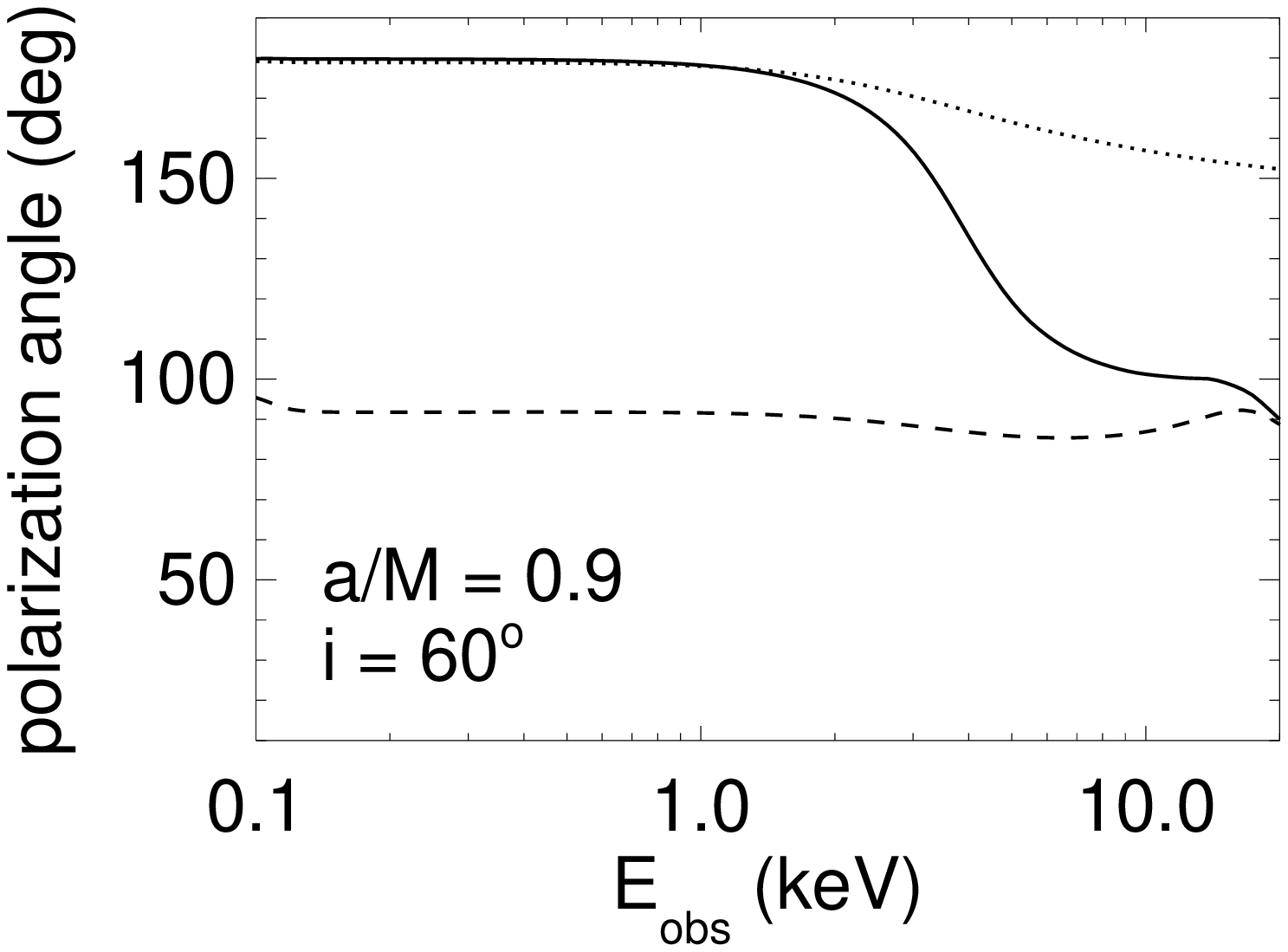}}\\
\scalebox{0.3}{\includegraphics{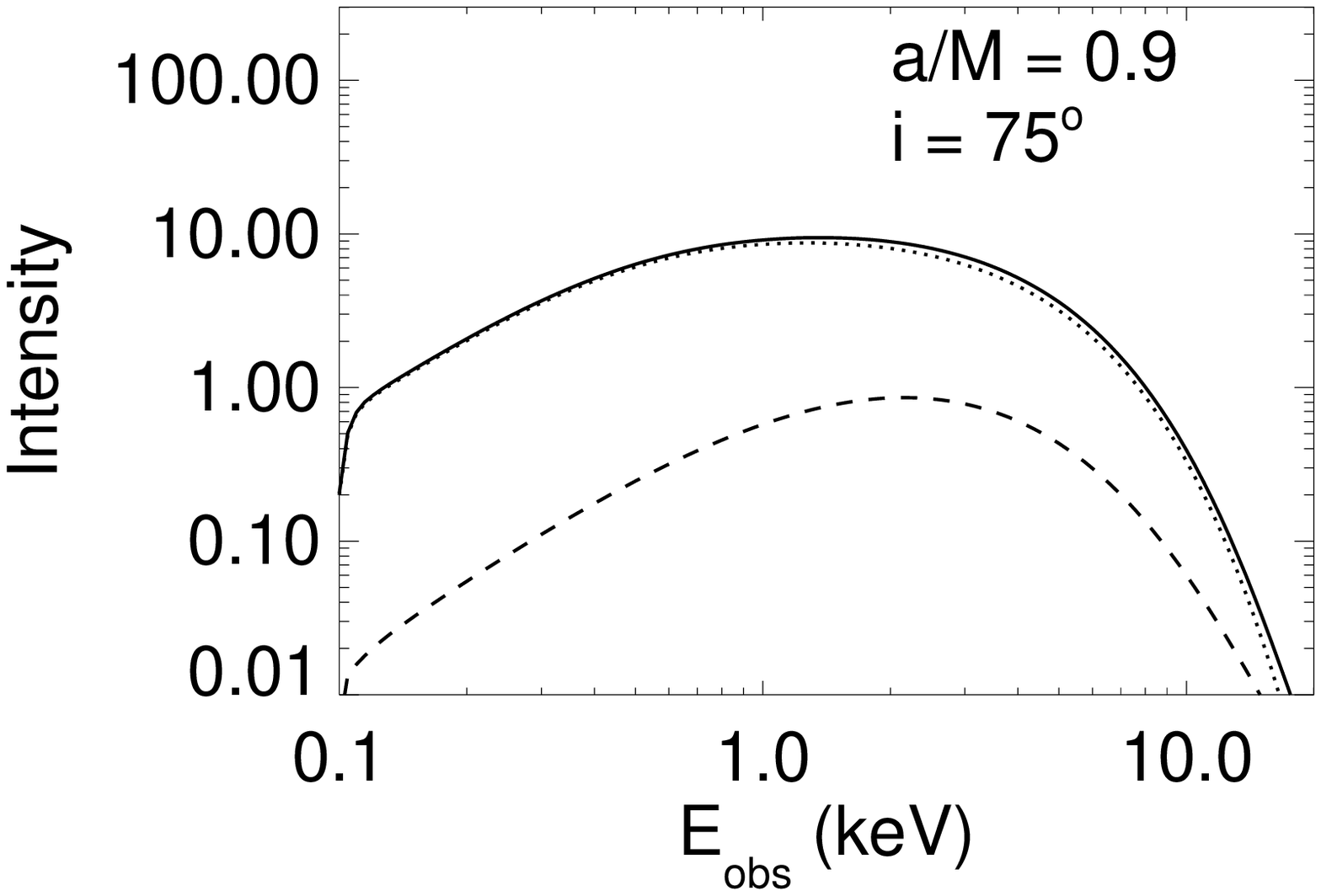}}
\scalebox{0.3}{\includegraphics{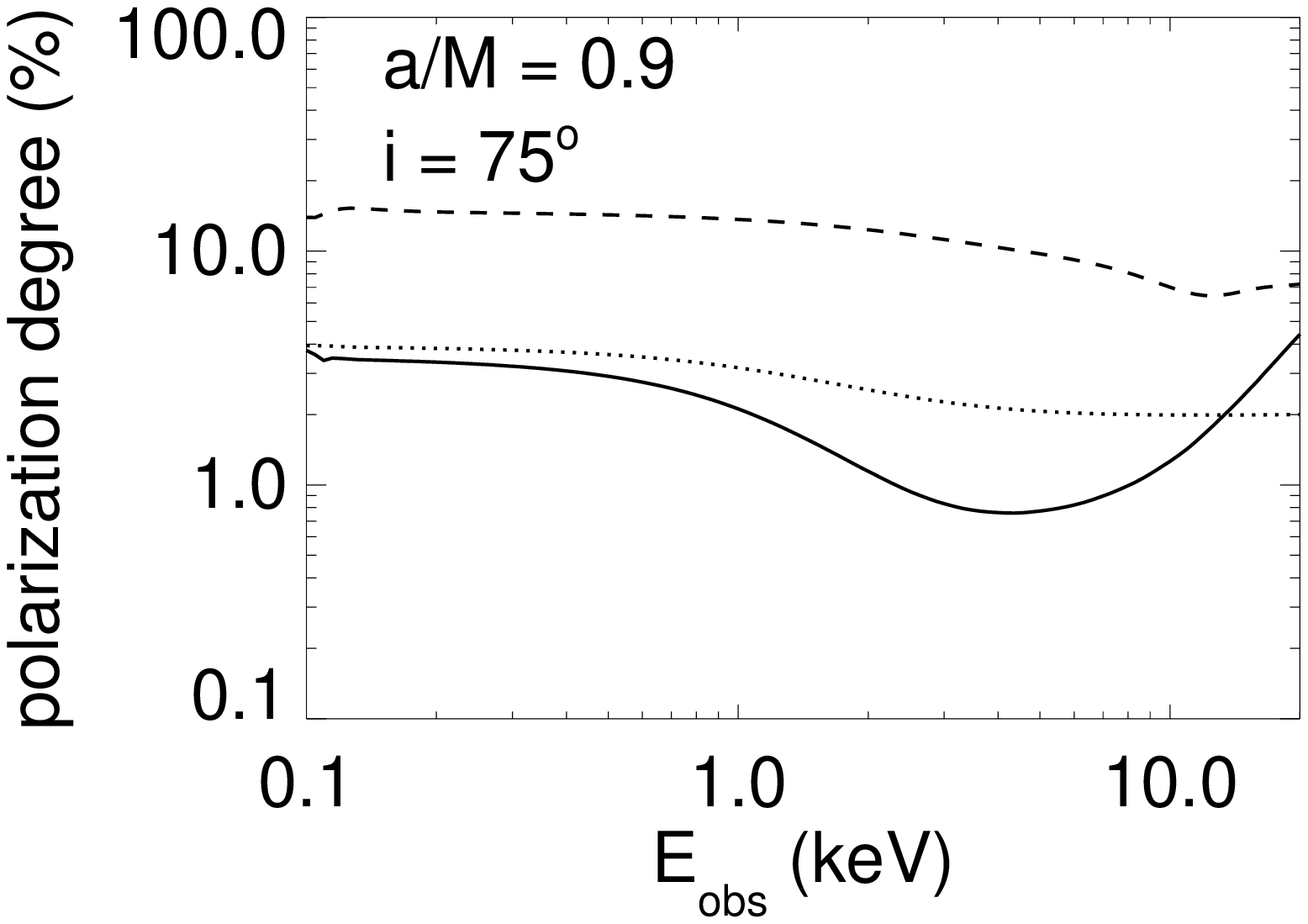}}
\scalebox{0.3}{\includegraphics{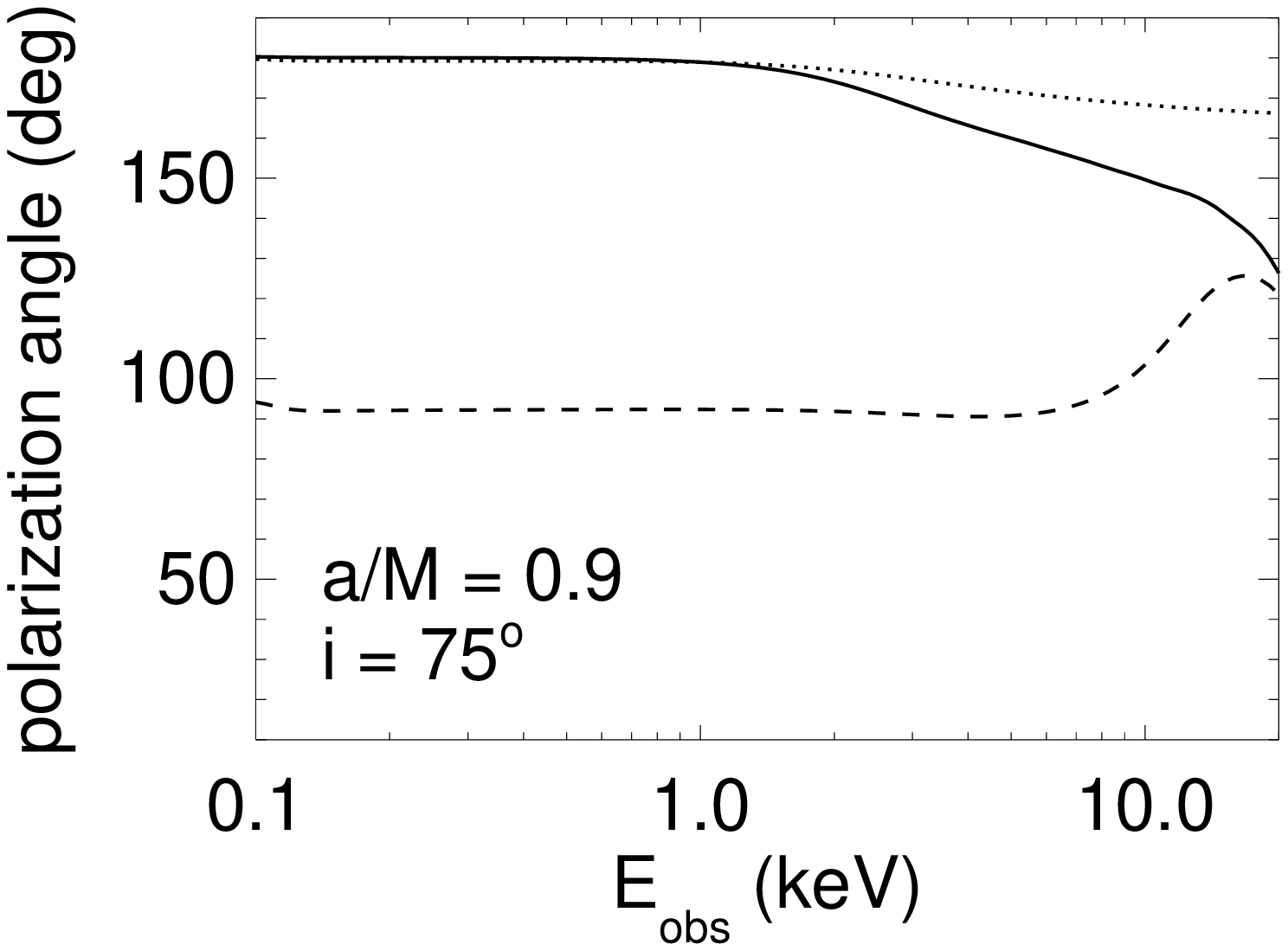}}
\end{center}
\end{figure}

\begin{figure}
\caption{\label{total_pol_a998} Intensity spectrum and polarization degree
  and angle from a thermal disk, as in Figure \ref{total_pol_a0}, but for
  a Kerr BH with $a/M=0.998$.}
\begin{center}
\scalebox{0.3}{\includegraphics{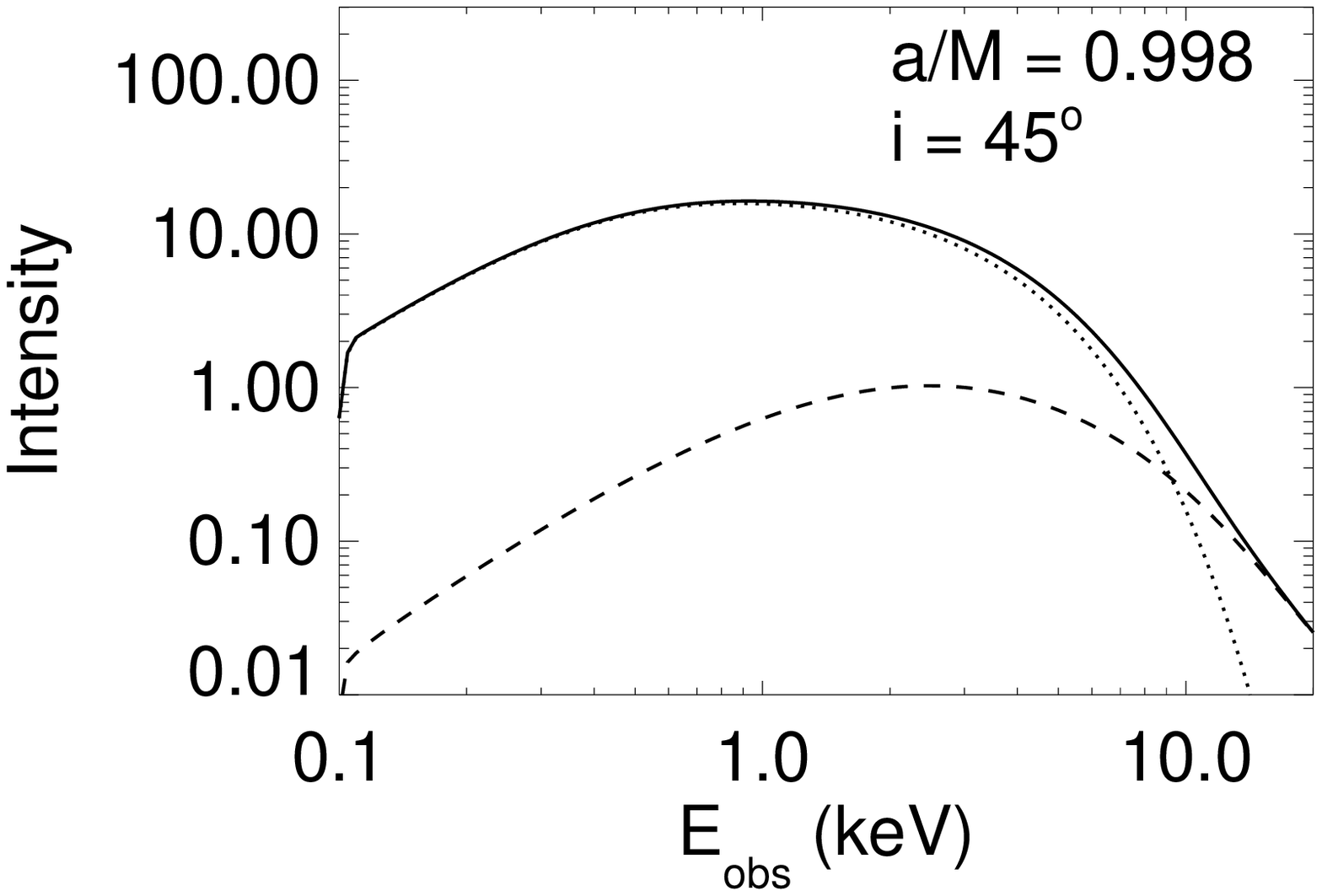}}
\scalebox{0.3}{\includegraphics{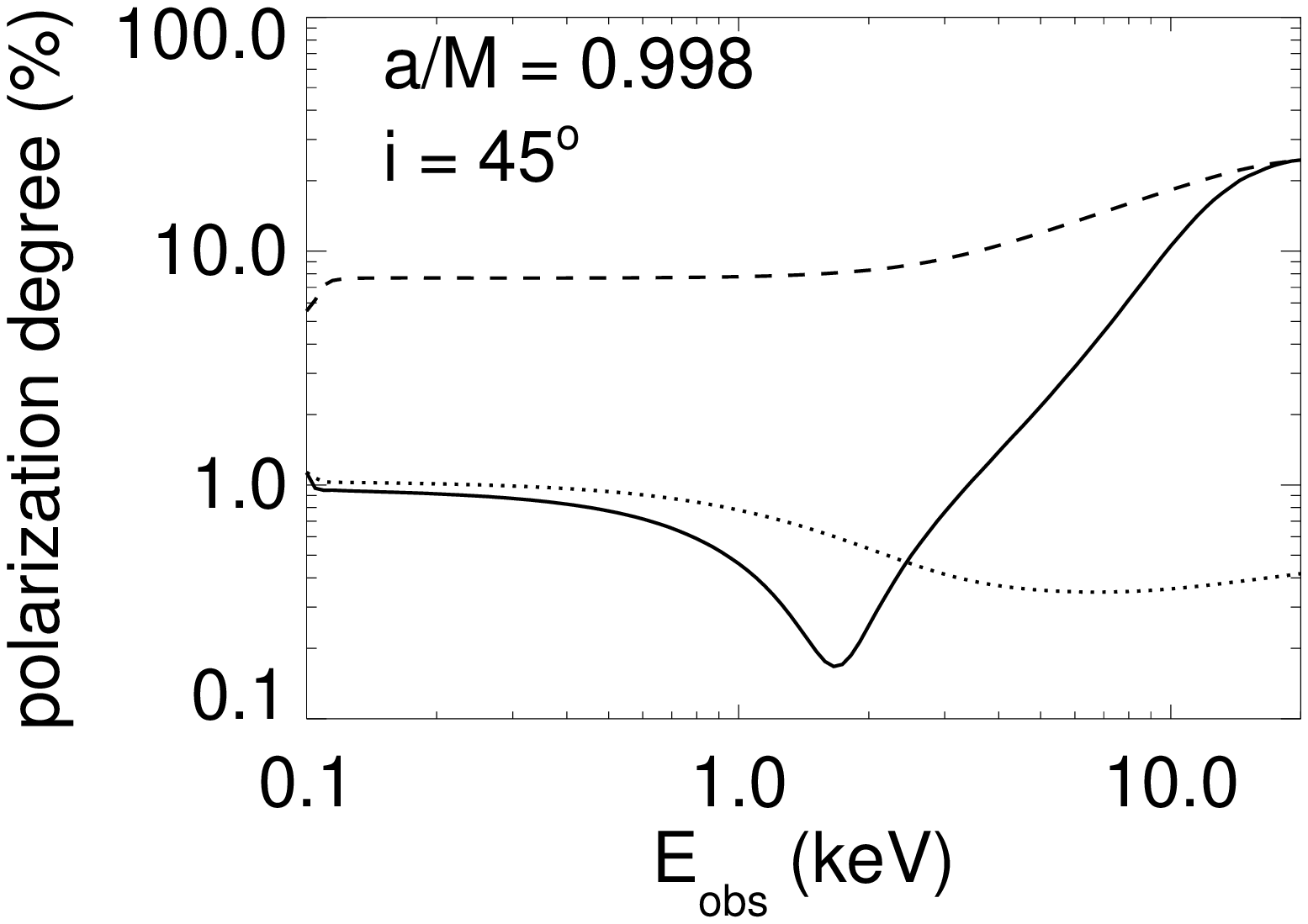}}
\scalebox{0.3}{\includegraphics{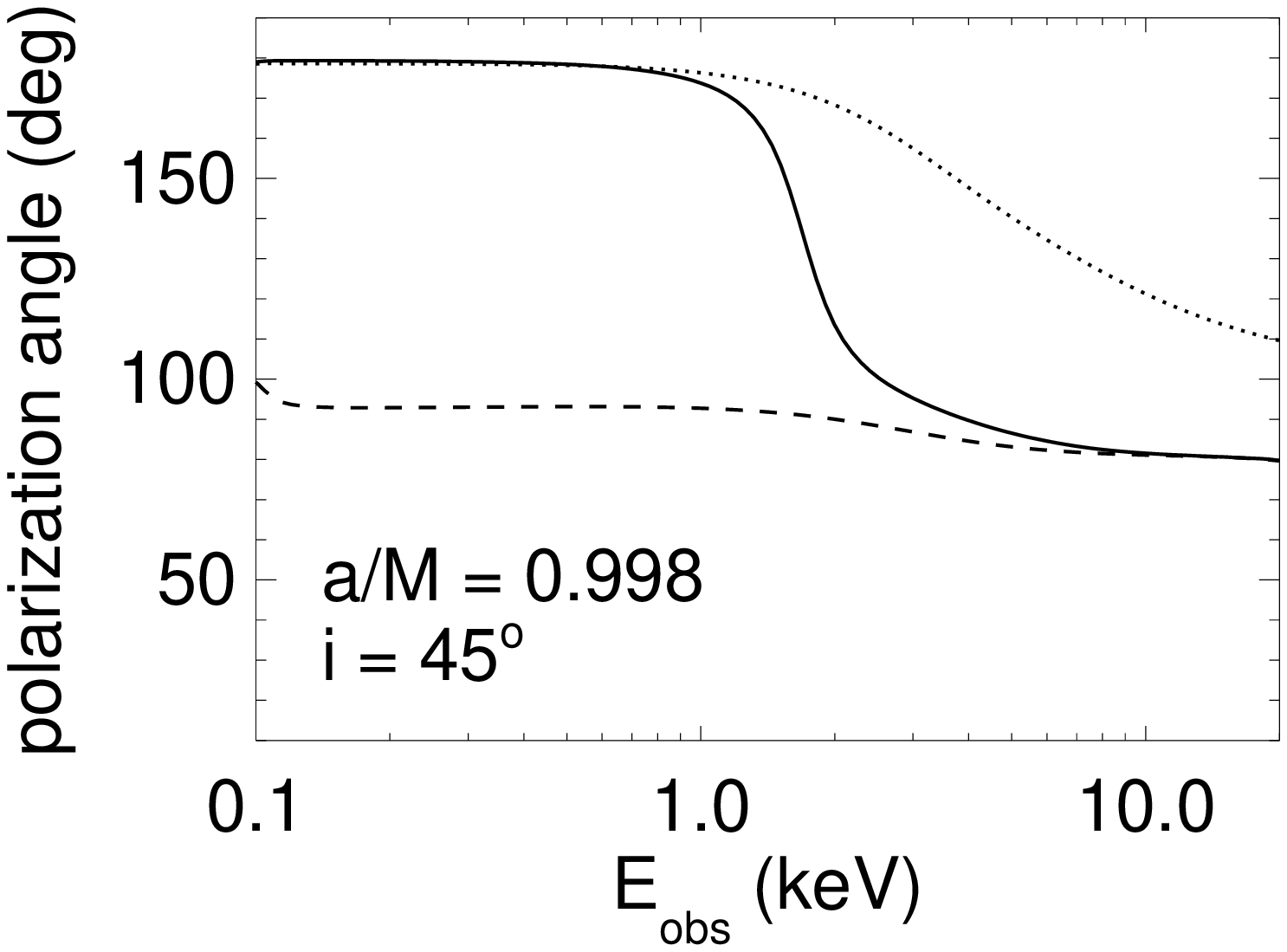}}\\
\scalebox{0.3}{\includegraphics{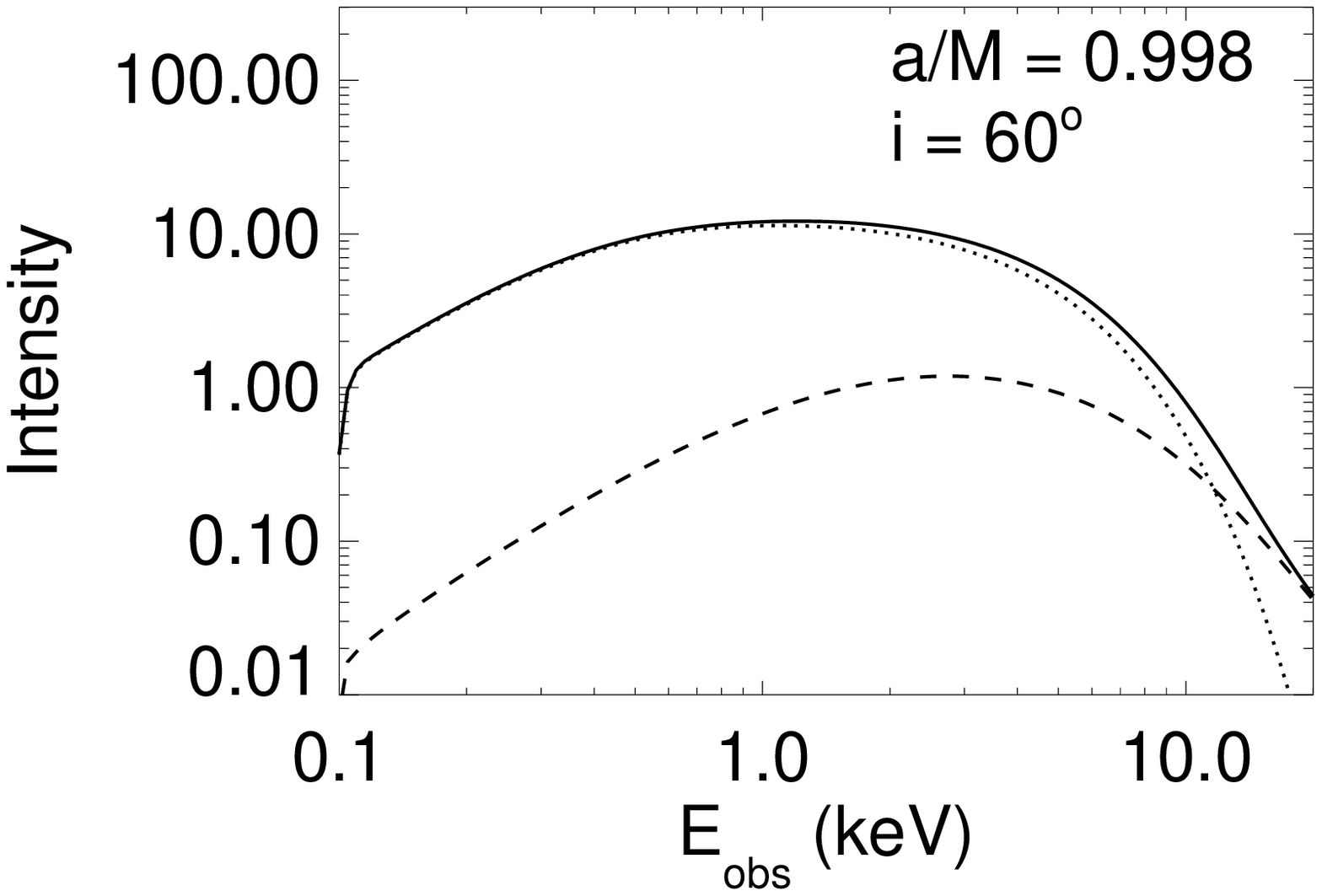}}
\scalebox{0.3}{\includegraphics{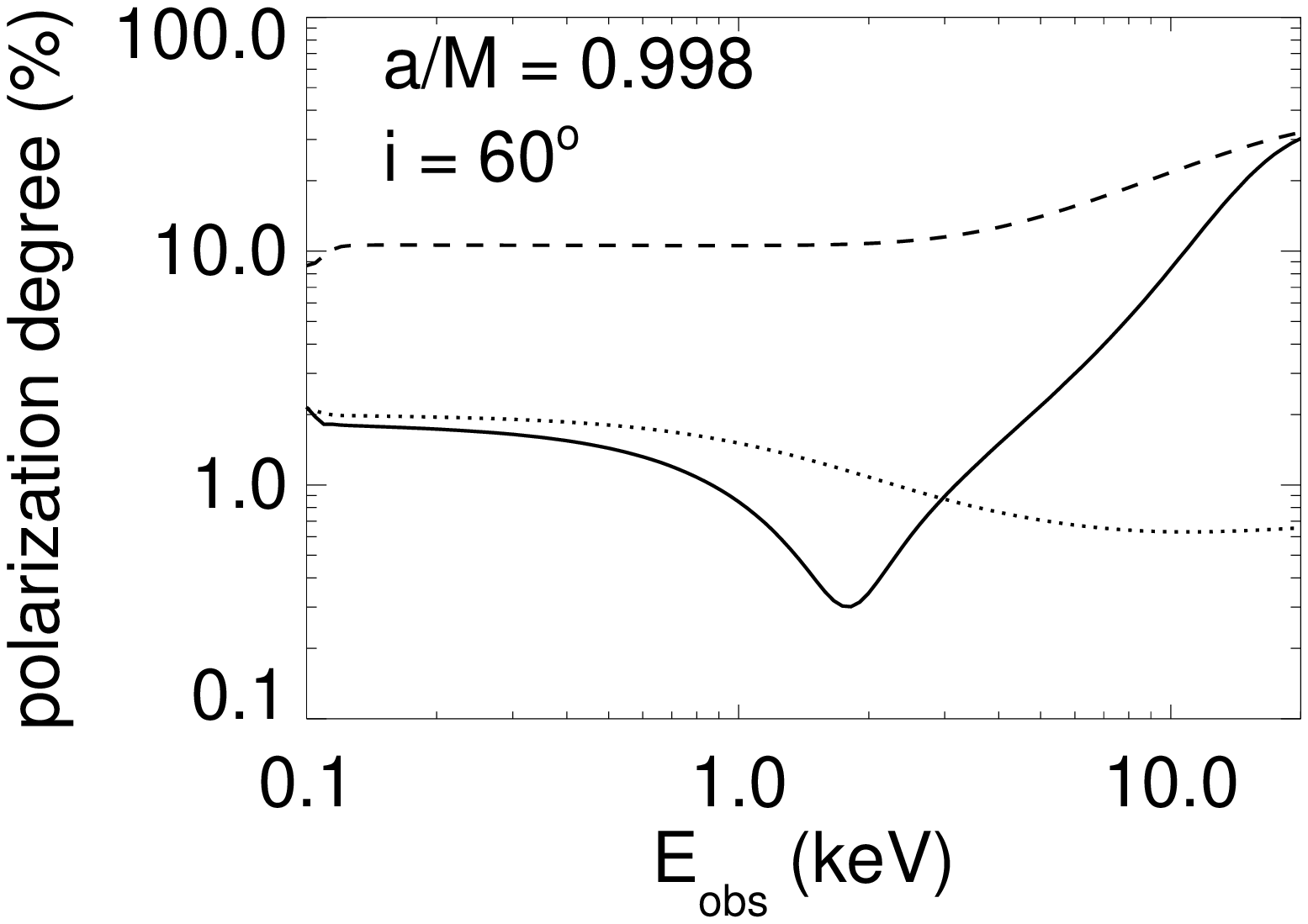}}
\scalebox{0.3}{\includegraphics{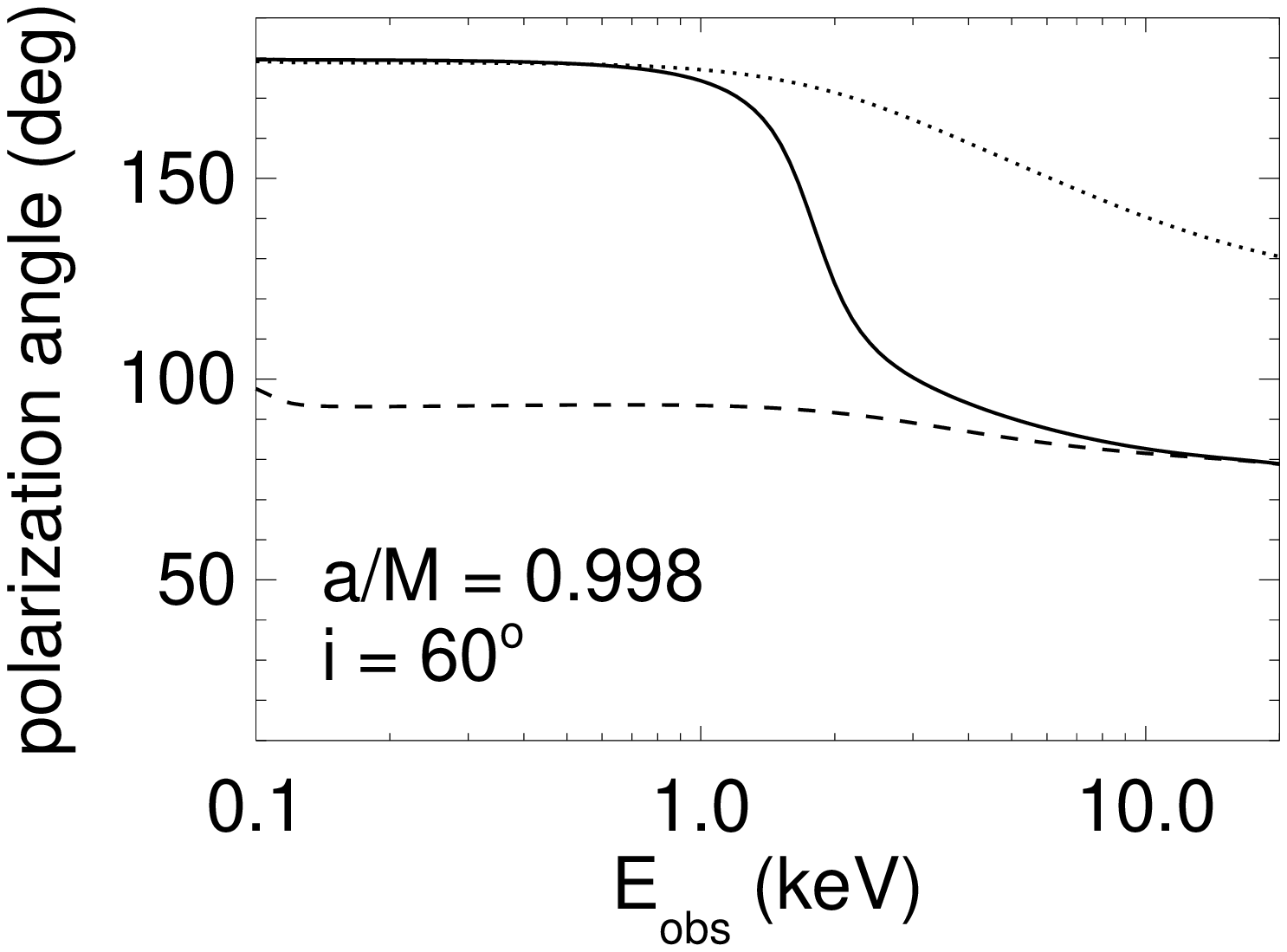}}\\
\scalebox{0.3}{\includegraphics{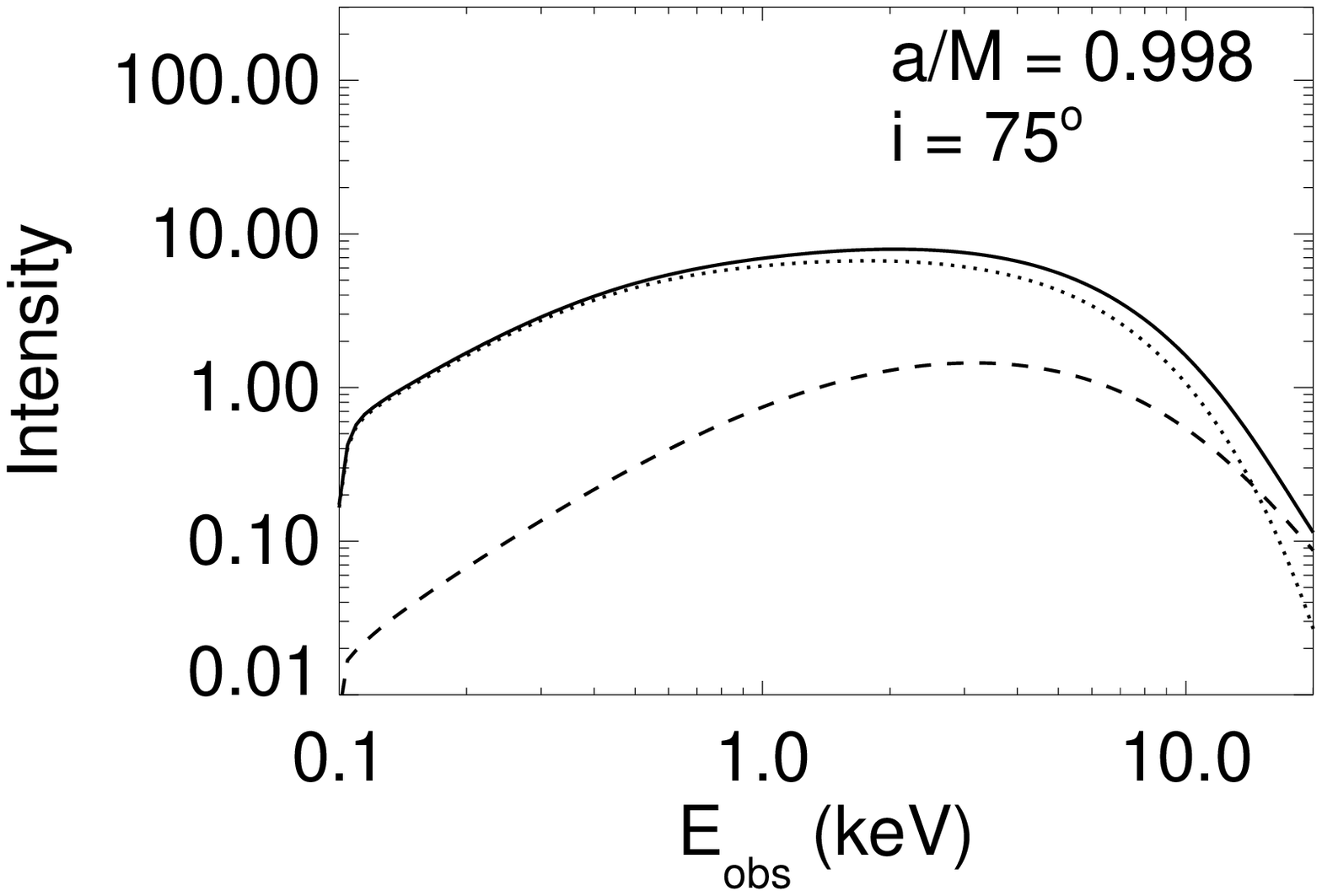}}
\scalebox{0.3}{\includegraphics{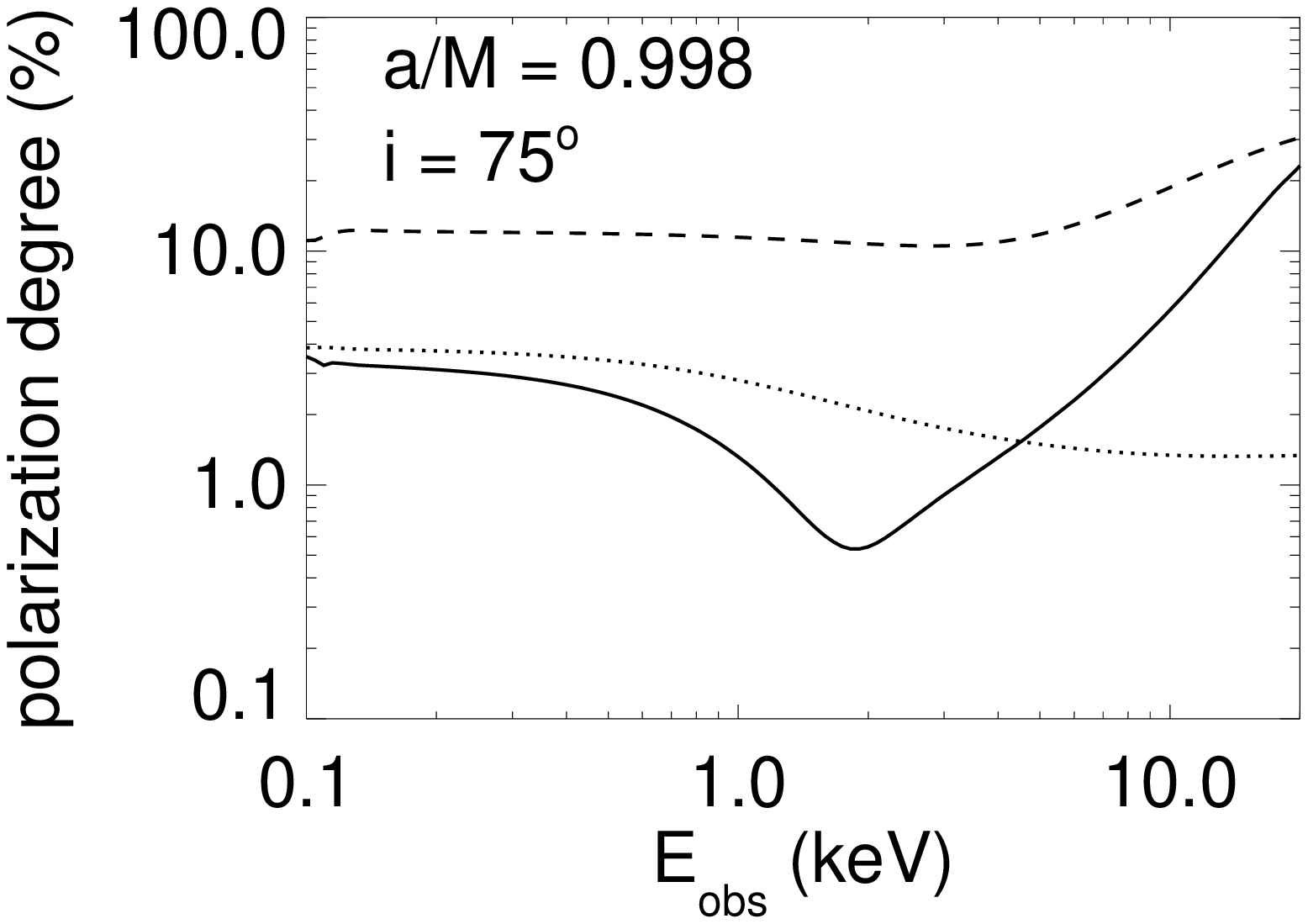}}
\scalebox{0.3}{\includegraphics{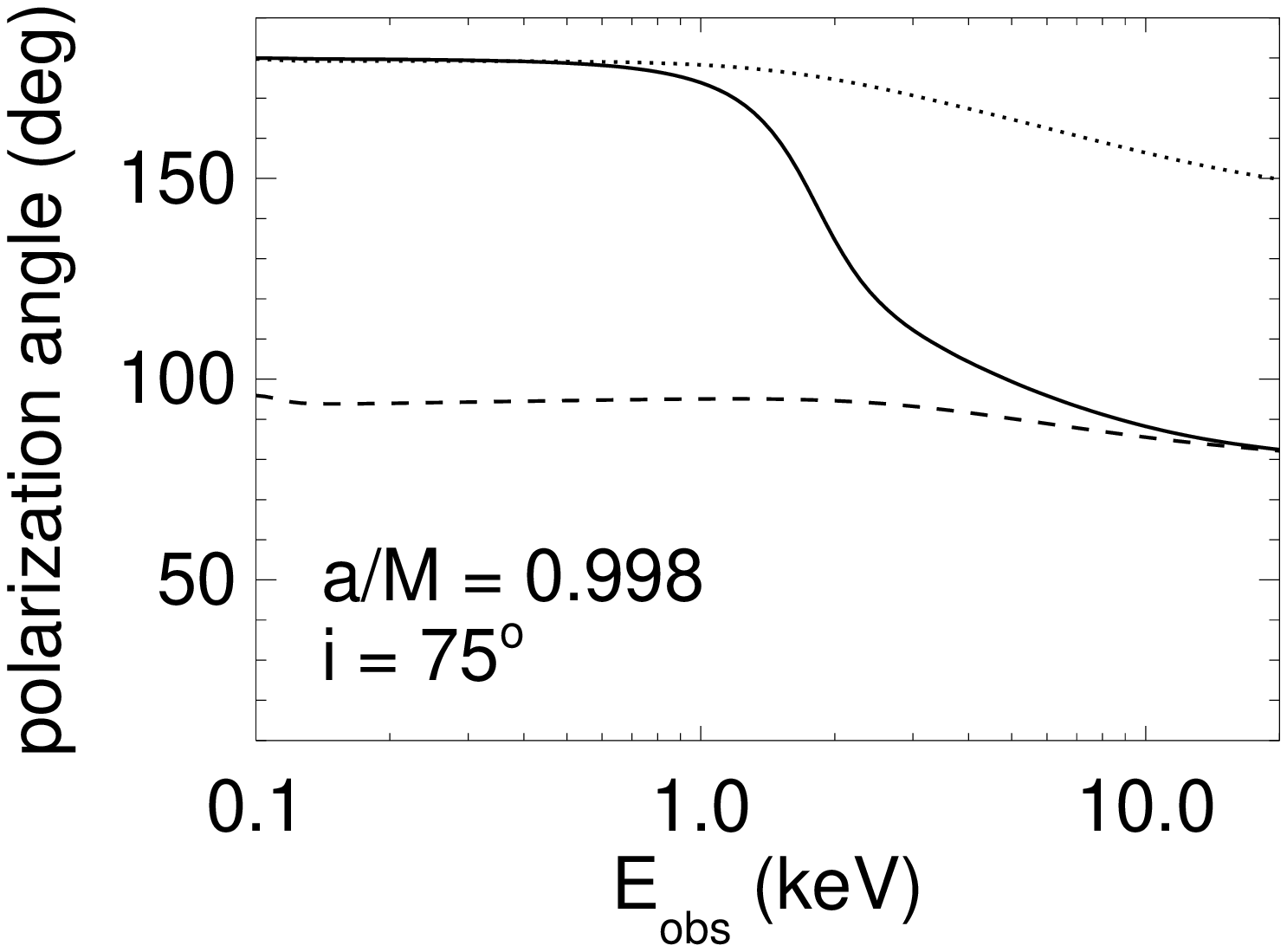}}
\end{center}
\end{figure}

In Figures \ref{total_pol_a0}--\ref{total_pol_a998} we show the
spectral intensity, degree, and angle of polarization as a function of
photon energy for a range of BH spins and inclinations. In each frame,
we plot the total flux (solid curves), as well as the direct flux from
the disk (dotted curves) and the returning radiation (dashed
curves). Beginning with the spectral intensity, we see the standard
broad thermal peak characteristic of BHs in the ``high soft''
state. For BH masses around $10 M_\odot$ and luminosities of $\sim
0.1L_{\rm Edd}$, the thermal spectrum peaks around 1 keV. For
a Novikov-Thorne (NT) emission profile, the direct radiation from a
non-spinning BH (peak emissivity at radius $R\sim 9M$) dominates over
the returning radiation 
by a factor of roughly $100$. As the spin increases and the
peak emission region moves closer to the horizon, the relative
fraction of observed flux that comes from returning radiation 
increases to $\sim 5\%$ for $a/M=0.9$ and $\sim 20\%$ for
$a/M=0.998$. Naturally, the relative
contribution from returning radiation increases with energy because the
highest energy photons come from the innermost regions of the disk.
In fact, for $a/M=0.998$, the returning radiation actually
dominates the spectrum above $\sim 10$ keV. 

For all spins, the
fraction of returning radiation increases with inclination for two
reasons: forward/backward scattering has a larger differential
cross-section than right-angle scattering, so returning photons, which
characteristically have large angles of incidence, are reflected at
large angles from the axis and preferentially reach
observers with high inclination;
secondly, the direct radiation suffers limb-darkening effects at high
inclination due both to the scattering atmosphere and also the
diminished solid angle of the disk, while both effects affect the
reflected radiation to a lesser degree (see Paper I for a detailed
calculation). For the examples shown in Figures
\ref{total_pol_a0}--\ref{total_pol_a998}, the fraction of observed
flux coming from returning radiation is larger for $i=75^\circ$ than
$i=45^\circ$ by a factor of about two. 

At low energies (below the thermal peak), the flux comes predominantly
from the outer regions of the disk, where relativistic effects are
negligible, so the classic results of \citet{chandra:60} are
reproduced: horizontal polarization with amplitude of a few
percent. The degree of polarization at low energies increases with
observer inclination, ranging from $\delta=1\%$ for $i=45^\circ$ to
$\delta=3.5\%$ for $i=75^\circ$.  Although the free-free absorption
opacity in the disk increases below $\sim 1$ keV, which should
reduce the net polarization, we have ignored this effect for
two reasons. First, large uncertainties in the vertical structure of
the disk make it difficult to arrive at a quantitative prediction for
the absorption profile, which is strongly dependent on local
temperature and density. Second, most proposed X-ray polarization
missions will be sensitive only to photons above 1 keV
\citep{bellazzini:06}, so we choose to focus on this energy range. 

Above the thermal peak, we see the relativistic effects described in
the previous Sections begin to dominate the polarization signal. The
direct radiation (dotted curves) undergoes lensing and beaming,
thereby reducing the polarization amplitude and rotating the
angle. The polarization properties of the returning radiation (dashed
curves) are roughly constant in the $1-10$ keV range, but as the
relative flux from returning radiation increases, we see the {\it total}
polarization makes a transition from following the direct behavior to following
the returning behavior. In the process of making this transition from
horizontal polarization at low energies to vertical polarization at
high energies, the degree of polarization goes through a minimum as
the two contributions cancel each other. This transition is sharpest
for lower inclinations,
where there is a smaller degree of horizontal
polarization to overcome. Yet even for large inclination angles, when
the BH spin is high enough, we still see a dramatic transition from
horizontal to vertical polarization close to the thermal peak, caused by
the enhanced level of returning radiation from the innermost disk. 

Due to this dependence on the relative contribution from returning
radiation, if the emissivity profile is NT, the energy at which 
the polarization transition takes place may be a powerful diagnostic
of BH spin. In Figure \ref{spin_dependence2} we plot the polarization
degree and angle as a function of energy for a range of spin
parameters. In all cases, the BH mass is $10 M_\odot$, $i=75^\circ$, and the
luminosity is $0.1 L_{\rm Edd}$ (due to differences in accretion
efficiency, we use a fixed luminosity, not mass accretion rate, when
comparing different spins). For inclinations of $i=45^\circ$ or
$60^\circ$, this same dependence on spin can be seen by comparing the top and
middle rows of Figures \ref{total_pol_a0}--\ref{total_pol_a998}. 
With even moderate energy resolution and
polarization sensitivity of $\lesssim 1\%$, observations of BHs in the
thermal state should be able to distinguish between Schwarzschild and
extremal Kerr with high significance---if one can safely assume
the NT surface brightness profile.

However, recent relativistic MHD simulations suggest that, contrary
to the NT assumption of zero stress and zero dissipation
inside the ISCO, there are in fact significant stress and
dissipation in the plunging region, leading to
steadily increasing surface brightness (as measured in the local fluid frame)
with decreasing radius \citep{noble:08}. For a simplistic treatment of
this flux, we apply a quasi-Newtonian emissivity profile with
$\mathcal{F}(R) \sim R^{-3}$ all the way to the horizon, regardless of
the BH spin
or ISCO location. The local spectral distribution is still
given by equations (\ref{T_eff}-\ref{Inu_dilute}). The resulting
polarization signals are plotted in Figure \ref{spin_dependence3},
showing a clear reduction in the sensitivity to spin parameter. While
there is still some variation between the different curves, it would
likely be impossible to measure $a/M$ with a first-generation X-ray
polarization mission (see below, Sec.\ \ref{parameters}).

The transition energy in the polarization angle is a strong
function of the luminosity, which in turn is a function of the
temperature of the inner disk.  Since the initial polarization and
subsequent scattering-induced polarization are independent of frequency
at these photon energies, we can think of the luminosity dependence
simply as a rescaling of the entire polarization spectrum with inner
disk temperature.  This scaling is illustrated in Figure~\ref{lum_dependence2},
which plots the polarization signature for a range of luminosities, in
all cases using a NT flux profile with a diluted thermal
spectrum and BH mass $10 M_\odot$, for $a/M=0$ (solid curves) and $a/M=0.9$ (dashed
curves). Since most galactic BHs are observed over a wide range in
luminosities, even within the thermal state, multiple polarization
observations could give even better constraints on the value of the
spin parameter. By reproducing the same polarization at a shifted energy,
such a measurement would also help confirm that the features
in the polarization spectrum are in fact coming from relativistic
effects in the inner disk. Again, if the underlying emission profile
increases continuously down to the horizon (see Fig.~\ref{lum_dependence3}),
we lose some sensitivity in determining the spin, but gain sensitivity
to $L/L_{\rm Edd}$ because the dependence on that variable sharpens.

\section{PARAMETER ESTIMATION}\label{parameters}

\begin{figure}
\caption{\label{harm3d} Local flux (as measured in the fluid frame)
  versus radius for $a/M=0.9$. In the plunging region, the
  flux is parameterized by a power law with index $\alpha$, normalized
  to match the NT profile ({\it solid curve}) just outside
  the ISCO. Also shown is
  the emissivity calculated by {\tt HARM3D} ({\it dotted curve}), a
  3-dimensional global MHD code in full relativity \citep{noble:08}.}
\begin{center}
\scalebox{0.8}{\includegraphics{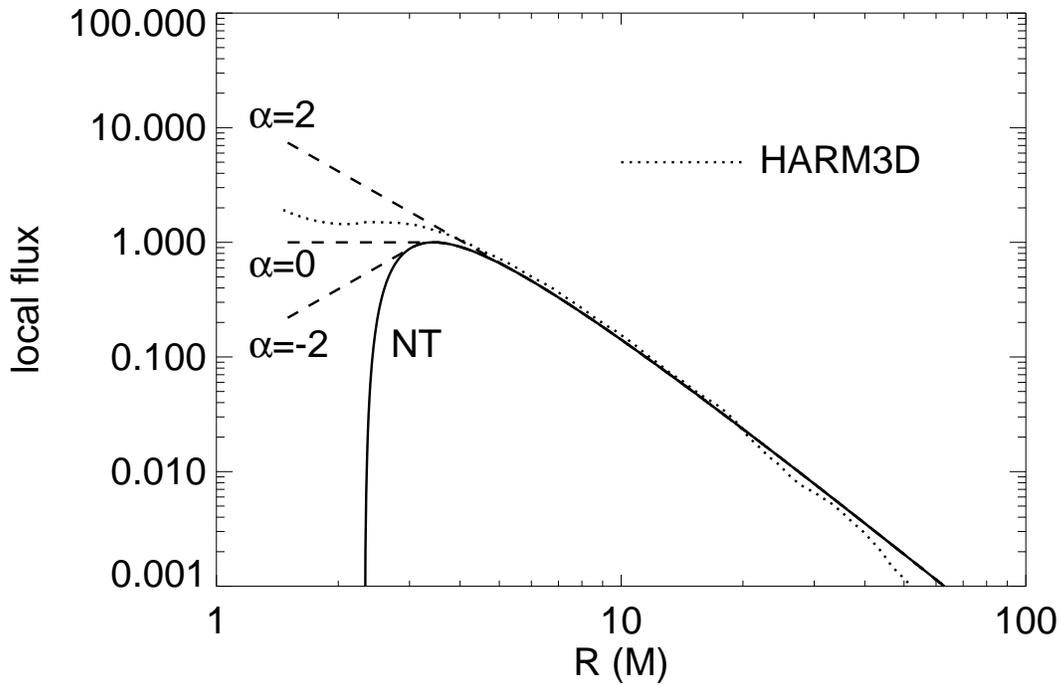}}
\end{center}
\end{figure}

As shown in the previous section, spectropolarimetry measurements of
BHs in the thermal state will give a powerful new way of measuring BH
spin, {\it if} the underlying emission follows a Novikov-Thorne
profile with zero dissipation inside of the ISCO. More generally,
polarization observations will probe the emissivity profile directly,
offering a new way to measure the temperature profile as a function of the
geometrized radius $(R/M)$. Well outside the ISCO, conservation laws
constrain all models of time-steady disks to follow the NT form;
however, nearer the ISCO and throughout the plunging region inside it, more
variation is possible.  Motivated by global MHD simulations such as those
presented in \citet{noble:08} and \citet{beckwith:08}, we
describe the range of possible
dissipation profiles by power-laws in radius in the plunging region
matched smoothly to the NT profile just outside the
ISCO. Figure \ref{harm3d} shows a sample of emission profiles for
$a/M=0.9$ and a range of power-law index $\alpha$. In the limit
$\alpha \to -\infty$, we reproduce NT;
$\alpha=3$ is the pseudo-Keplerian case shown above in Figures
\ref{spin_dependence3} and \ref{lum_dependence3}. Also shown for
comparison is the dissipation profile calculated by {\tt HARM3D}
for the case of $a/M = 0.9$ and a disk aspect
ratio $H/R\simeq 0.1$, which corresponds
to $\alpha \approx 0.5-1$ \citep{noble:08}.

To estimate the ability of polarization observations to
accurately determine $a/M$ and $\alpha$ for a typical galactic BH binary,
we consider two theoretical X-ray polarimeters: a ``first-generation''
instrument with
modest energy range $1-10$ keV, resolution $\Delta E/E \approx 1$,
and minimum polarization sensitivity of $\delta \gtrsim 1\%$ with
$1\sigma$ confidence. A ``next-generation'' detector is characterized
by a broader energy range $0.1-100$ keV, better resolution $\Delta E/E
\approx 0.1$, and ten times the collecting area,
giving $\delta \gtrsim 0.3\%$ for the same source. We then select a
few sample ``target'' models
with different emission profiles and generate simulated polarization
data for these cases using the ray-tracing code described
above. Scanning over a range of model parameters, we
calculate a large number of ``template'' polarization spectra to be
matched against the targets.

For each point in parameter space, we calculate the overlap between
a template spectrum and the target spectrum by calculating $\chi^2$ for
the specified detector sensitivity and source luminosity. Rather than
attempting to combine the absolute luminosity, distance, detector
response function and collecting area into an error estimate for a
generic spectro-polarimeter, we simply describe the instrument and
source together in terms of the minimum error achievable in a
measurement of $\delta$ for a single energy bin. For a broad-band
detector with constant sensitivity at all energies, the minimum error
will correspond to the peak of the observed intensity spectrum, i.e.\
the bin with the most photon counts. The error in any other bin simply
scales like the square root of the counts in that bin. We use for our
observables the Stokes parameters $Q_\nu$ and $U_\nu$, both normalized
to the {\it total} observed flux $I_{\rm tot}=\int I_\nu d\nu$. For a
peak sensitivity (minimum error) of $\delta_{\rm min}$, the error on
each value of $Q_\nu$ and $U_\nu$ is given by
\begin{equation}\label{delta_nu}
\Delta Q_\nu = \Delta U_\nu = \delta_{\rm min}\, I_\nu\, \sqrt{\frac{I_{\rm
      peak}}{I_\nu}}. 
\end{equation}
With the target values of $Q_\nu$ and $U_\nu$ and their associated
measurement errors at each energy, it is straightforward to calculate
the $\chi^2$ fit for the match between the target ``observation'' and
any given template spectrum. By generating multiple templates to cover
the parameter space, we can estimate the
significance with which the model parameters can be determined from
future data. 

In practice, this search through parameter space would be
prohibitively expensive computationally if each template were
constructed using the full Monte Carlo calculation (typically $\sim
10^8$ rays traced per run). Fortunately, we are able to utilize
transfer functions similar to those used in calculating the
relativistic effects on broad-band continuum spectra (e.g.\
\citet{davis:05}). The full details are described in Paper I, but a
short summary of the method is as follows: for each value of the spin
parameter (roughly 20 points spaced evenly in $\log (1-a/M)$),
a single complete Monte Carlo simulation is carried out as above, with
the emission spectrum at each radius given by a delta-function in
energy. The initial polarization is still given by a
scattering-dominated atmosphere, and the returning radiation is
treated as before, typically adding a strong vertical component to the
observed polarization. When each ray reaches an observer at infinity,
the Stokes parameters $I_\nu$, $Q_\nu$, and $U_\nu$ are recorded, each
looking essentially like a delta function shifted in energy and
amplitude by the relativistic effects of the ray-tracing. 

\begin{figure}
\caption{\label{lum_dependence2} Polarization degree and angle 
  for a range of luminosities for $a/M=0$ (solid curves) and
  $a/M=0.9$ (dashed curves). All systems have inclination
  $i=75^\circ$, BH mass $10 M_\odot$, and Novikov-Thorne radial emission profiles.} 
\begin{center}
\scalebox{0.8}{\includegraphics{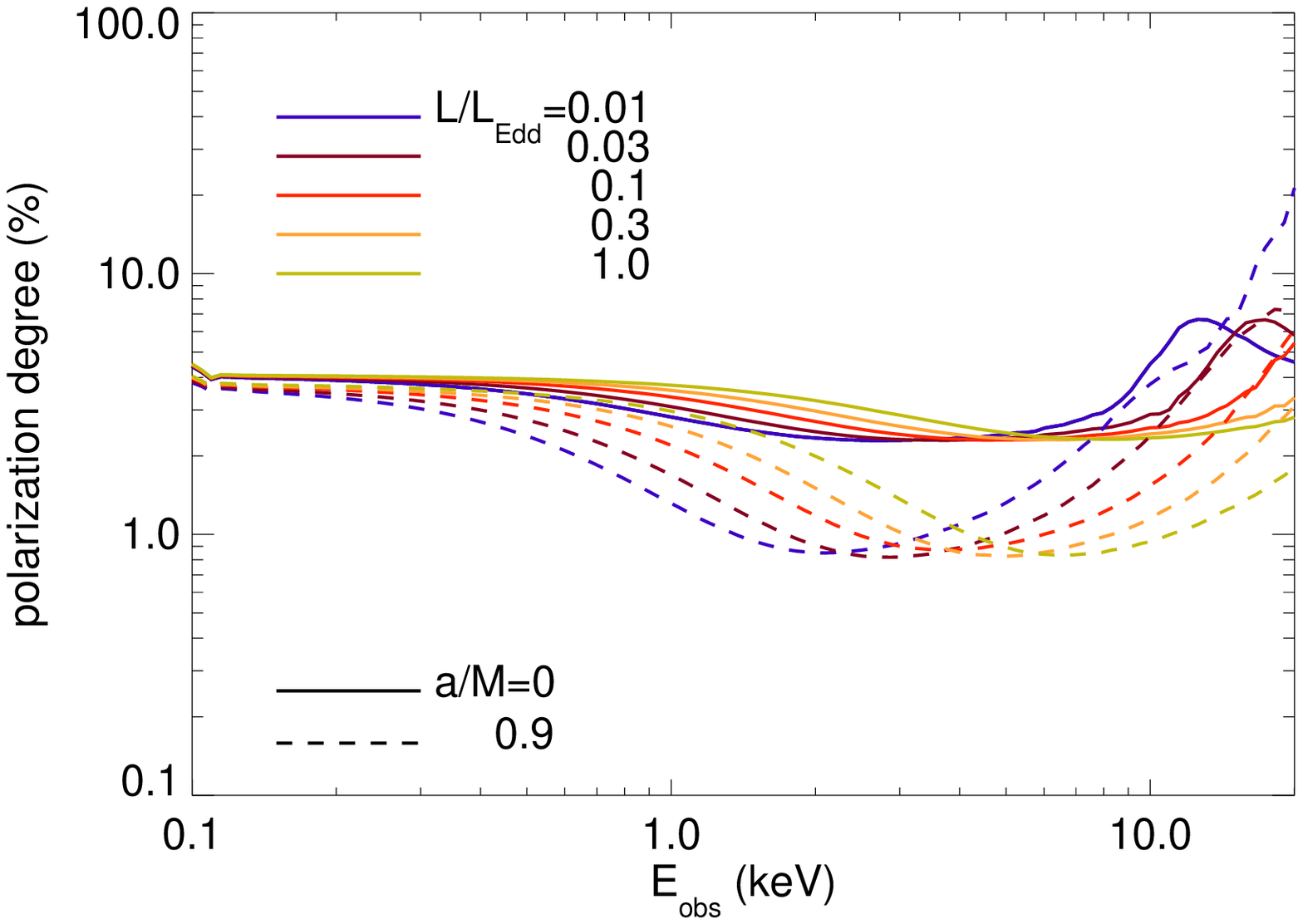}}
\scalebox{0.8}{\includegraphics{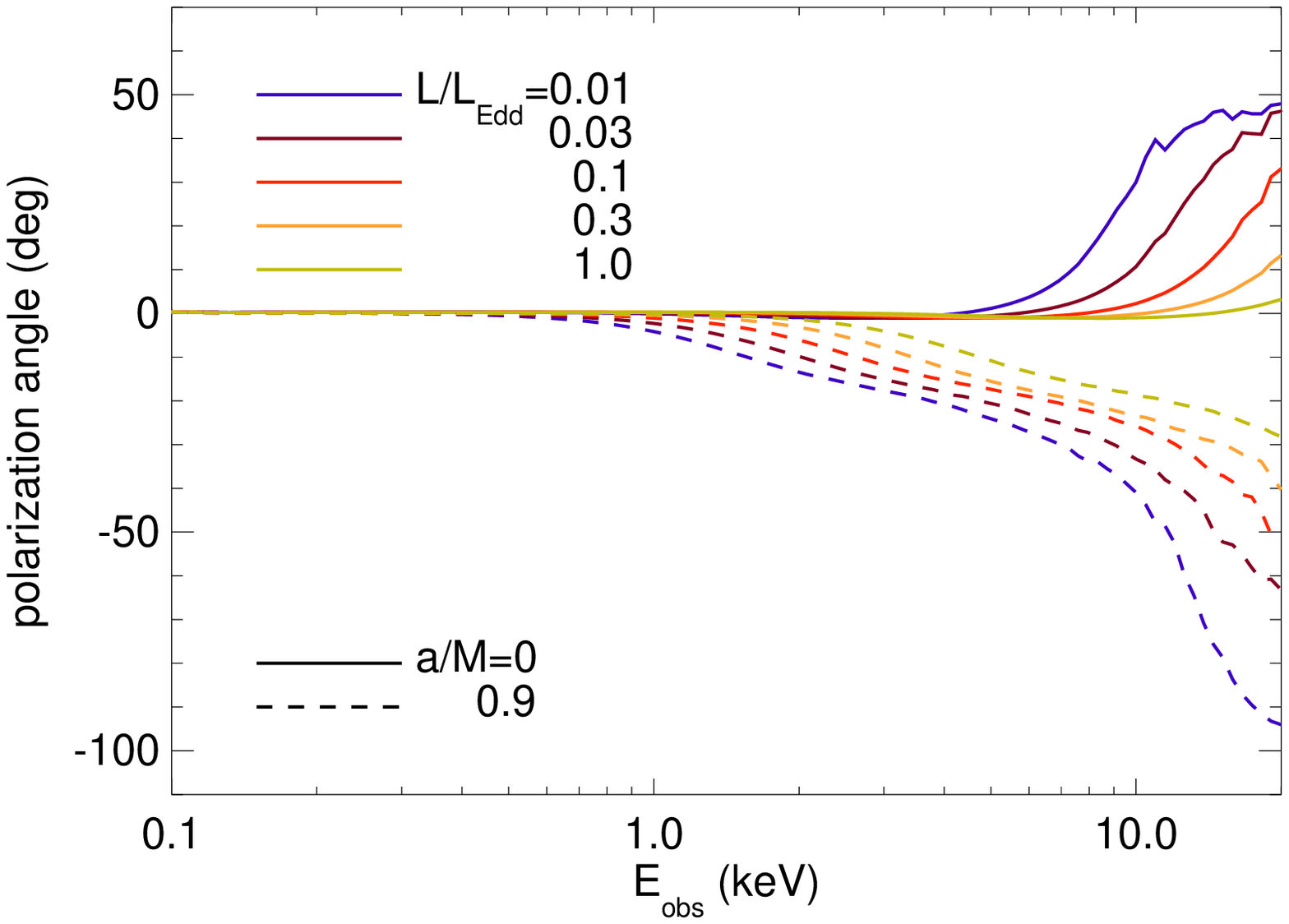}}
\end{center}
\end{figure}

\begin{figure}
\caption{\label{lum_dependence3} Same as Figure
  \ref{lum_dependence2} but for a power-law emissivity profile $\mathcal{F}(R)
  \sim R^{-3}$ all the way to the horizon.}
\begin{center}
\scalebox{0.8}{\includegraphics{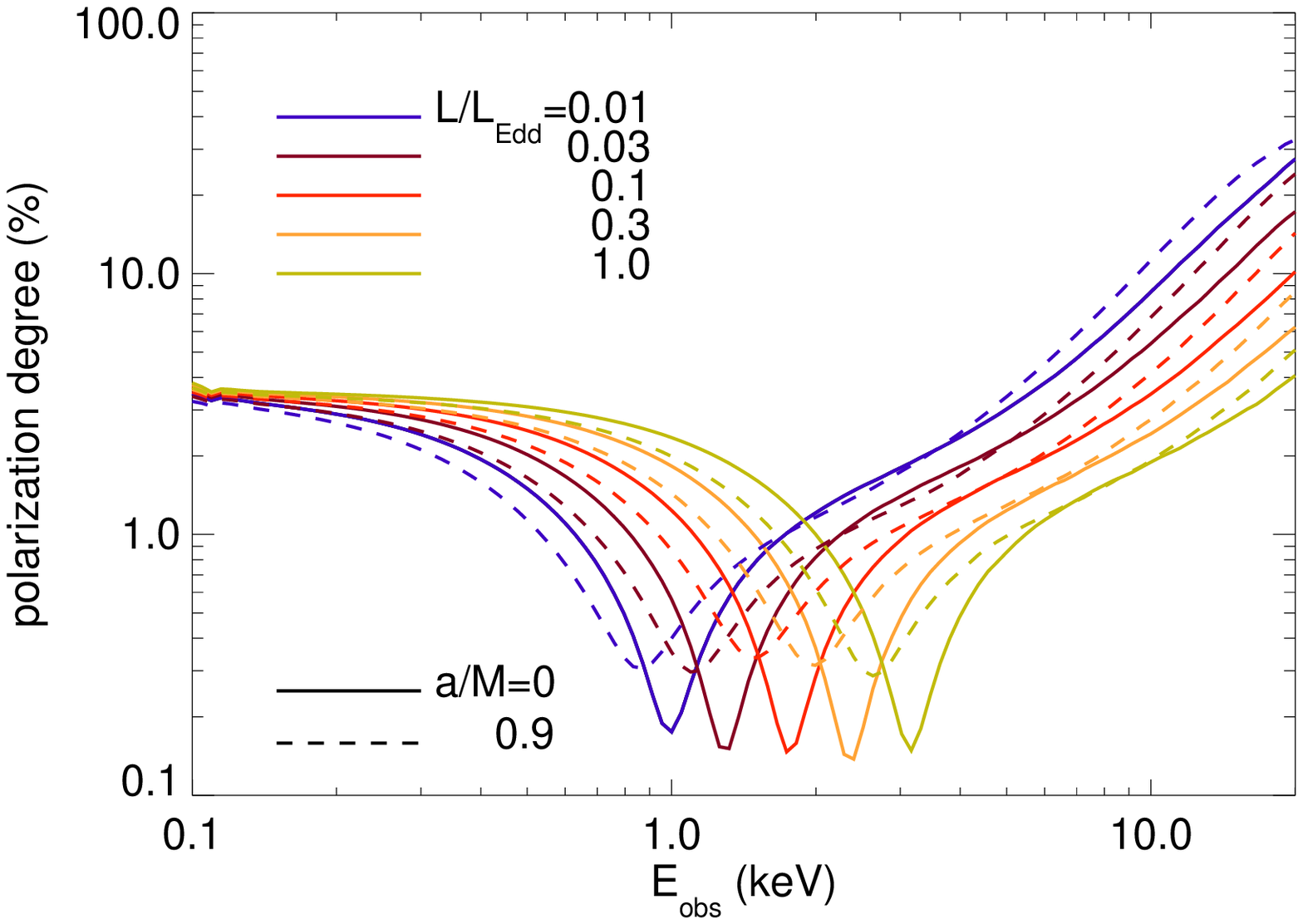}}
\scalebox{0.8}{\includegraphics{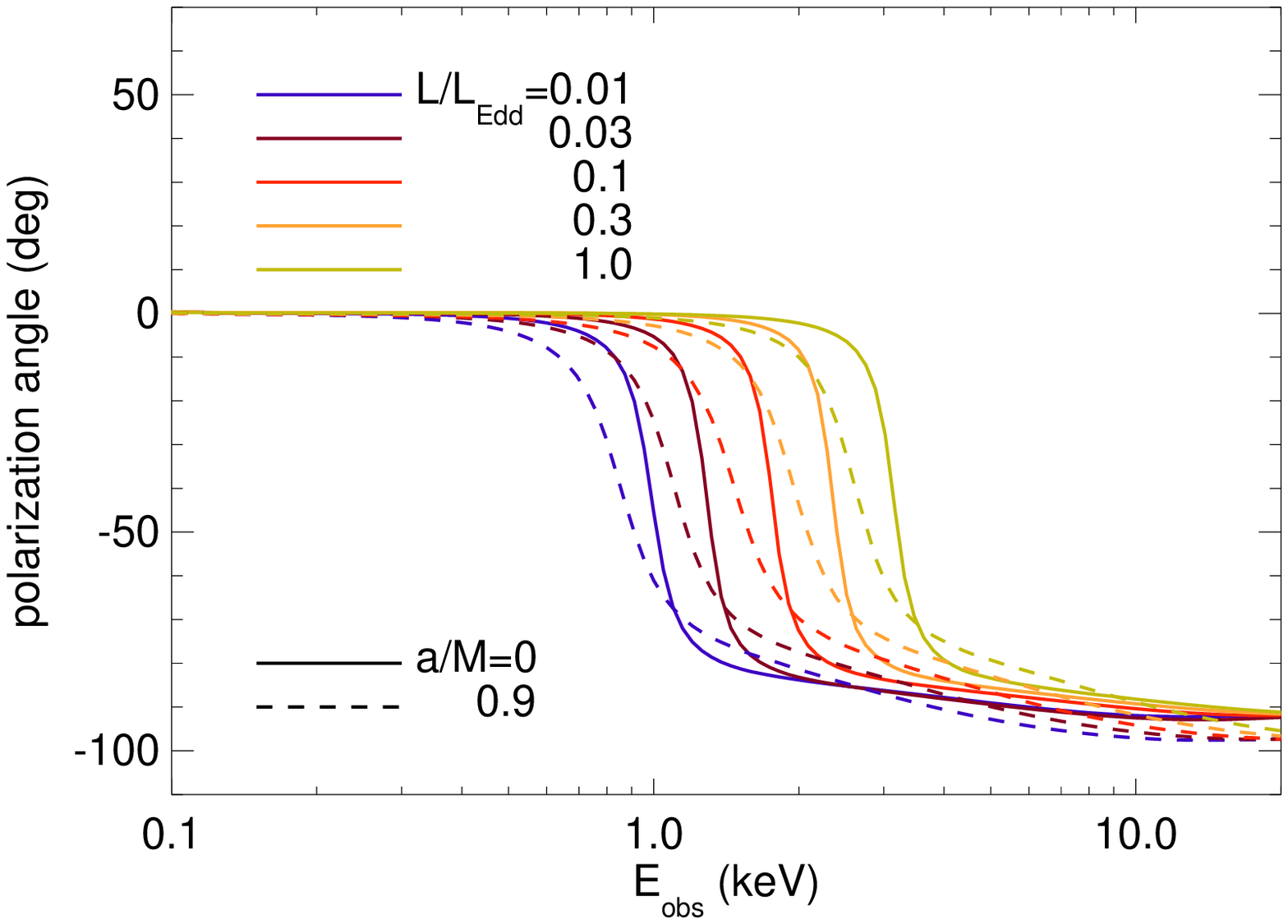}}
\end{center}
\end{figure}

Sorting all the rays by observer inclination angle, we construct a
transfer function $g(a/M)$ that takes
the input emission profile $I_\nu^{\rm em}(R)$ and generates the
observed polarization in terms of
$I_\nu^{\rm obs}(i)$, $Q_\nu^{\rm obs}(i)$, and $U_\nu^{\rm
  obs}(i)$. For a given spin, a single transfer function can be used
to generate polarization spectra for any set of model parameters $M$,
$L/L_{\rm Edd}$, and $\alpha$, all of which combine to give
$I_\nu^{\rm em}(R)$. The transfer function can also be used to
calculate the transition radius $R_{\rm trans}$ described above in
Section \ref{returning}. For emission profiles with steep
gradients around $R_{\rm trans}$ (as would
be produced by large values of
$a/M$ or $\alpha$; see Figs.\ \ref{spin_dependence3} and
\ref{lum_dependence3}), the observed polarization spectrum displays
a sharp swing around the photon energy corresponding to the
disk temperature at that radius: $E_{\rm trans} \approx 3 T(R_{\rm
trans})$. On the other hand, moderate spin systems
with NT emission profiles have a shallow temperature gradient around
$R_{\rm trans}$, giving a broad swing in the polarization angle
(Figs.\ \ref{spin_dependence2} and \ref{lum_dependence2}), or
none at all in low-spin cases where there is simply not enough
emission from inside $R_{\rm trans}$ to overcome the horizontal
contribution from emission outside of $R_{\rm trans}$.

For reasonably high resolution spectra, we find
the transfer function method gives a speed-up factor of better than
10,000 compared to the direct Monte Carlo calculation. This method is
particularly straightforward in the thermal state, where the
scattering cross sections and polarization transport are essentially
independent of energy. However, even when including inverse-Compton
coronal scattering, the transfer function method is still
applicable, with the added dimension that the transfer function is
dependent on the seed photon energy and accretion geometry (again, see
Paper I for details).

\begin{figure}
\caption{\label{gen1_cont} Contour plots of fit quality in matching
  emissivity profiles to simulated data for a first-generation X-ray
  polarimeter. The left column corresponds to simulated data from a
  non-spinning Schwarzschild BH with a NT emissivity
  profile $(\alpha=-\infty)$, the center column has $a/M=0.97$ and
  $\alpha=1$, and the right column has $a/M=0.998$ and
  $\alpha=-\infty$. The top row shows the fit quality regions assuming
  the luminosity in Eddington units and the disk inclination are
  both known; the
  middle row assumes we know the disk inclination but not the
  accretion rate, and the bottom row assumes we know neither. The
  color coding is $\le 1\sigma$ (white), $2\sigma$ (blue),
  $3\sigma$ (purple), $4\sigma$ (red), $5\sigma$ (orange), and $>
  5\sigma$ (yellow), where $\sigma$ is the variance of the $\chi^2(\nu)$
  distribution, as defined in the text.}
\begin{center}
\scalebox{0.35}{\includegraphics*[52,415][540,720]{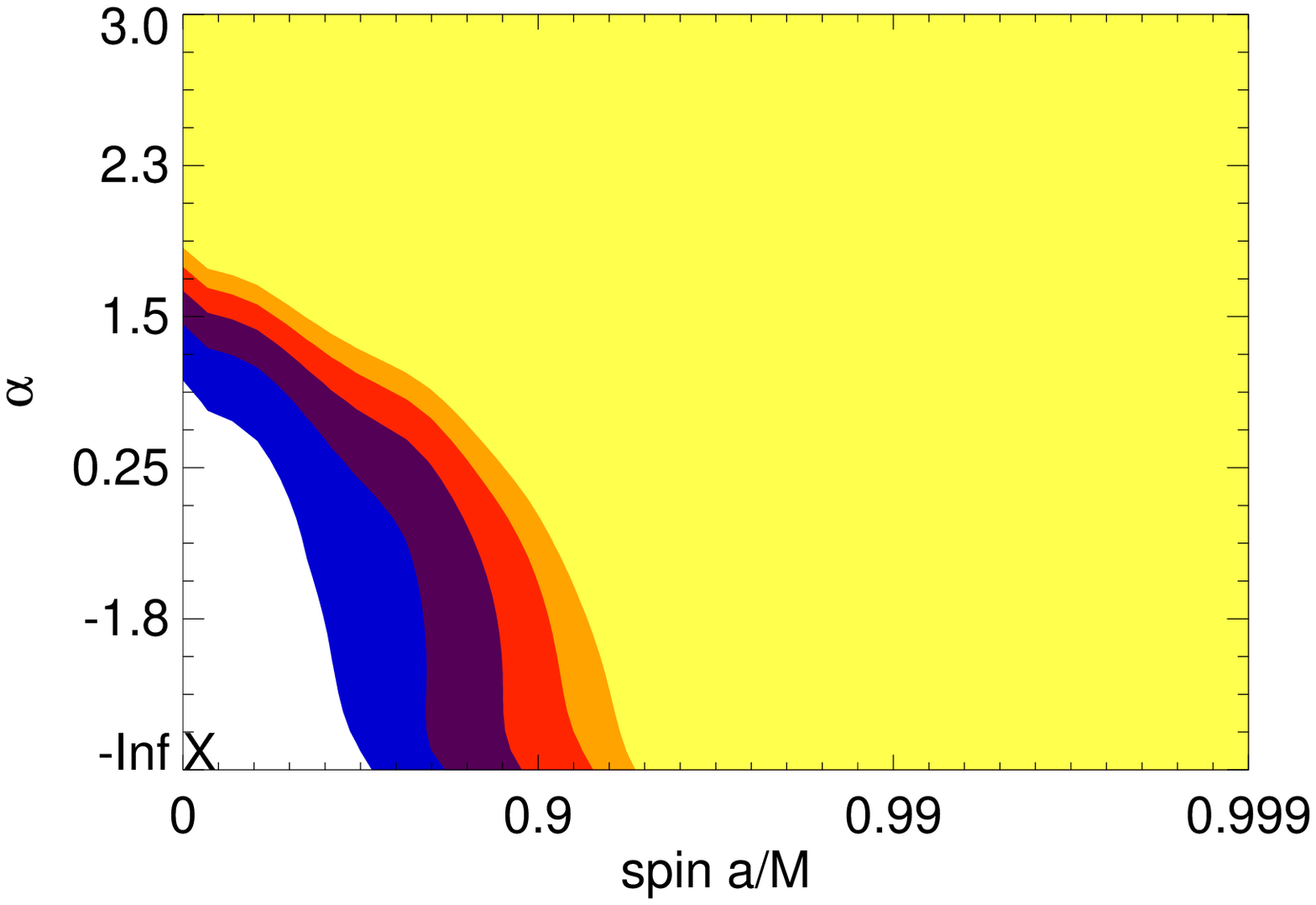}}
\scalebox{0.35}{\includegraphics*[145,415][540,720]{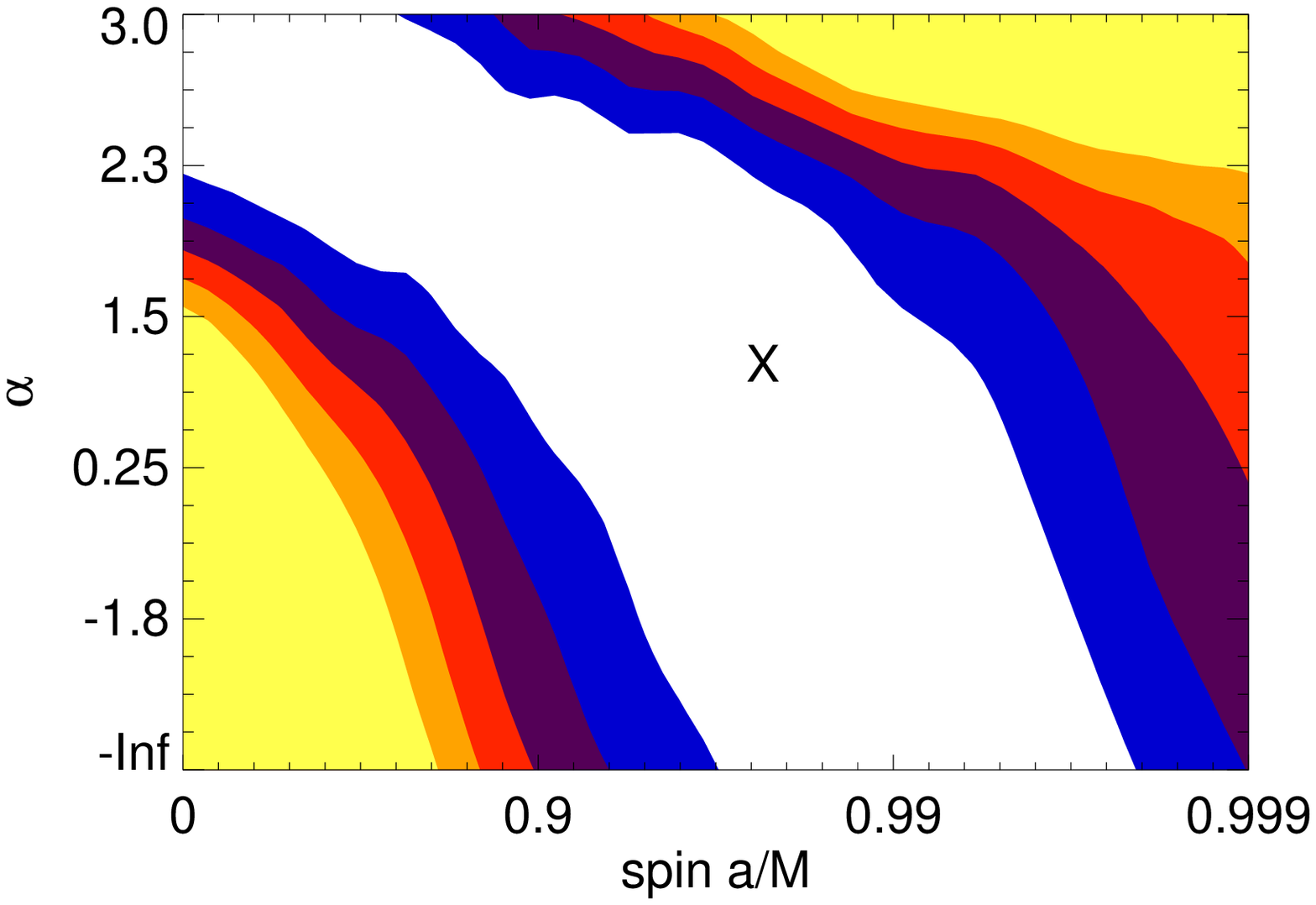}}
\scalebox{0.35}{\includegraphics*[145,415][540,720]{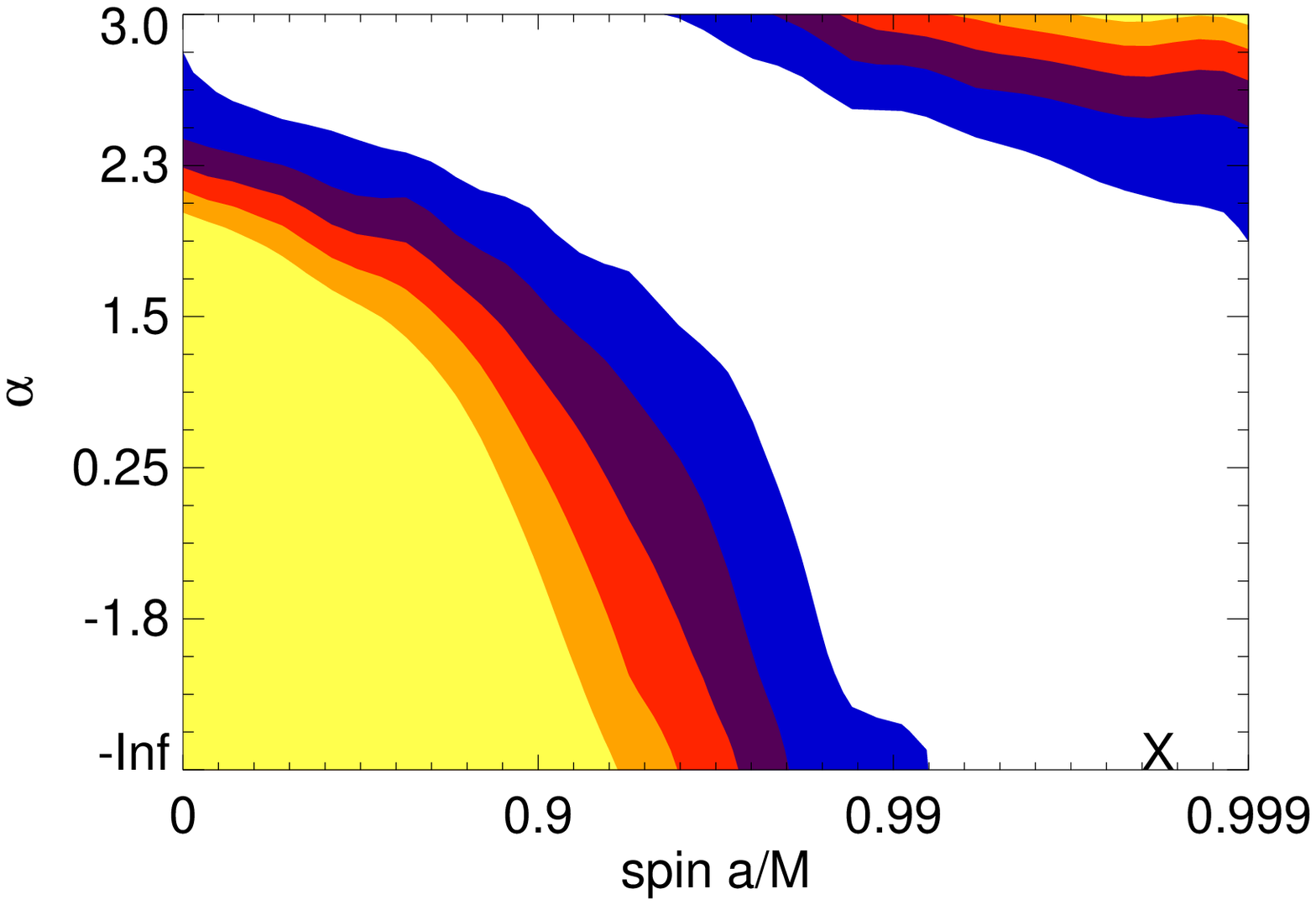}}\\
\scalebox{0.35}{\includegraphics*[52,415][540,720]{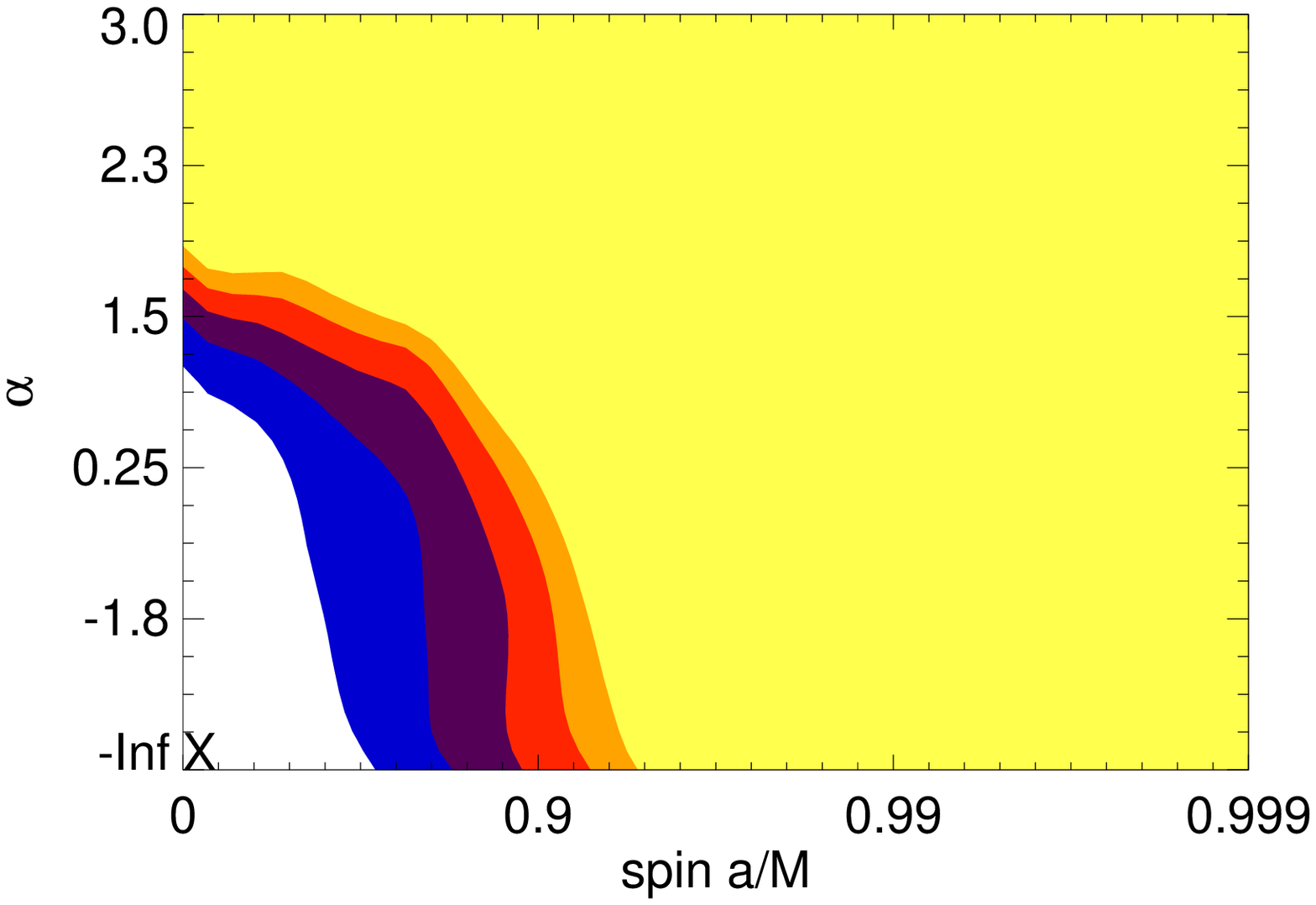}}
\scalebox{0.35}{\includegraphics*[145,415][540,720]{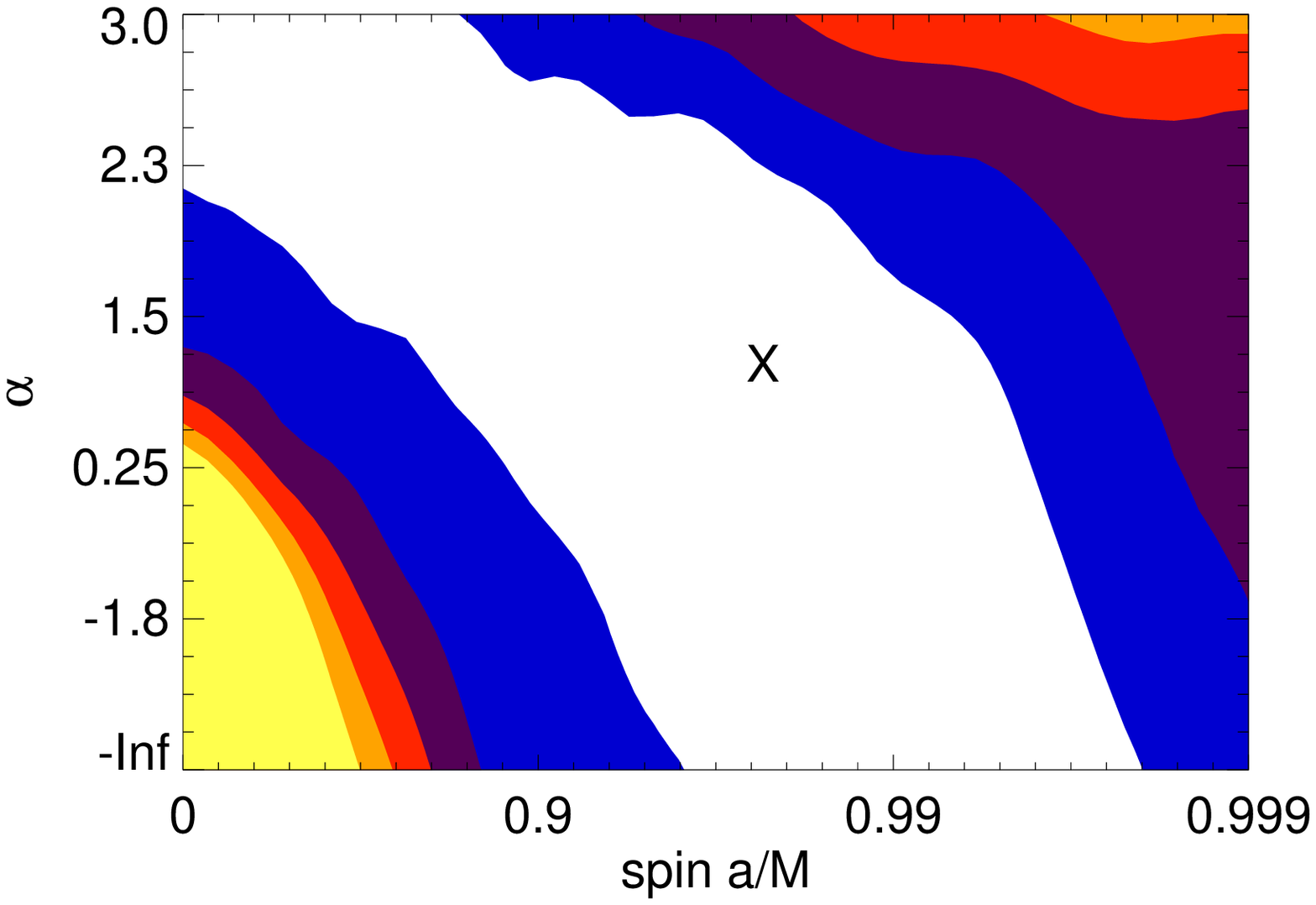}}
\scalebox{0.35}{\includegraphics*[145,415][540,720]{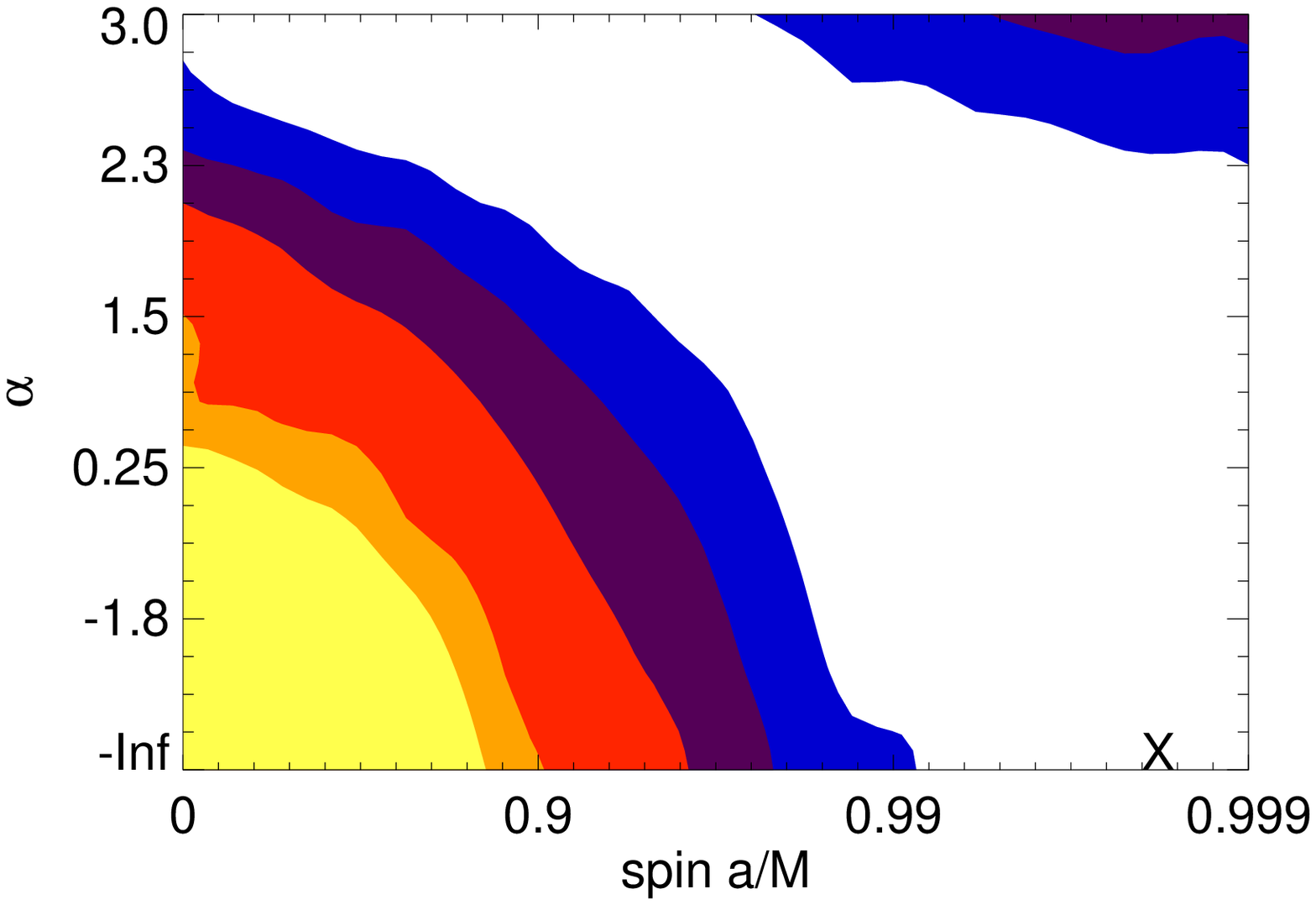}}\\
\scalebox{0.35}{\includegraphics*[52,360][540,720]{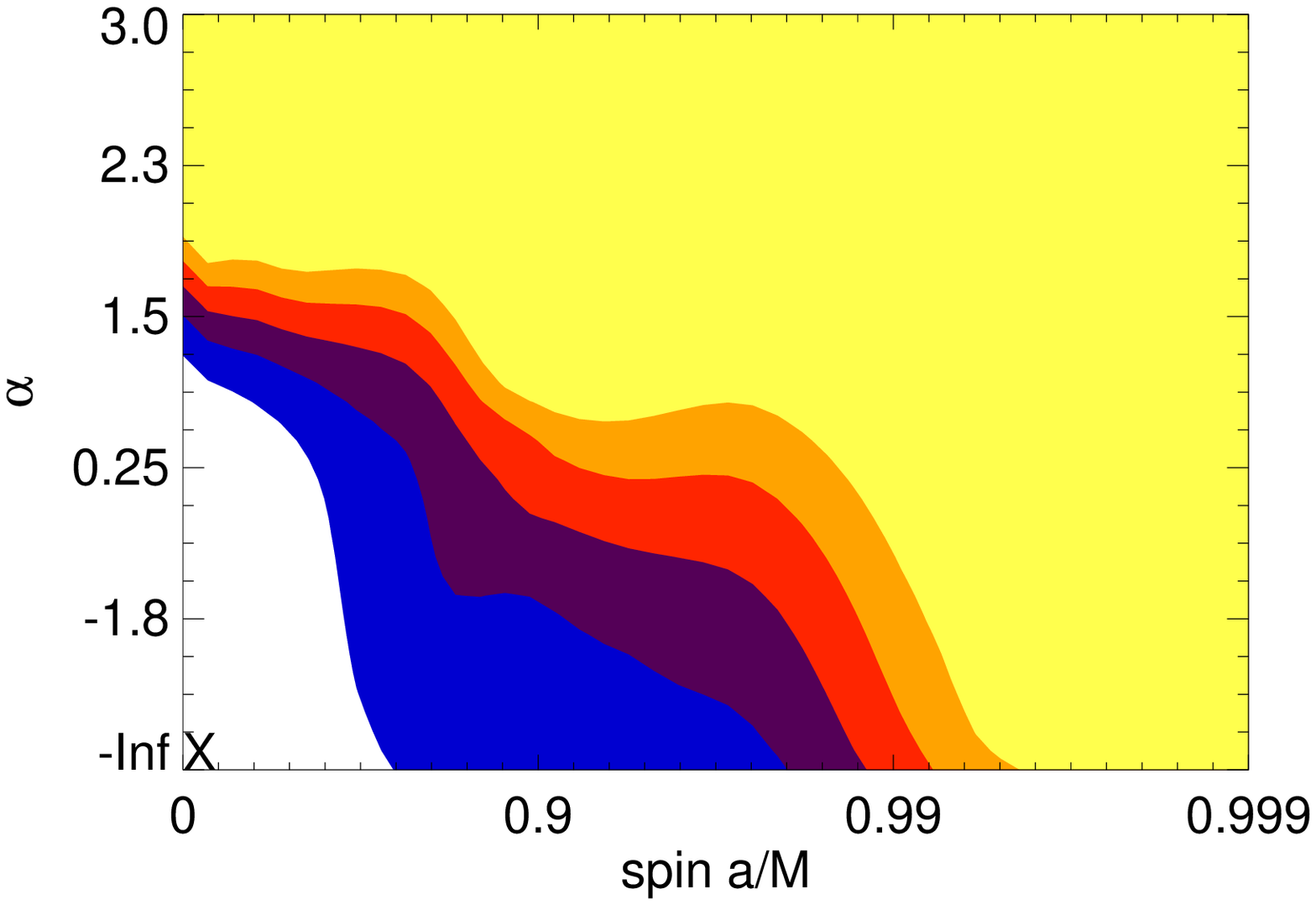}}
\scalebox{0.35}{\includegraphics*[145,360][540,720]{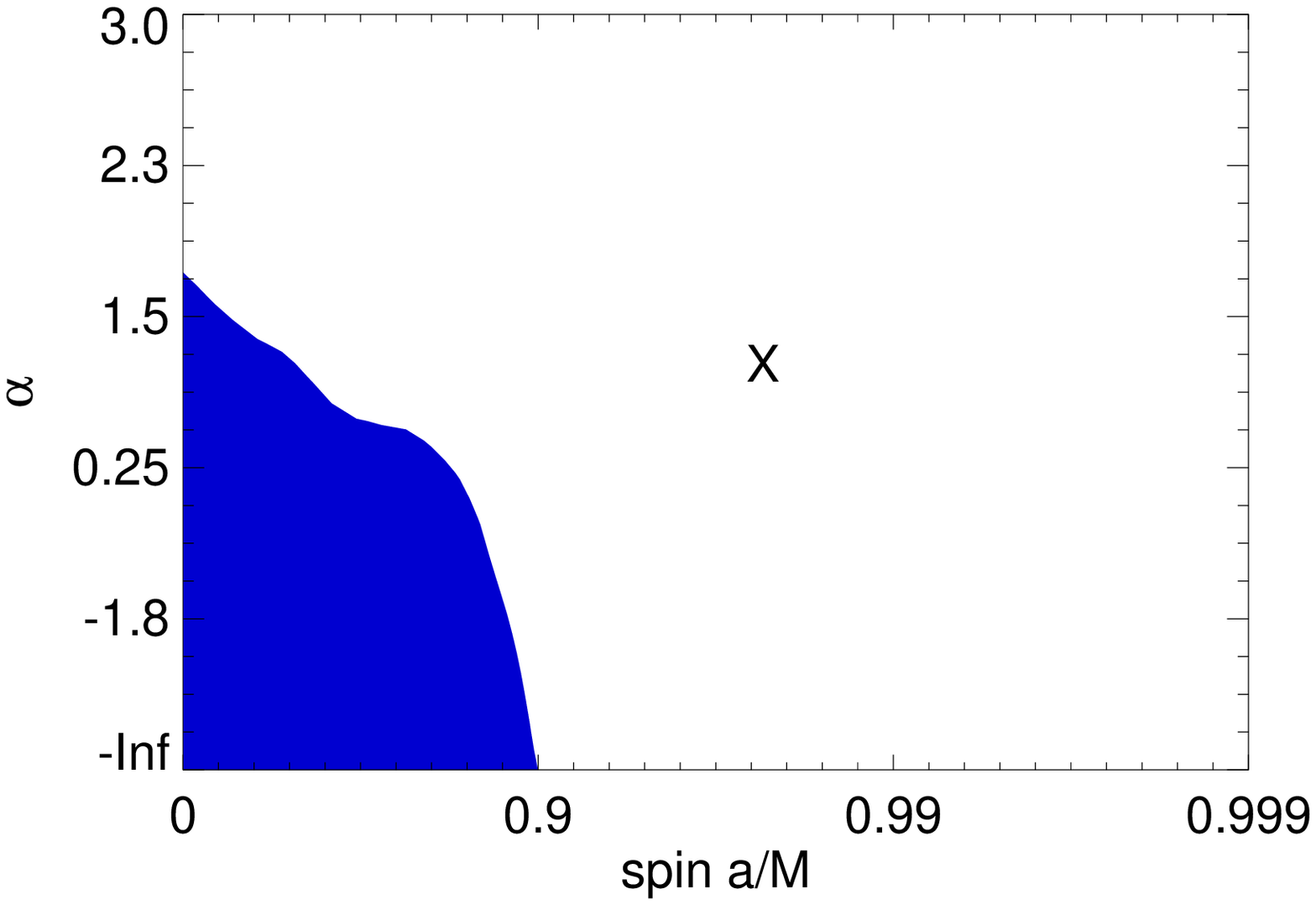}}
\scalebox{0.35}{\includegraphics*[145,360][540,720]{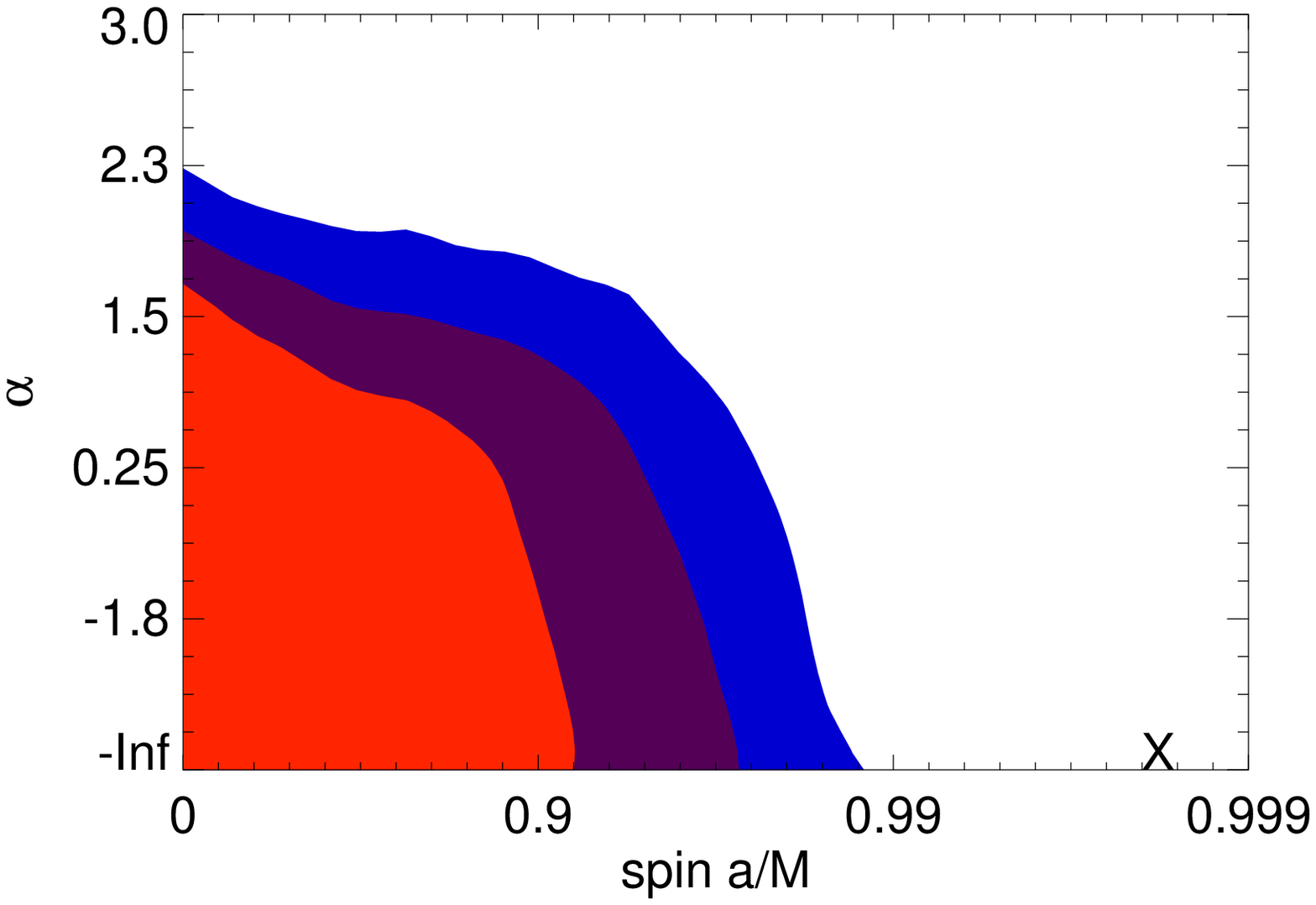}}
\end{center}
\end{figure}

In Figure \ref{gen1_cont}, we show quality-of-fit contours for matching
the target parameters of three different models: two NT
radial profiles with $a/M=0$ and $a/M=0.998$, and one case with
additional emissivity in the plunging region with $\alpha=1$ and
$a/M=0.97$. All three cases have $M=10 M_\odot$, $L/L_{\rm Edd}=0.1$,
and $i=75^\circ$. The top row of figures corresponds to cases where the disk
inclination and Eddington-scaled luminosity (i.e.\ the BH mass and distance)
are known {\it a priori}, and the target data is fit by scanning over $a$ and
$\alpha$. The middle row repeats this calculation, but assumes prior
knowledge only of the disk inclination, and then for each value of $a$
and $\alpha$ minimizes $\chi^2$ over $L/L_{\rm Edd}$. Finally, the
bottom row assumes no prior knowledge about the system and minimizes
$\chi^2$ with respect to $L/L_{\rm Edd}$ and inclination $i$. In each
frame, the target parameters are marked with an 'X' and the contours
show confidence intervals in significance units:
$\le 1\sigma$ (white),
$2\sigma$ (blue), $3\sigma$ (purple), $4\sigma$ (red), $5\sigma$
(orange), and $> 5\sigma$ (yellow). Here $\sigma$ is given by a
standard $\chi^2$ distribution: $\sigma=\sqrt{2\nu}$, with $\nu$ the
number of data points minus the number of free parameters. For large
values of $\nu$, these confidence intervals correspond very closely to
those of a normal distribution: (68\%, 95.5\%, 99.7\%, etc.). 

As seen in Figure \ref{lum_dependence2}, the Schwarzschild case
closely resembles the Chandrasekhar limit of horizontal polarization
with amplitude of a few percent. When the emission cuts off at the
ISCO, relativistic effects play a much smaller role in rotating the
polarization of the direct radiation. Furthermore, the fraction of
returning radiation is much smaller, minimizing the contribution from
vertically polarized photons. The only way to generate
such a polarization spectum is by producing the observed flux at large
radii, from where the photons can propagate through nearly flat space
to the distant observer, arriving with their polarization
essentially unchanged. Thus if the observed
polarization is relatively strong ($\gtrsim 1\%$), horizontally
oriented, and constant with energy, then the emission likely comes
from large radii, in turn constraining the spin to be relatively
small and strongly limiting the amount of dissipation inside the
ISCO. This is evident from the upper-left frame in Figure
\ref{gen1_cont}, where a polarization observation of a Schwarzschild
BH with a NT dissipation profile rules out the possibility
of a high value for $a/M$ or $\alpha$. 

On the other hand, when matching the polarization from a near-maximal
spin BH with $a/M=0.998$ (upper-right frame of Fig.\ \ref{gen1_cont}),
any system with a NT emission profile and
spin $a/M \lesssim 0.97$ is ruled out at the $3\sigma$ confidence
level. Yet as described above in Section \ref{results}, a non-spinning
BH can appear to be rapidly spinning if there is significant emission
inside the ISCO. Additionally, since the plunging region is so small
for large values of $a/M$, templates with a range of $\alpha$ (i.e.\ the
emission profile inside the ISCO) are nearly degenerate, as seen in
Figure \ref{spin_dependence3}. These systems are characterized by a
sharp transition in the polarization: it changes from strong,
horizontal polarization at low energies to strong, vertical
polarization above the thermal peak. 

For intermediate cases with moderately high spin 
and $\alpha$ [($0.9\lesssim a/M \lesssim 0.99$); ($-2\lesssim \alpha
\lesssim 2$)], there is a broad minimum in the polarization amplitude
between 1 and 10 keV (see, e.g., the red and orange curves in Fig.\
\ref{spin_dependence2}). If the inclination is known to be relatively
high, as in our test cases, then this broad minimum can still constrain
$a/M$ and $\alpha$ (or at least some combination of the two
parameters) reasonably well, as seen in the upper-middle plot
of Figure \ref{gen1_cont}. However, if we have no prior knowledge of
the disk inclination, we could reproduce such a feature
by assuming a low-inclination system, whose light would have
little or no polarization near the thermal peak
regardless of the emission profile. Therefore, as shown in the lower-middle
panel of Figure \ref{gen1_cont}, when minimizing $\chi^2$ over all
model parameters, it is always possible to get a decent fit to an
unpolarized source, and the emission profile is poorly constrained. At
the same time, if a non-zero polarization {\it is} detected, it can
give a measure not only of the emission profile, but also of the disk
inclination.

There are also a number of BH binary systems for which
the inclination is
known relatively well, but the BH mass $M$ and/or distance $D$ are not known.
This is the situation illustrated in the middle row of
Figure~\ref{gen1_cont}.  As those panels show, substantial constraining
power is retained even without knowledge of $L/L_{\rm Edd}$ (and as the bottom
panels show, even when the inclination angle is also unknown,
spectropolarimetry still provides some limit on the possible range of
$\alpha$ and $a/M$).  The reason is that
with a {\it single} spectropolarimetry observation, we
can measure, using the shape and energy of the polarization transition,
the temperature profile $T(r)$ in the inner disk as a
function of radius in geometric units $r\equiv R/M$. From this
temperature profile, we can constrain the system's ratio of bolometric
luminosity to mass through the relation:
\begin{equation}
L \propto M^2 \int r\, T(r)^4\, dr\, .
\end{equation}
Using the net flux gives $L/D^2$, so that we have two
constraint equations for three unknowns. If either the BH mass
or the distance can be measured by some other means,
all three variables ($L$, $M$, and $D$) will be known. 

Viewed in this way, spectropolarimetry measurements can be
seen to have significant advantages relative to pure spectral measurements.
When the inclination, mass, and distance are all known,
detailed continuum-fitting can be used to measure the temperature
profile in the inner disk, giving constraints on 
$\alpha$ and $a/M$ that are comparable to what is shown in the top row of
Figure~\ref{gen1_cont}.  However, when one or more of these quantities
(inclination, distance, or mass) is not
known, the continuum-fitting method becomes less able to
constrain the radial emission profile and thus the BH spin.

At the qualitative level, one can understand the greater power
of spectropolarimetry as simply the result of having
more observables: $\delta(E_{\rm obs})$
and $\psi(E_{\rm obs})$ in addition to $F(E_{\rm obs})$, giving more
constraints on the model parameters. However, there are more
specific reasons as well.  On the one hand, inferring parameters
from continuum spectra can be very ambiguous because
there are multiple ways to reproduce an observed spectrum
by varying $M$, $D$, $i$, and $L/L_{\rm Edd}$. On the other
hand, especially for high-inclination systems, the polarization
tends to constrain individual parameters more directly: for
example, its magnitude at low energies (where relativistic effects
are weak) directly relates to the inclination. At higher energies, the
distinctive swing in polarization angle is a direct probe of extreme
gravitational lensing and returning radiation, giving a sensitive
indicator of strongly relativistic effects in the inner disk.

\begin{figure}
\caption{\label{gen2_cont} Contour plots of fit quality in matching
  emissivity profiles to simulated data for a next-generation X-ray
  polarimeter with broader energy range and larger collecting
  area. The different frames and their color codes are the same as in
  Figure \ref{gen1_cont}.}
\begin{center}
\scalebox{0.35}{\includegraphics*[52,415][540,720]{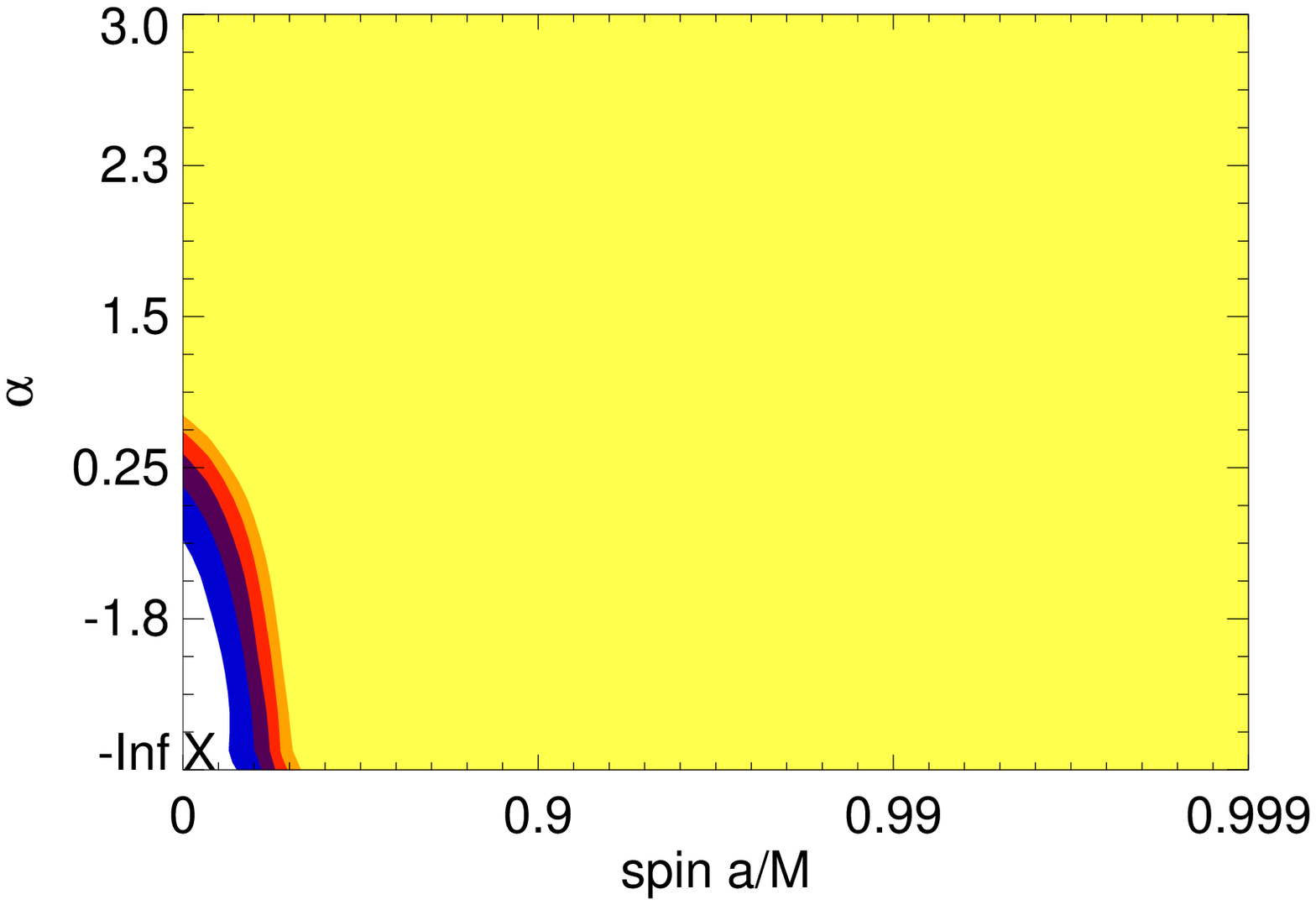}}
\scalebox{0.35}{\includegraphics*[145,415][540,720]{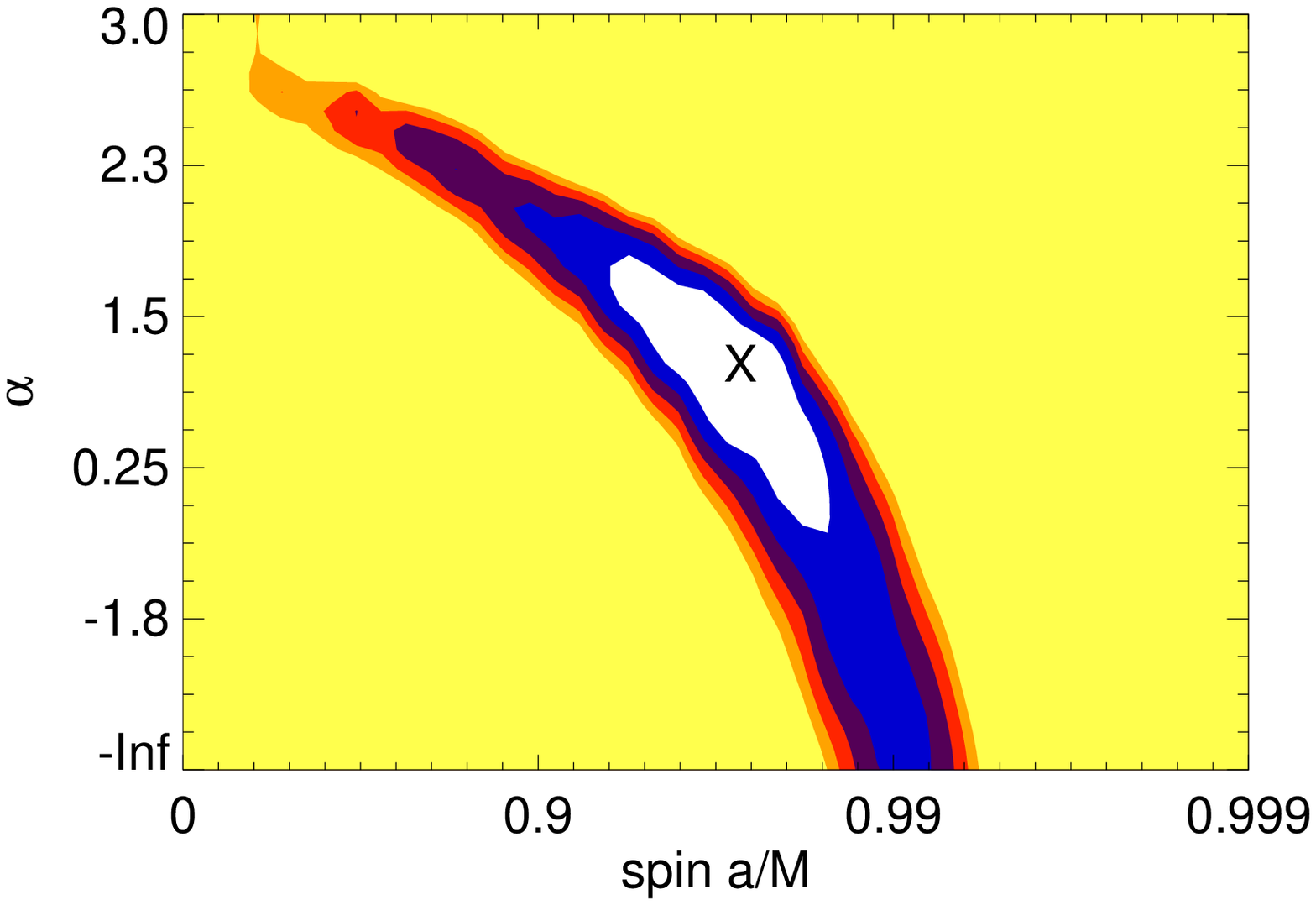}}
\scalebox{0.35}{\includegraphics*[145,415][540,720]{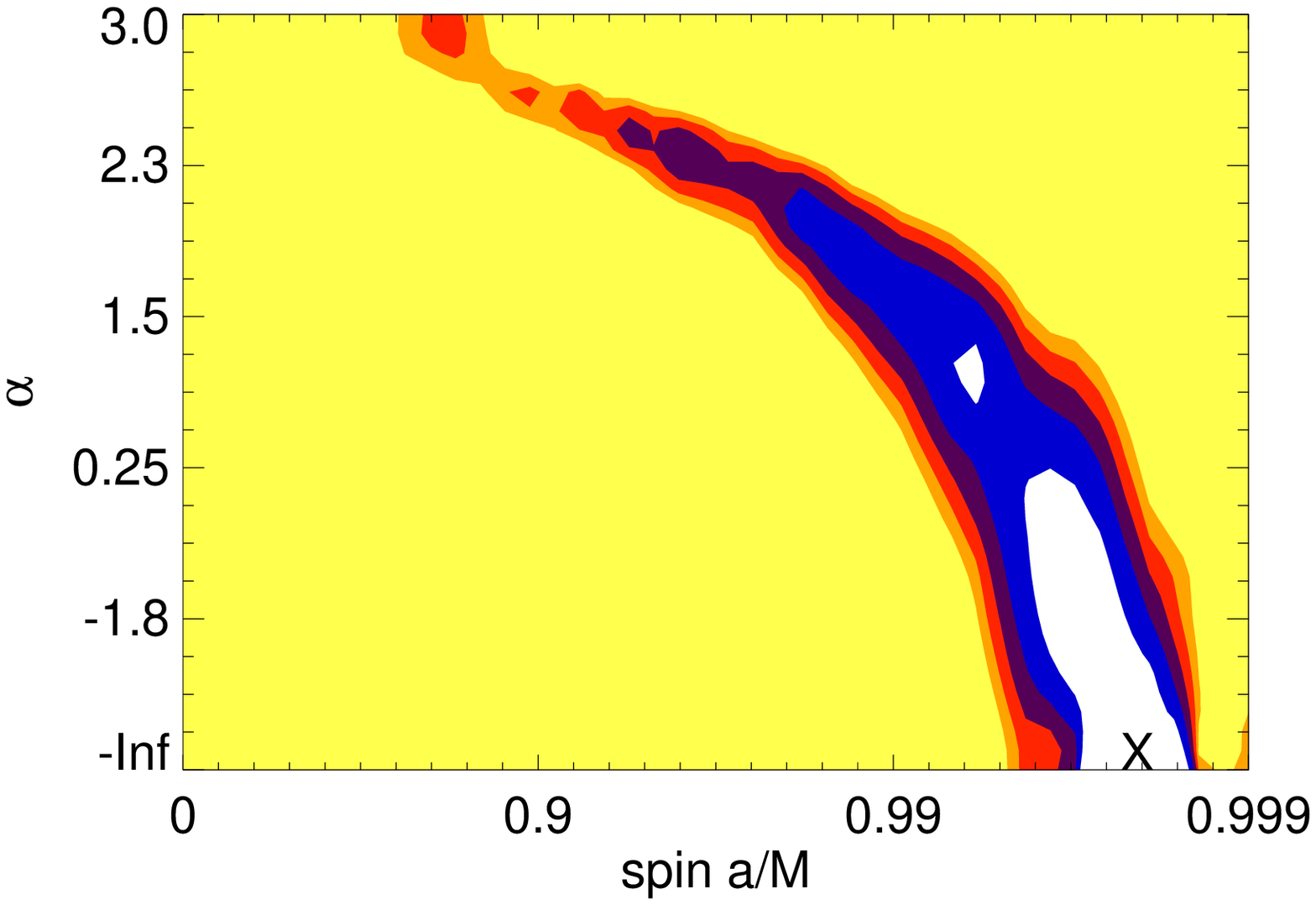}}\\
\scalebox{0.35}{\includegraphics*[52,415][540,720]{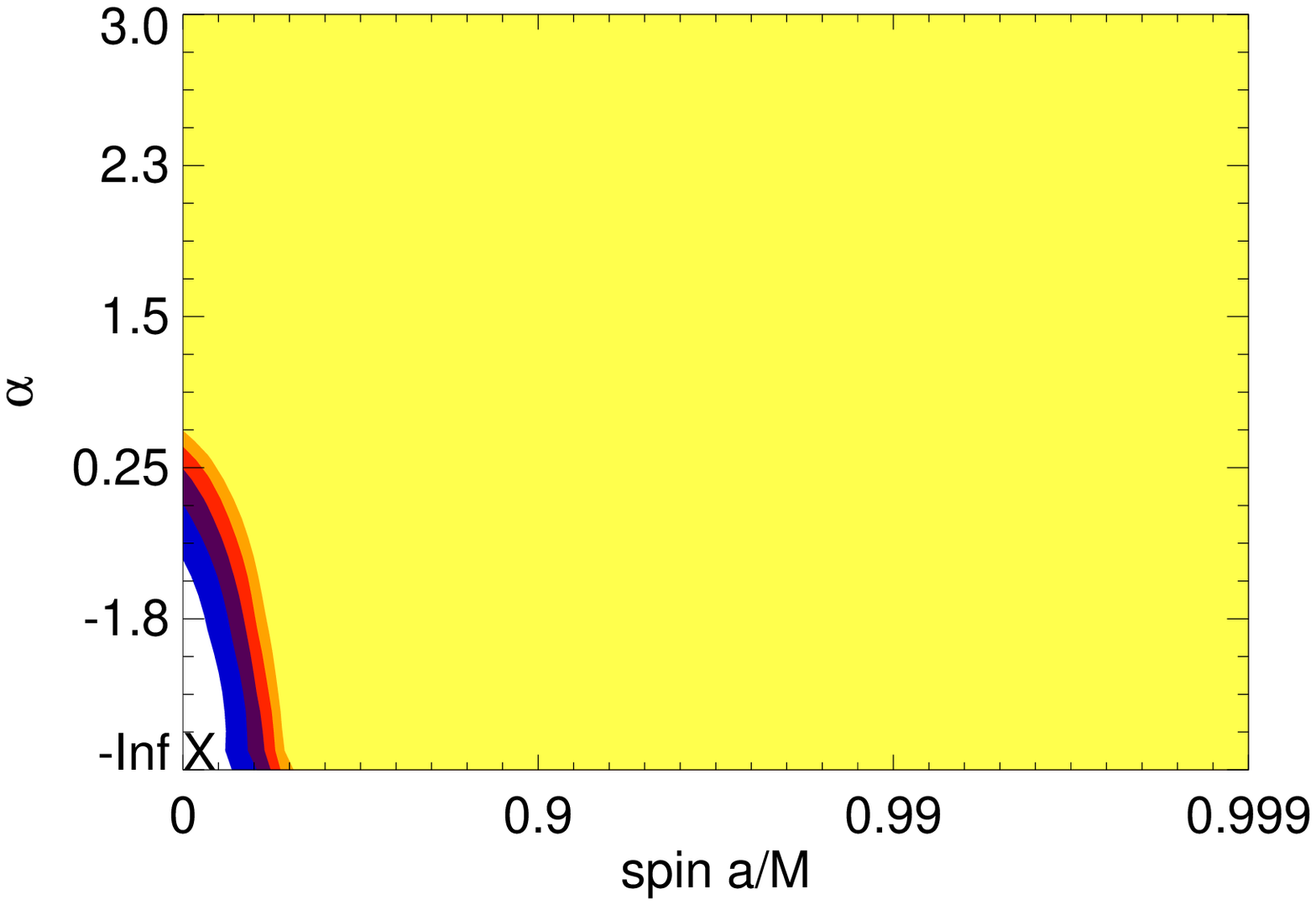}}
\scalebox{0.35}{\includegraphics*[145,415][540,720]{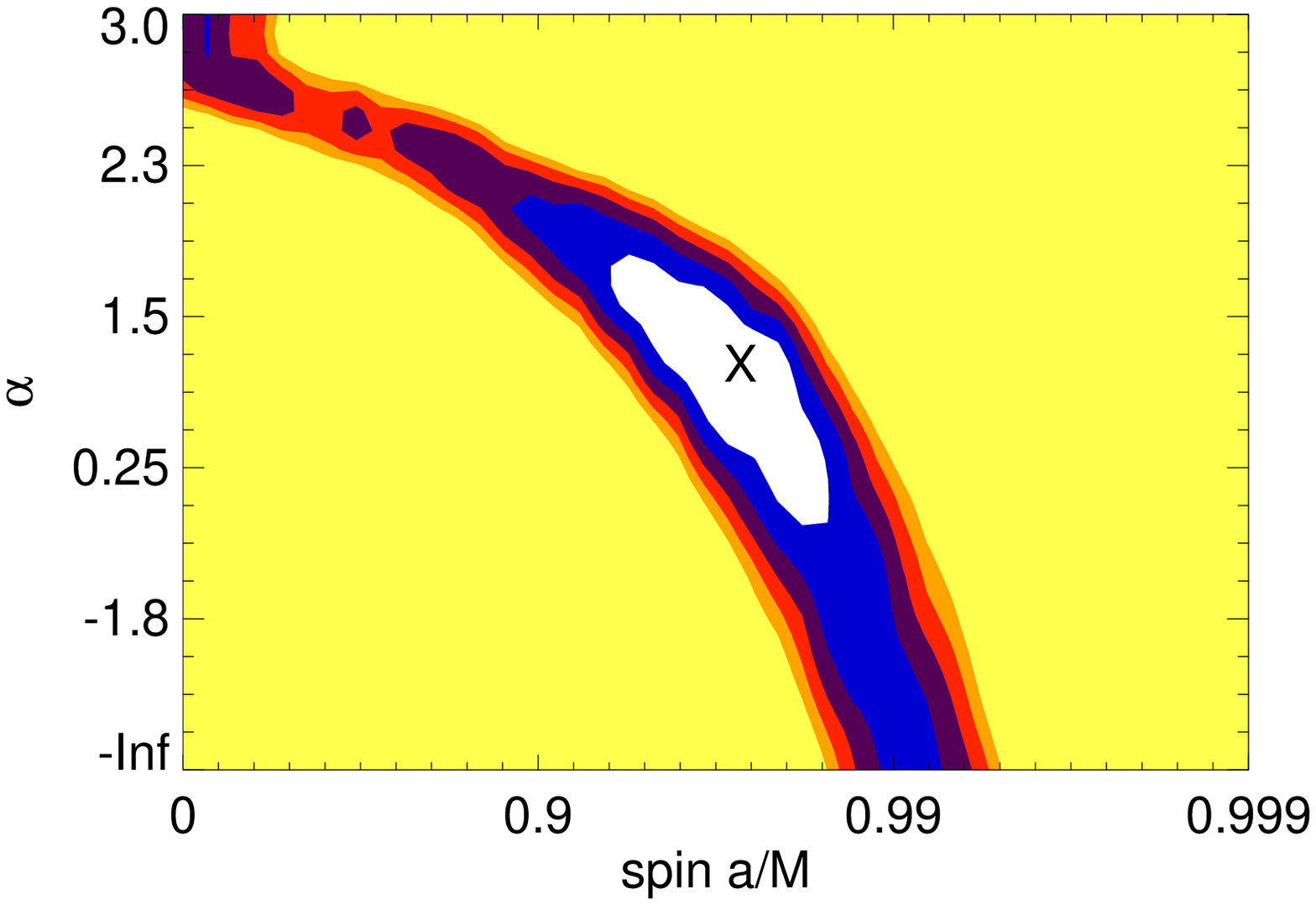}}
\scalebox{0.35}{\includegraphics*[145,415][540,720]{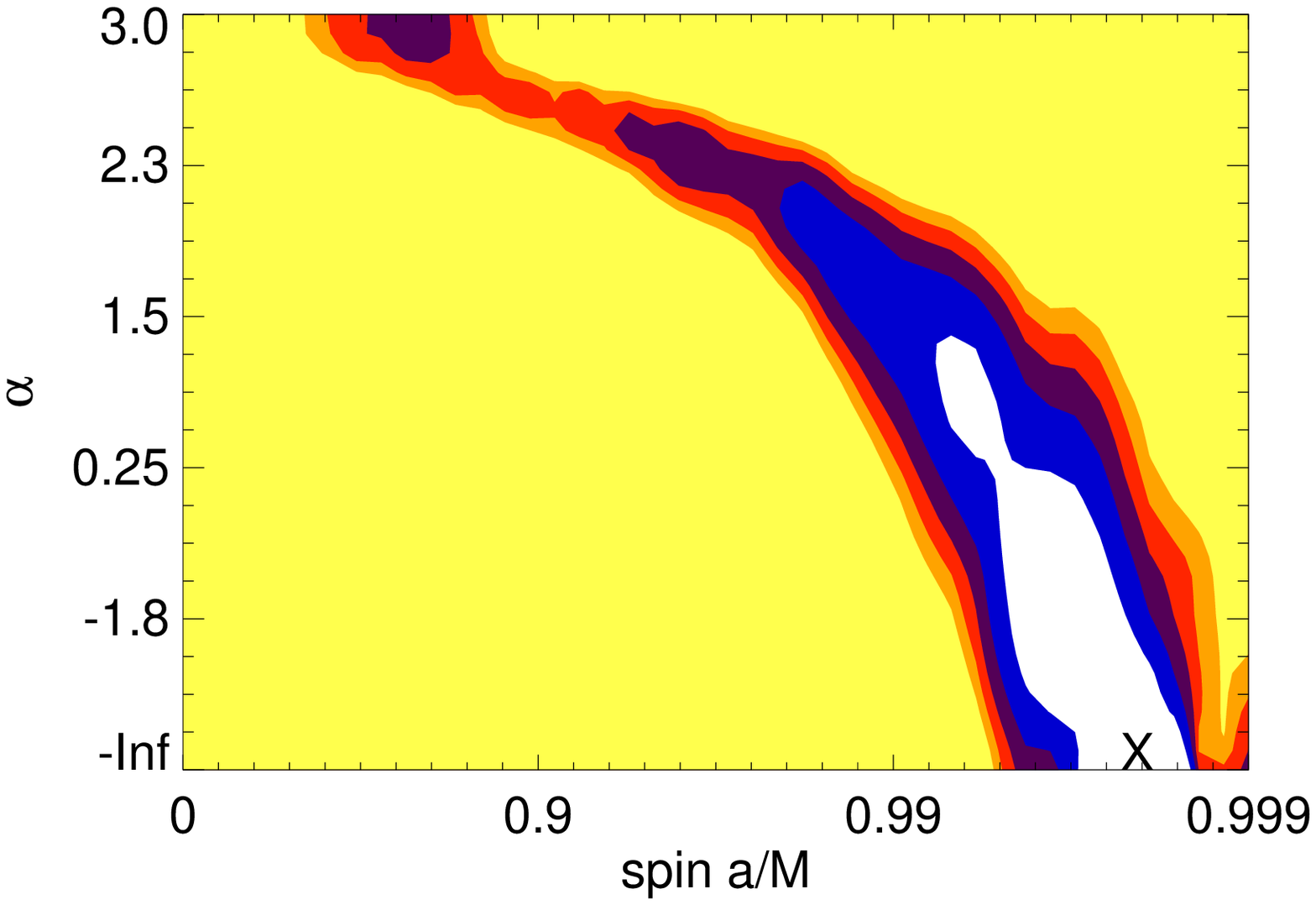}}\\
\scalebox{0.35}{\includegraphics*[52,360][540,720]{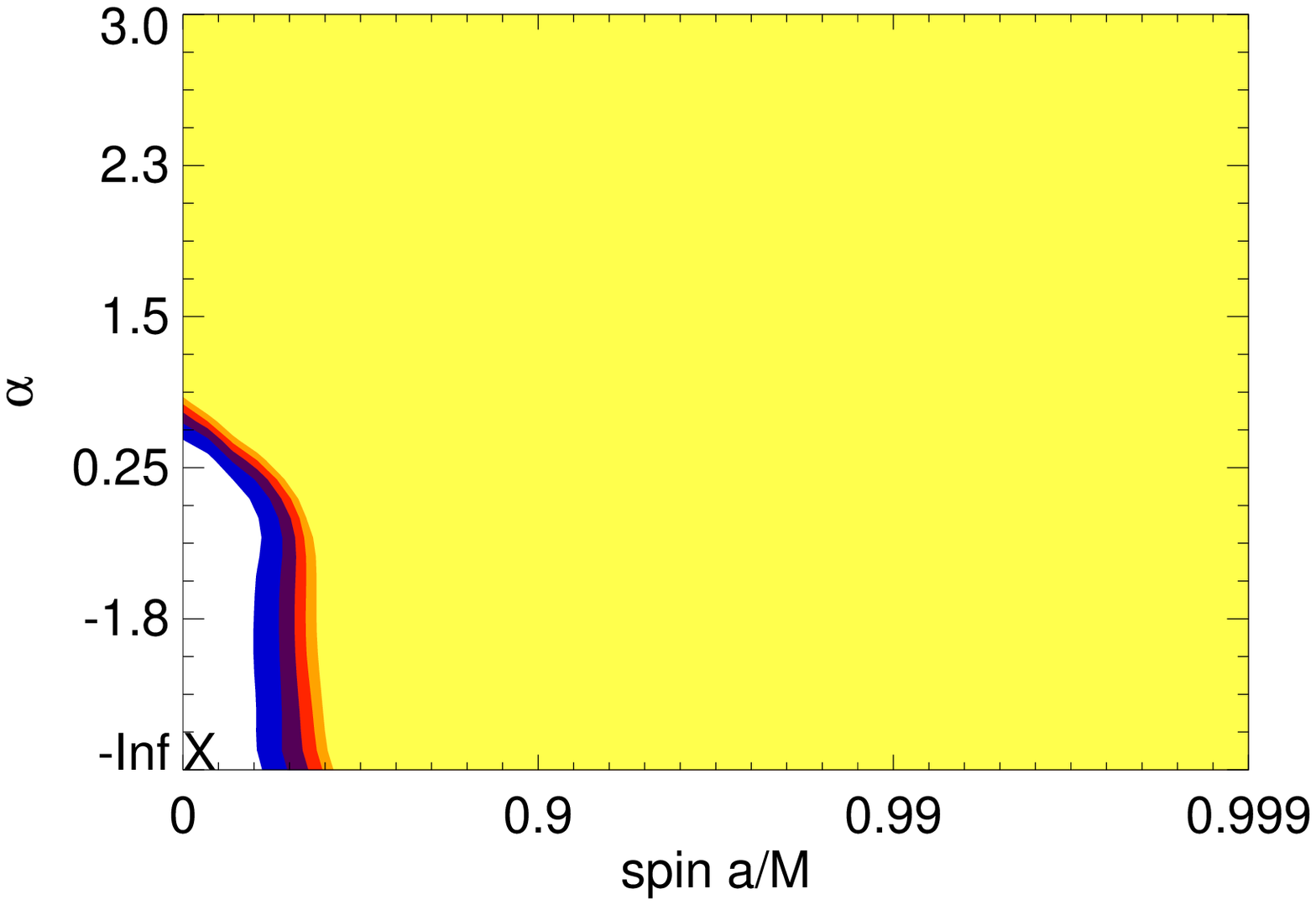}}
\scalebox{0.35}{\includegraphics*[145,360][540,720]{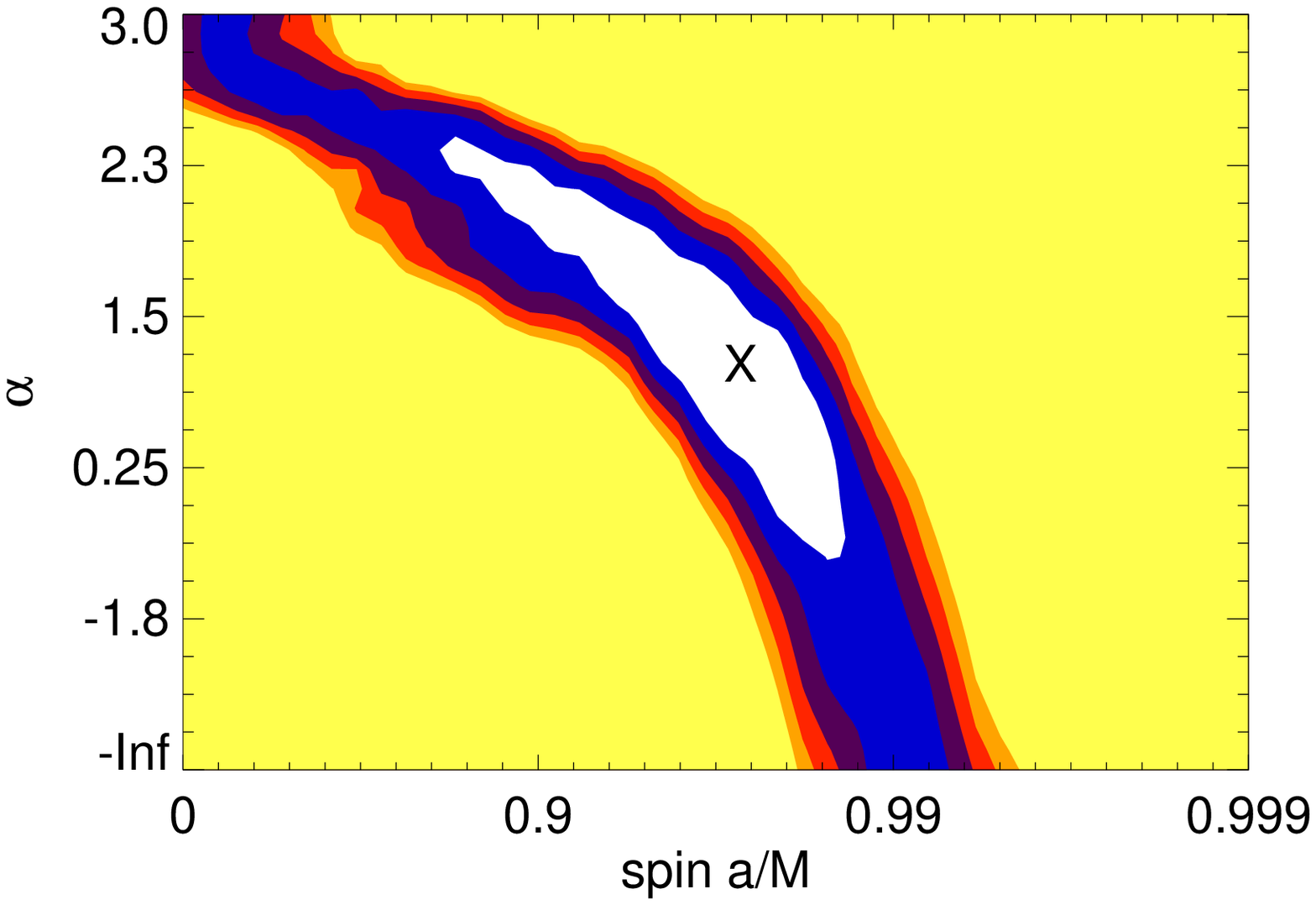}}
\scalebox{0.35}{\includegraphics*[145,360][540,720]{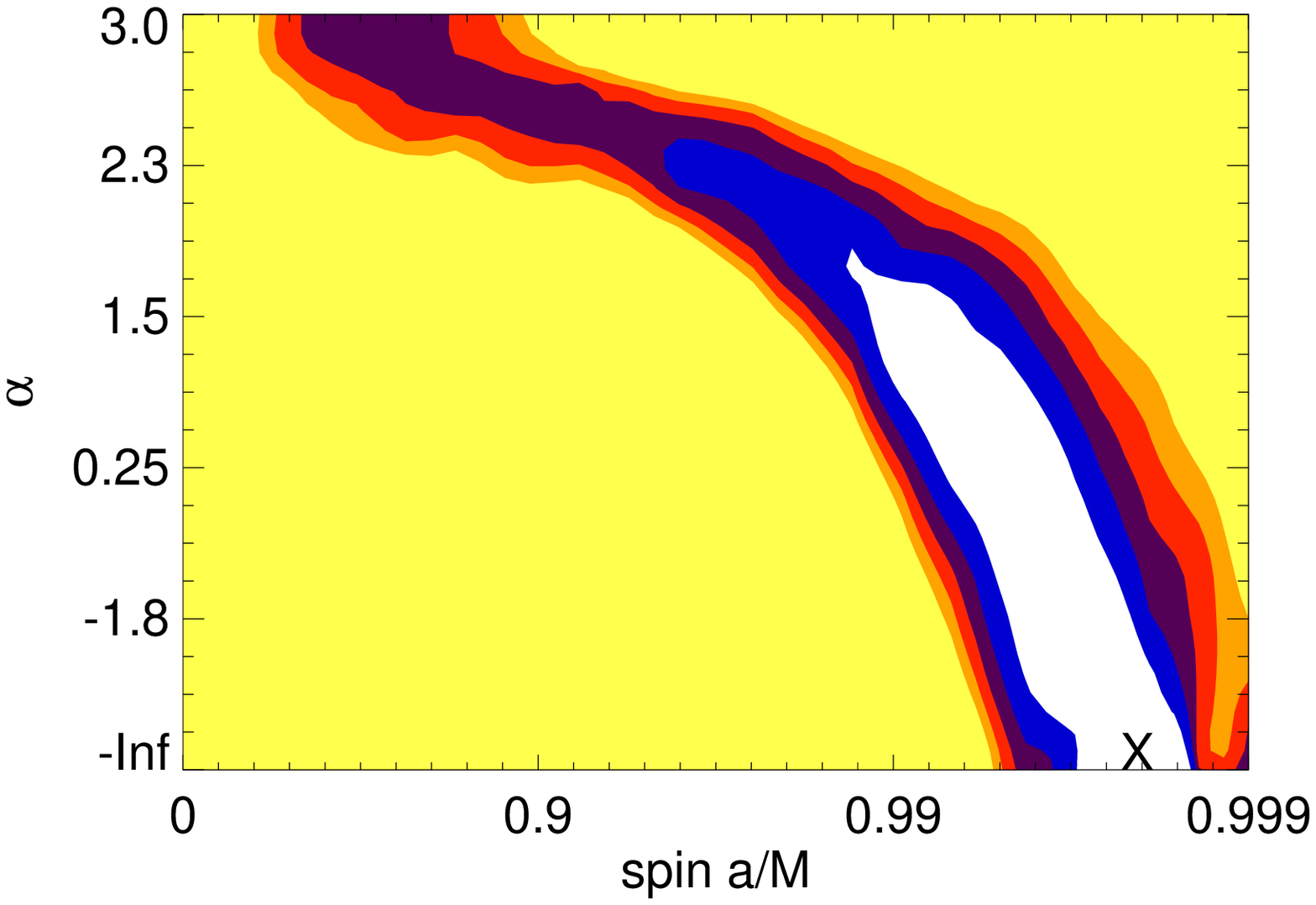}}
\end{center}
\end{figure}

Not surprisingly, a next-generation polarimeter with roughly a factor
of ten improvement in energy bandwidth, resolution, and collecting
area can do a much better job at measuring all these system
parameters. Figure \ref{gen2_cont} shows the same confidence contours
as Figure \ref{gen1_cont}, now for the next-generation
detector. Almost all degeneracy has been removed, and even when
we have no prior knowledge of the binary parameters, $a/M$, $\alpha$,
$i$, $L/M^2$, and $M/D$ can be determined with high precision. For the
three cases considered in the bottom row of Figure \ref{gen2_cont},
the inclination is recovered within $\sim 5^\circ$ and the accretion rate
within $\sim 20\%$:
\begin{subequations}
\begin{equation}
70^\circ \lesssim i \lesssim 80^\circ
\end{equation}
and
\begin{equation}
0.08 \lesssim \left(\frac{L}{L_{\rm Edd}}\right)\left(\frac{10
  M_\sun}{M}\right) \lesssim 0.12 \, .
\end{equation}
\end{subequations}

In fact, at these levels of polarization sensitivity, the ability to
measure the intrinsic properties of the system likely becomes more
dependent on the accuracy of the underlying emission model, rather
than on the signal-to-noise of the
observation. For example, the simplified form of the emission spectrum
may be modified by a more careful treatment of the disk atmosphere
\citep{davis:05}. Additionally, the low-energy polarization may be
strongly modified or reduced by absorption in the disk
\citep{laor:90}, Faraday rotation as the photons
pass through turbulent magnetic fields (S.\ Davis, private
communication), and even a small amount of inverse-Compton scattering
that is sometimes seen in the thermal state. 

\section{DISCUSSION}\label{discussion}

We have presented here the first results of a new Monte Carlo
ray-tracing code for calculating the X-ray polarization from
black holes. This code, described in detail in Paper I, is
most notable for its emitter-to-observer paradigm of radiation
transport, which allows for the inclusion of returning radiation and
electron scattering in a completely general accretion geometry. In
this paper we focus on polarization signatures of the thermal state
in stellar-mass black holes, a condition in which the disk is
optically thick and geometrically thin, and its opacity is
scattering-dominated. The emitted radiation has a
diluted thermal spectrum and is weakly polarized parallel to the disk
surface. The integrated polarization spectrum seen by a distant
observer contains
distinct energy-dependent features, as the high-energy photons from
the inner disk are modified by relativistic effects such as Doppler
boosting and gravitational lensing near the BH.

\begin{figure}
\caption{\label{spin_dependence2} Polarization degree and angle
  for a range of BH spin parameters.
  All systems have inclination
  $i=75^\circ$, BH mass $10 M_\odot$, luminosity $L/L_{\rm Edd}=0.1$, and
  Novikov-Thorne radial emission profiles.} 
\begin{center}
\scalebox{0.8}{\includegraphics{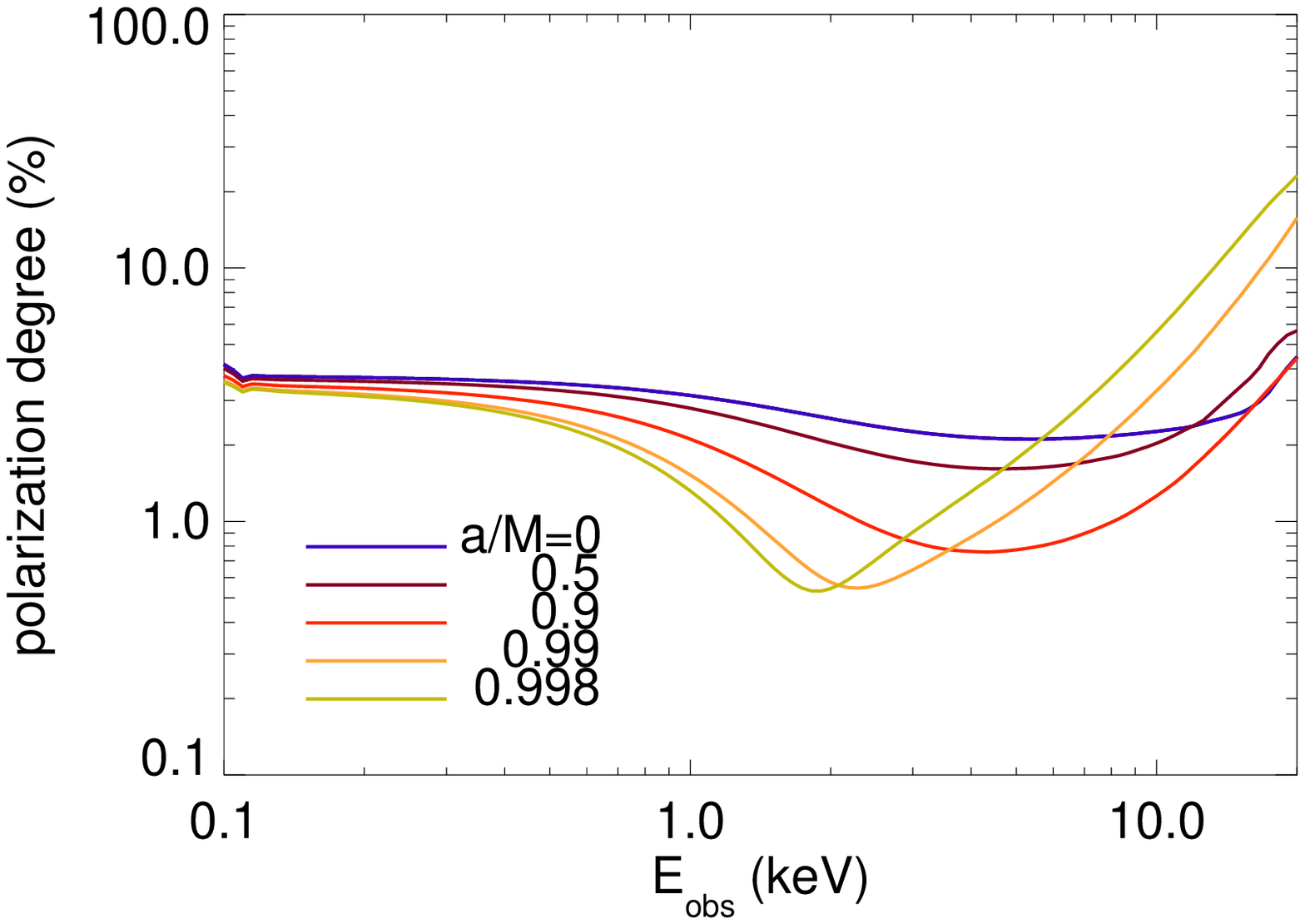}}
\scalebox{0.8}{\includegraphics{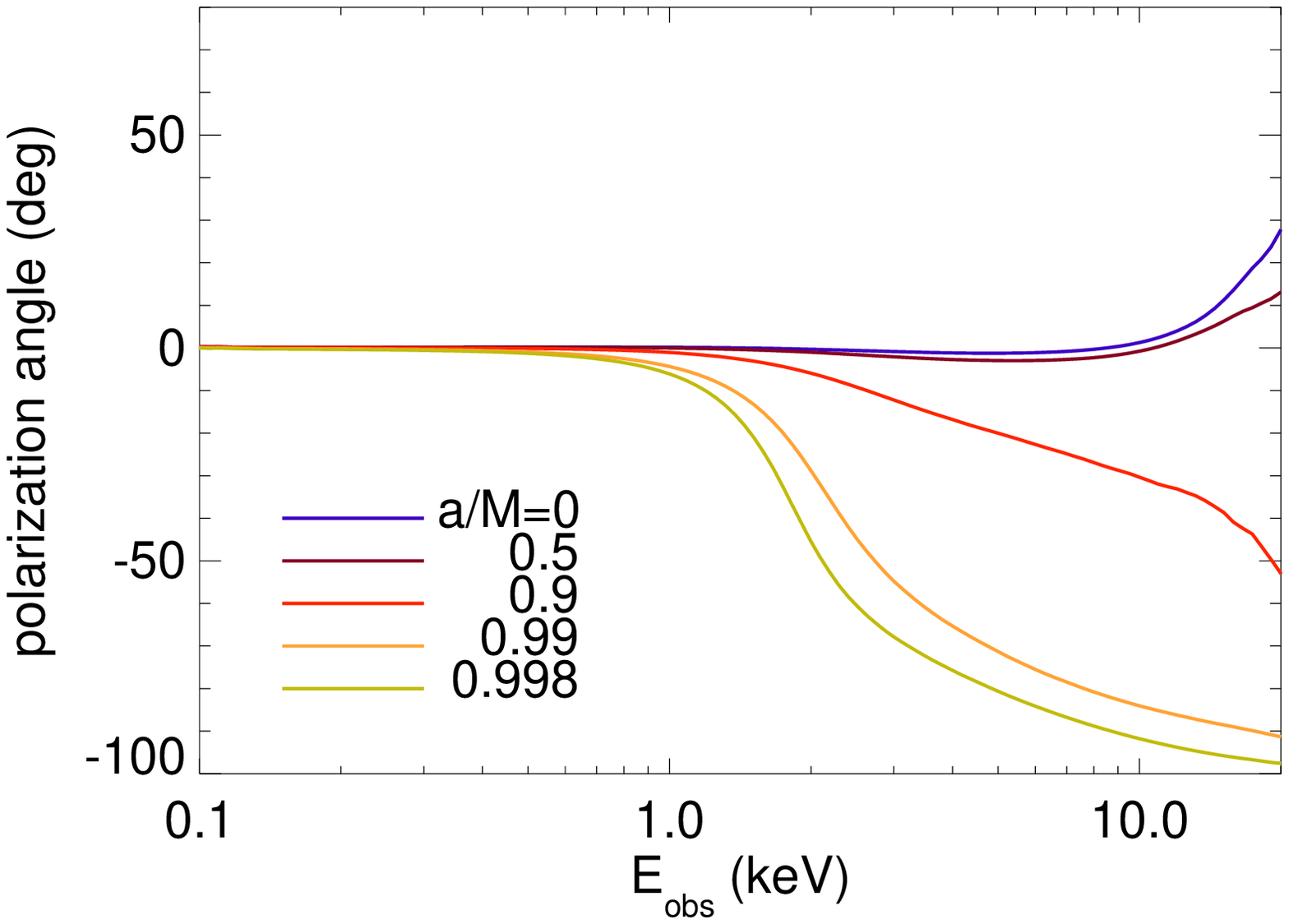}}
\end{center}
\end{figure}

\begin{figure}
\caption{\label{spin_dependence3} Same as Figure
  \ref{spin_dependence2} but for a power-law emissivity profile $\mathcal{F}(R)
  \sim R^{-3}$ all the way to the horizon.}
\begin{center}
\scalebox{0.8}{\includegraphics{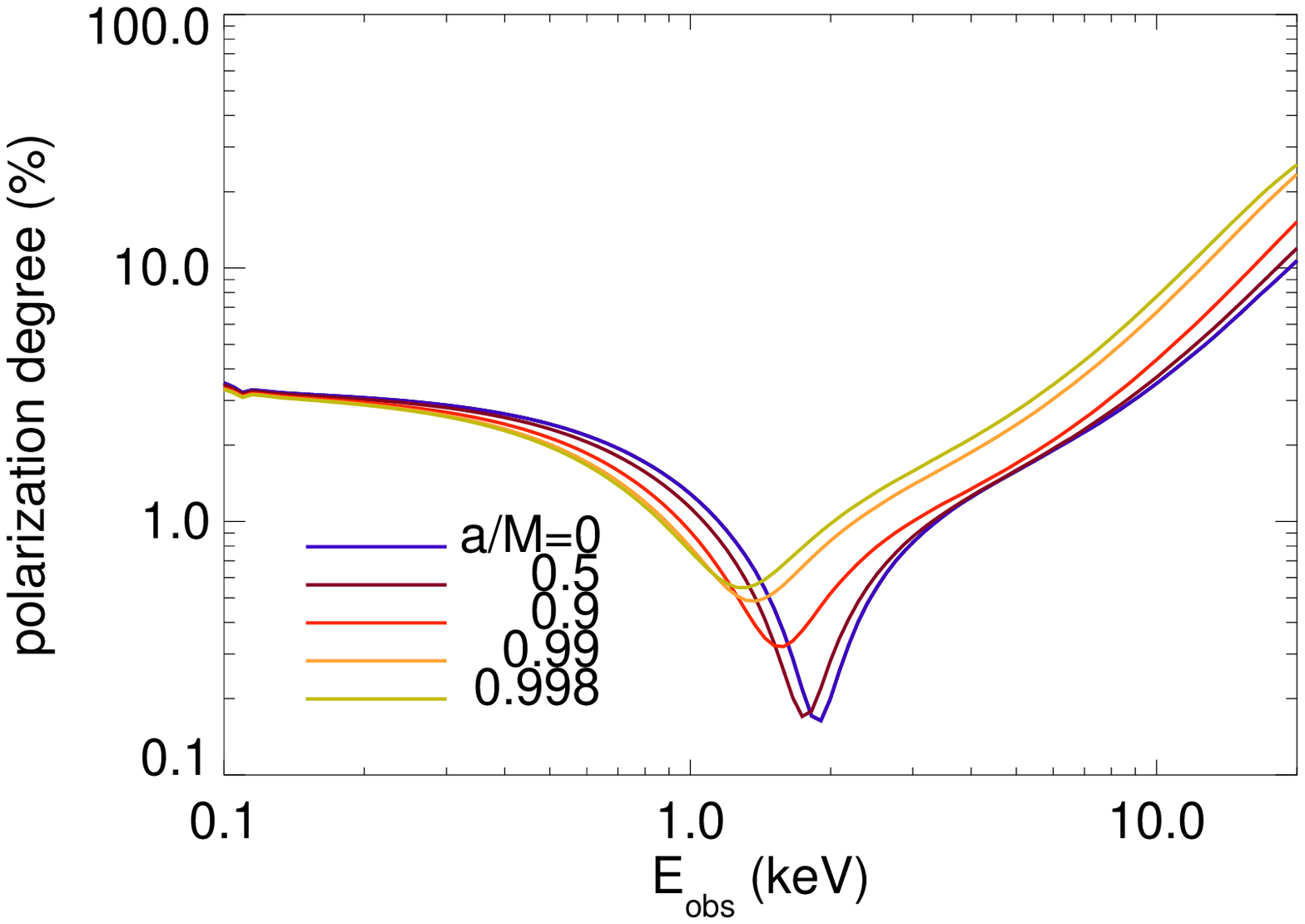}}
\scalebox{0.8}{\includegraphics{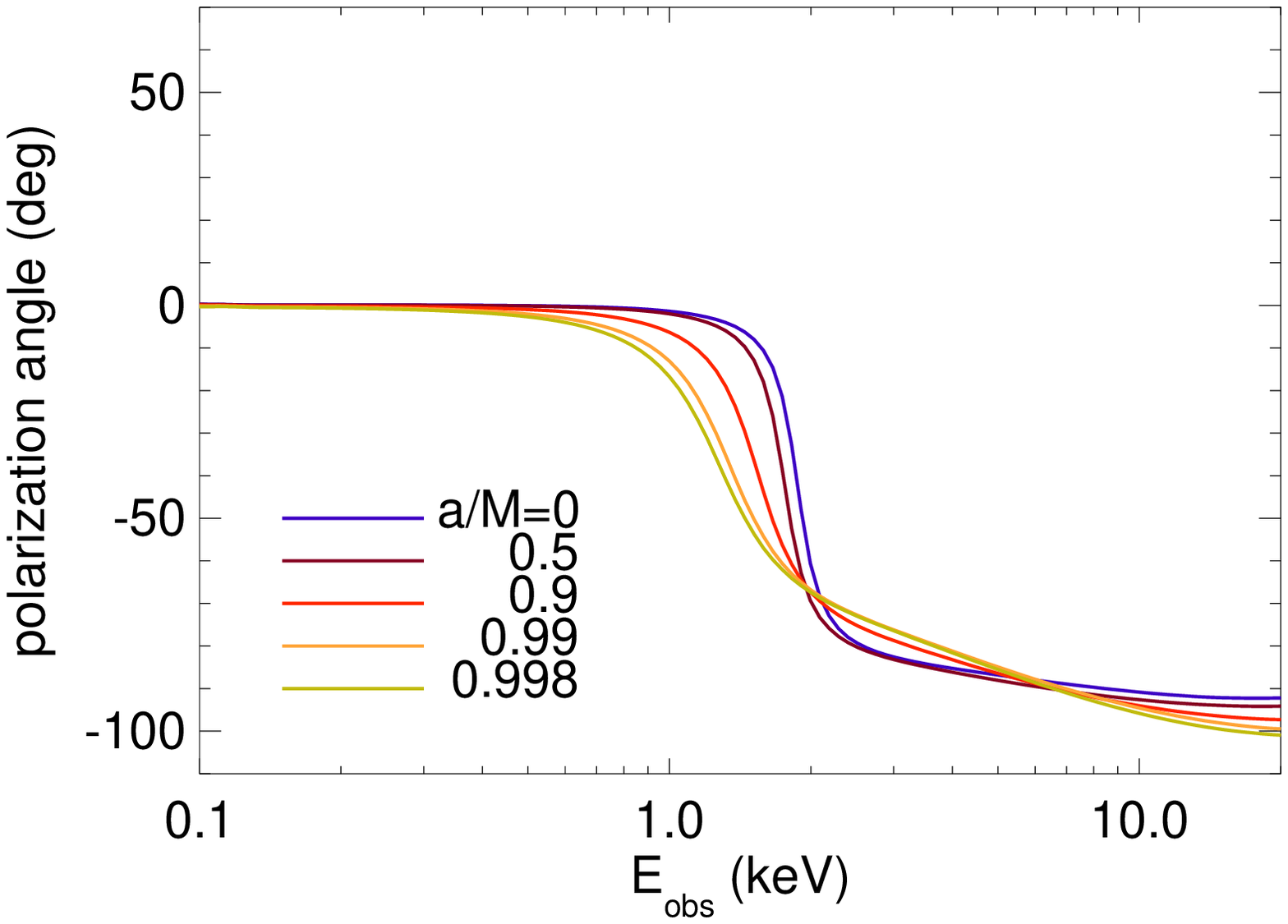}}
\end{center}
\end{figure}

For radiation originating very close to the BH, the most important
relativistic effect is the strong gravitational lensing that causes
the photons to get bent back onto the disk and scatter towards the
observer.  This scattering can induce very high levels of polarization,
especially at large inclination angles, and leads to a distinct
transition in the polarization angle from horizontal at low energies
to vertical above the thermal peak. Observing such a swing in
the polarization angle would give the most direct evidence to date for
the extreme relativistic light bending predicted around black
holes. Furthermore, by measuring the location and shape of this
transition, we will be able to constrain the temperature profile of the
inner disk.  If we assume a NT disk
with zero emission from inside the ISCO, the polarization transition
energy gives a direct measurement of the BH spin
(see Fig.~\ref{spin_dependence2}).  Alternatively, if we relax 
this assumption, the polarization can constrain models
for the dissipation profile and provide compelling evidence for
strong-field gravitational effects.

For BH systems where the mass, inclination, and distance are known
from other observations, polarization measurements would be comparable
in power to the continuum fitting method
\citep{gierlinski:01,davis:06,shafee:06} for the purpose of inferring the
emissivity profile of the inner disk.  However, when we lack
knowledge of any one of these priors, polarization provides a
significantly stronger tool than continuum fitting for constraining both
the shape of the emissivity profile and the unknown parameter(s).

In this paper, we have presented the first results from our
new black hole X-ray polarization analysis code, corresponding to
thermal emission from a geometrically thin accretion disk.  In future
work, we will extend its application to cases in which a
hot corona partially covers a cooler disk (as may be the case in AGN)
and to other accretion geometries, appropriate to other spectral
states of Galactic black hole binaries.

\newpage

\end{document}